\documentclass[aps,amsmath,twocolumn,prb,twocolumn,superscriptaddress,floatfix,citeautoscript]{revtex4-2} 

\usepackage{amsthm,amssymb,amsfonts,graphicx,verbatim, xcolor,bm,physics} 
\usepackage[colorlinks=true]{hyperref}
\usepackage[utf8]{inputenc} 
\usepackage{siunitx} 
\usepackage{cleveref}
\usepackage{multirow}
\usepackage{colortbl}

\newcommand{\beqn}{\begin{eqnarray}}
\newcommand{\eeqn}{\end{eqnarray}}

\newcommand{\br}{\bm r}
\newcommand{\bk}{\bm k}
\newcommand{\bq}{\bm q}

\newcommand{\bb}{\bm b}
\newcommand{\bR}{\bm R}

\usepackage{hyperref}

\usepackage{dsfont}

\usepackage{csquotes}
\MakeOuterQuote{"}

\begin{document}
\title{Helical trilayer graphene in magnetic field: Chern mosaic and higher Chern number ideal flat bands}
\author{Anushree Datta}
\affiliation{Universit\'e Paris Cit\'e, CNRS,  Laboratoire  Mat\'eriaux  et  Ph\'enom\`enes  Quantiques, 75013  Paris,  France}
\affiliation{Laboratoire de Physique des Solides, Universit\'e Paris Saclay, CNRS UMR 8502, F-91405 Orsay Cedex, France}

\author{Daniele Guerci}
\affiliation{Center for Computational Quantum Physics, Flatiron Institute, 162 5th Avenue, NY 10010}

\author{Mark O. Goerbig}
\affiliation{Laboratoire de Physique des Solides, Universit\'e Paris Saclay, CNRS UMR 8502, F-91405 Orsay Cedex, France}

\author{Christophe Mora}
\affiliation{Universit\'e Paris Cit\'e, CNRS,  Laboratoire  Mat\'eriaux  et  Ph\'enom\`enes  Quantiques, 75013  Paris,  France}

\begin{abstract}
Helical trilayer graphene (hTG) exhibits a supermoiré pattern with large domains centered around stacking points ABA and BAB, where two well-separated low-energy bands appear with different total Chern numbers at each valley, forming a Chern mosaic pattern. In the chiral limit, the low-energy bands become exactly flat at zero energy for magic-angle twists.
Here we investigate these zero-energy flat bands and their topological properties in the presence of a perpendicular magnetic field.
We show that hTG retains the precise flatness of the zero-energy bands, even at finite magnetic fields. We find topological phase transitions at fields corresponding to unit and half magnetic flux leading to an emergence of higher Chern number flat bands. Consequently the Chern mosaic gets modified for finite magnetic fields. We further find the analytical forms of zero-energy wave functions and identify a set of hidden wave functions, which gives crucial insights into both the topological transitions and enhancement of Chern numbers across them. We also find topological transitions away from the chiral limit with finite corrugations and at different magic angles.

\end{abstract}

\maketitle

Twisted moiré materials have opened a new avenue for studying the interplay of correlations and topology within their narrow energy bands \cite{Andrei2020}. In addition to observations of various correlated phases such as superconductivity \cite{Cao2018_1, Park2021,doi:10.1126/science.abg0399,Cao2021,Park2022}, correlated insulators \cite{Cao2018_2,Lu2019,Choi2019,Wong2020,Zondiner2020,Nuckolls2020,Liu2022}, heavy-fermionic phases \cite{PhysRevLett.129.047601,Datta2023,PhysRevLett.131.166501,rai2023dynamical,Guerci_2023,Ramires2021,Jiabin2023,merino2024evidence}, density waves \cite{Polshyn2022,Jiang2019}, moiré flat bands (FB) are also gaining significant attention as a platform to realize
 topologically non-trivial phases, like fractional Chern insulating phases \cite{Zeng2023,Cai2023,Xie2021,Jiabin2024,moralesduran2024} and quantum anomalous Hall states~\cite{doi:10.1126/science.aaw3780,Chen2020,Repellin2020}. 
Though magic-angle twisted bilayer graphene (TBG) stands out as the most studied moiré material, there is a recent increasing focus on exploring other configurations of twisted multilayer graphene due to their potential for enhanced tunability \cite{Uri2023,Chen2019,PhysRevLett.125.116404,Park2021, doi:10.1126/science.abg0399,Polshyn2022,lai2023imaging,Chen2021,PhysRevLett.123.026402,PhysRevB.107.125423,PhysRevLett.128.176403,PhysRevB.100.085109,PhysRevResearch.6.013165,PhysRevResearch.4.L012013,PhysRevLett.127.166802,ren2023tunable,xie2024strong,Zhang2022}.

One such configuration that has attracted recent attention both theoretically \cite{doi:10.1126/sciadv.adi6063,guerci2023chern,popov2023,PhysRevB.109.125141} and experimentally \cite{xia2023helical,PhysRevLett.127.166802}, is helical trilayer graphene (hTG). In hTG, three layers are twisted in the same direction relative to each other resulting in two incommensurate moiré patterns that interfere with each other and form a moiré of moiré or a supermoiré pattern~\cite{PhysRevB.107.125423, PhysRevLett.125.116404}. This supermoiré pattern exhibits local regions centered around different high symmetry stacking points (ABA, BAB, AAA), characterized by distinct local topological properties~\cite{popov2023}. 

Previous studies have revealed that the ABA and the BAB centered local regions host two central topological narrow bands with total valley Chern numbers $C_{tot}=1$ and $C_{tot}=-1$, respectively, even with finite corrugations \cite{guerci2023chern,doi:10.1126/sciadv.adi6063}. These regions are also favored by lattice relaxations allowing them to expand at the expense of the AAA centered regions forming a Chern mosaic pattern with large triangular domains around the ABA and BAB stackings \cite{doi:10.1126/sciadv.adi6063,PhysRevResearch.6.013165,nakatsuji2023multiscale,yang2023multimoire}. These Chern mosaic structures can be experimentally probed by measuring the local orbital magnetization, a technique already illustrated in the context of TBG nearly aligned with an hBN layer \cite{Grover2022,PhysRevB.102.121406,Cea2020}. Remarkably, in a special chiral limit, hTG demonstrates another characteristic feature, where the narrow bands become exactly flat at zero-energy separated by a large gap from the remote energy bands, when twists are tuned to magic-angle. 
These FB exhibit an ideal quantum geometry~\cite{PhysRevResearch.2.023237,PhysRevLett.127.246403,PhysRevResearch.3.023155,ledwith2022vortexability} and a correspondence with lowest Landau levels which extends to color entangled wavefunctions for higher Chern numbers~\cite{guerci2023chern,jiewang2023,dong2023}. 

On a different side, twisted moiré materials have also opened avenues for studies with magnetic fields \cite{Yu2022,Saito2021,Xie2021,Pierce2021,Kometter2023,Dean2013,sheffer2021,PhysRevB.106.L121111,singh2024topological}. 
This is because of relative large moiré length scale, 
which reduce the unit magnetic flux quantum by several orders of magnitude. For instance, in magic-angle TBG the unit flux quantum per moiré unit cell with a length scale of $\sim 12$ nm, corresponds to a magnetic field of $\sim 25$T\cite{PhysRevLett.129.076401,herzog-arbeitman2022}, in contrast to above $\sim 10 \, 000$ T in materials with typical length scales of a few Angstroms. This opens up the possibility for experimentally probing the Hofstadter spectra and their novel topological properties in twisted moiré materials.  

Motivated by these properties, we explore an interplay of a perpendicular magnetic field with the ideal FB of hTG and their local topological properties. We observe that the zero-energy ideal FB persist even in the presence of a finite magnetic field, similar to a finding in TBG \cite{sheffer2021,PhysRevB.103.155150}. We find that the magnetic field induces topological phase transitions in these zero modes, leading to emergence of higher Chern numbers ideal FB. 
By focusing on local stacking configurations of the supermoiré, ABA and BAB, we interestingly identify two distinct critical fields for the topological transitions occurring at unit and half flux quantum per moiré unit cell. 
For a given direction of the magnetic field, the two transitions appear at the ABA and BAB stackings, which modify the valley Chern mosaic pattern, and the transitions interchange when the direction of the magnetic field is reversed.  
We also find the exact analytical forms of the zero-energy wave functions
and identify a set of hidden wave functions.
We further demonstrate that these hidden wave functions play a key role in the underlying mechanism of the topological transitions and enhancement of Chern numbers in the FB. 

\begin{figure*}
\centering
    \includegraphics[width=0.9\textwidth]{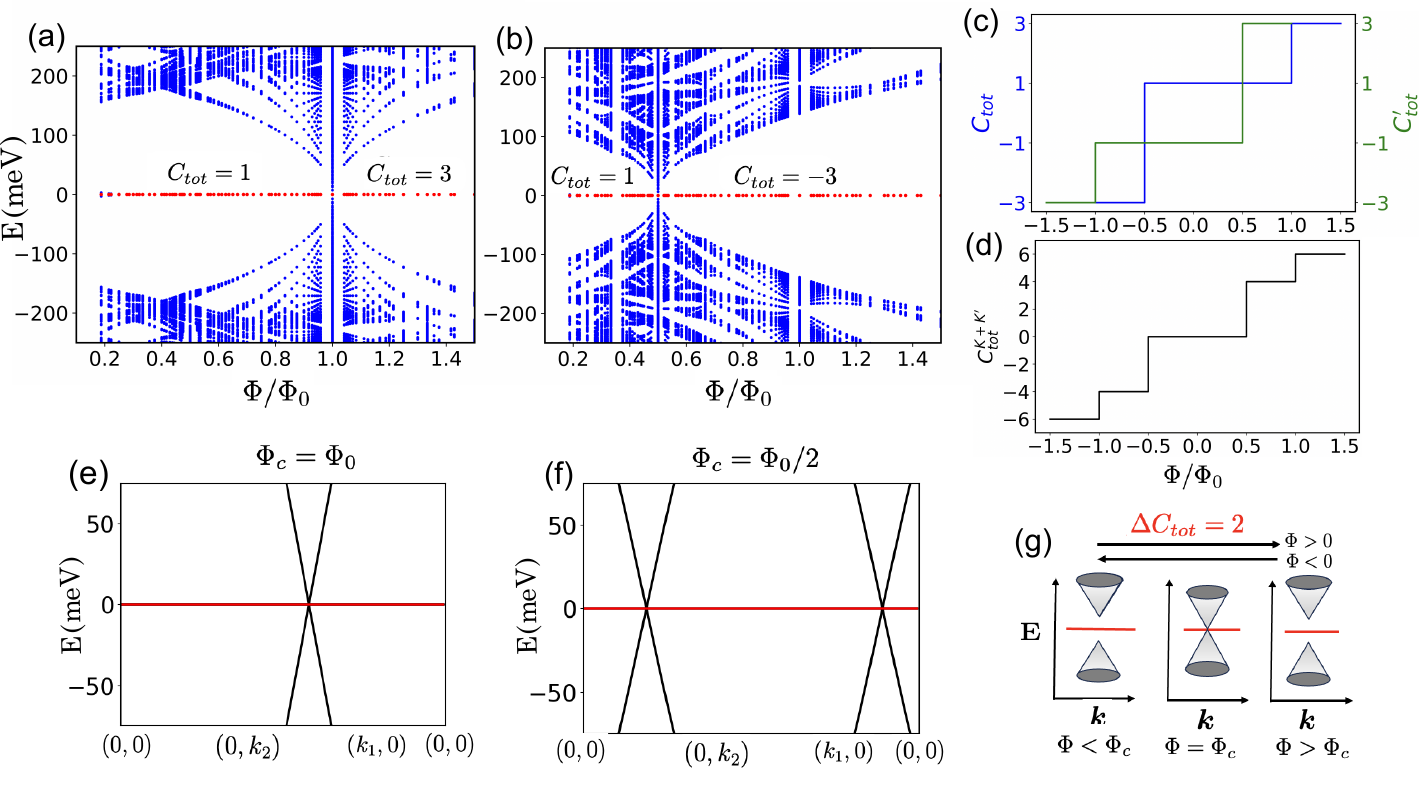}
\caption{Hofstadter spectra at the ABA stacking configuration defined with $\phi=(0, 2\pi/3, -2\pi/3)$ for positive (a) and negative (b) magnetic flux in valley $K$. (c,d) Change of total Chern numbers for valley $K$ ($C_{tot}$), $K^{\prime}$ ($C^{\prime}_{tot}$) (c) and including both valleys (d). Energy bands in the momentum space along $(0,0)\to (0,k_2)\to (k_1,0)\to (0,0)$ at $\Phi_c$ corresponding to panel a (e) and to panel b (f), with $k_1=\sqrt{3}k_{\theta}/2$ and $k_2 =3k_{\theta}/2$ defining the magnetic Brillouin zone. The zero-energy flat bands are represented by the red lines. (g) Schematic of gap closing through Dirac cones across topological transitions.}
\label{fig:fig1}
\end{figure*}

\section{Model}

hTG is described by a local continuum model where the moir\'e of moir\'e modulation is treated parametrically \cite{PhysRevB.107.125423}. Within this continuum model, the valley index is an emergent quantum number and a preserved quantity. We therefore restrict the following discussion to a single valley ($K$) since the model and spectra for the $K'$ valley are obtained from the former by time-reversal symmetry.
Introducing the magnetic field within a minimal coupling of the corresponding magnetic vector potential $\bm{A(\br)}$, the resulting local Hamiltonian for $K$ valley reads
\begin{eqnarray}\label{Ham0}
    \mathcal{H}(\br,\bm{\phi}) = \mathbf{1}_{3\times 3}\otimes  v_{F} [\hat{\bk} + e {\bm A}(\br)/\hbar ]\cdot{\bm \sigma} \nonumber \\+ \begin{pmatrix}
    0   & T(\br, \bm{\phi}) & 0\\
    h.c. & 0 & T(\br, -\bm{\phi}) \\
    0 & h.c. & 0
    \end{pmatrix}~,
\end{eqnarray}
in a basis $\Psi=\left(\psi_{1}, \chi_{1}, \psi_{2}, \chi_{2}, \psi_{3}, \chi_{3}\right)^{T}$, where $\psi_{l}$, $\chi_{l}$ live on A and B graphene sublattices with $l$ being the layer index. In Eq.~\eqref{Ham0} $v_F$ is the Fermi velocity in untwisted graphene layers, $\bm\sigma$ represents Pauli matrices in the sublattice space, and $\hat{\bk}=-i\grad_{\br}$. $T(\br, \bm{\phi})=\sum^{3}_{j=1} T_{j}e^{-i\bq_j.\br}e^{-i\phi_j} $ describes the interlayer moiré tunneling with $T_{j+1}=w_{AA}\sigma_{0}+w_{AB}[\cos\left({2\pi j/3}\right)\sigma_{x}+\cos\left({2\pi j/3}\right)\sigma_{y}]$, $w_{AB}=110$ meV, 
and $\bq_{j+1}= ie^{2i\pi j/3}$ are moiré pattern wave vectors in the unit of $k_{\theta}$. $k_{\theta}=2K_{D}\sin(\theta/2)$, where $\theta$ is the twist angle between successive layers and $K_{D}$ is the Dirac momentum. Here $\bm{\phi}=\left(\phi_1, \phi_2, \phi_3\right)=\left(0, \phi_0, -\phi_0\right)$, with $\phi_0$ being the parameter that maps out different local stacking configurations ABA, BAB, AAA with values $2\pi/3$, $-2\pi/3$ and $0$, respectively \cite{guerci2023chern}. Due to lattice relaxations in hTG, the regions around ABA and BAB stacking points tend to expand in size, while significantly shrinking the regions around AAA stacking \cite{doi:10.1126/sciadv.adi6063,nakatsuji2023multiscale}. Hence ABA and BAB stackings play the most crucial roles for determining the properties of hTG. In this work, we mainly focus on ABA stacking with $\phi=(0, 2\pi/3, -2\pi/3)$ in Eq.~\eqref{Ham0} in most of our discussions unless mentioned. The corresponding properties of BAB stacking regions can be obtained by using the $C_{2z}\mathcal{T}$ symmetry along with a reversal of magnetic field orientation: $C_{2z}\mathcal{T} \mathcal{H}(\br, {\bm\phi}_{ABA}, \bm{A})(C_{2z}\mathcal{T})^{\dagger}=\mathcal{H}(-\br, {\bm\phi}_{BAB},-\bm{A})$ (See Appendix.~\ref{symmetryhTTG}).
We also demonstrate our findings mostly in valley $K$, noting that the other valley follows from time-reversal symmetry up to a sign change in the magnetic field direction (Eq.~\eqref{time_reversal_K}).
Furthermore, we neglect the Zeeman effect and therefore assume spin degeneracy in the model.

In the absence of a magnetic field, the zero-energy FB are obtained in the chiral limit, defined by $w_{\rm AA}=0$~\cite{PhysRevLett.122.106405,Oskar_RG_2020}, for a series of magic angles~\cite{guerci2023nature, popov2023magic,doi:10.1126/sciadv.adi6063}.
For finite magnetic fields, we first solve the Hamiltonian Eq.~\eqref{Ham0} in the chiral limit numerically by projecting it to low energy Landau levels~\cite{PhysRevB.84.035440,PhysRevB.100.035115,PhysRevB.106.L121111} (see Appendix.~\ref{sec:numericaldetails}) to analyze their energy bands and their topological properties (Sec.~\ref{numerics}), focusing on the first magic angle $\theta=1.687^{\circ}$~\cite{guerci2023chern,doi:10.1126/sciadv.adi6063}. The chiral limit also allows us to analytically probe the wave functions of zero-energy states. Hence we analytically find out the zero-energy wave functions in Sec.~\ref{analytics} to gain further insights into their topological properties, again at the first magic angle. We then discuss the departure from the chiral limit in the presence of finite corrugations in Sec.~\ref{finitecorr}. Furthermore, noting that the even and odd index magic angles show distinct properties \cite{guerci2023nature}, we also discuss the case of the second magic angle in Sec.~\ref{2ndmagic}.

\section{Numerical results}\label{numerics}
\subsection{Energy spectrum and topological phase transitions}
We begin by demonstrating the characteristics of the energy spectra of Eq.~\eqref{Ham0} obtained numerically. In Fig.~\ref{fig:fig1} (a,b), we show the Hofstadter spectra showcasing the energy eigenvalues ($\rm E$) for ABA stacking region in valley $K$ with varying magnetic flux $\Phi$ per moiré unit cell, relative to unit quantum flux $\Phi_0=h/e$. Panel (a) and (b) depict the spectrum for positive and negative flux, respectively. We see that the zero-energy states, denoted by red dots, persist for finite magnetic flux values. They originate from the zero-energy FB at zero flux. Moreover, we notice that the spectra in (a) and (b) are clearly different.

Now focusing on the remote energy bands, separated from the zero-energy states by approximately $\Delta\approx\pm 130$ meV at low $\Phi$, we see that for both positive and negative fluxes, they progressively shift towards zero energy with increasing $\Phi$. 
The remote band gap closes at a critical flux $\Phi_c$ in both (a) and (b). For $\Phi>\Phi_c$, the remote band gap reopens.
Since the magnetic field introduces an additional length scale $l_B$, the magnetic length, the shift in remote band energy can be semi-quantitatively estimated to be of the order of $\Delta A_M/l^{2}_B$, where $A_M$ is the moiré unit cell area. 
Therefore at gap closings the shift $\Delta A_M/l^{2}_B\sim \Delta$, implying that $\Phi_c$ should be of the order of $\Phi_0$. Notably, we find that for positive flux $\Phi_c=\Phi_0$ and for negative flux $\Phi_c=\Phi_0/2$. 

To investigate the topological properties of the zero-energy states, we compute the associated Chern numbers. 
In the chiral limit all zero-energy states are sublattice polarised \cite{PhysRevLett.122.106405}. Hence Chern numbers associated to these states can be classified as $C_A$ and $C_B$ for A and B sublattices, respectively.
For $\Phi<\Phi_c$, we find that the zero energy states maintain Chern numbers of $C_{A}=2$ and $C_{B}=-1$, identical to those at zero flux~\cite{guerci2023chern} as they remain separated by the remote band gap.
However, for $\Phi>\Phi_c$, we find $C_{A}=3$ ($C_B = -3$) and no B(A)-polarized states
for the positive (negative) flux direction.
Hence the total Chern number $C_{tot}=C_{A}+C_{B}$ is $C_{tot}=1$ for $\Phi<\Phi_c$ ($=\Phi_0$) and $C_{tot}=3$ for $\Phi>\Phi_c$ ($=\Phi_0$) in (a), and $C_{tot}=1$ for $\Phi<\Phi_c$ ($=\Phi_0/2$) and $C_{tot}=-3$ for $\Phi>\Phi_c$ ($=\Phi_0/2$) in (b). 
The changes in $C_{tot}$ thus indicate topological transitions happening at $\Phi_c$. It is remarkable that we find two distinct topological transitions in (a) and (b), i.e., at two directions of the magnetic field, as also shown in (c). In the opposite valley $K^{\prime}$, the response to the magnetic field is opposite with respect to the field's orientation. The sign of the Chern numbers also reversed. The sign of the Chern numbers also reversed: $C_{A/B}\to -C_{A/B}$. So, for the positive flux direction, the phase transition at $\Phi_{0}/2$ happens with the total Chern number $C^{\prime}_{tot}$ changing from $C^{\prime}_{tot}=-1$ to $C^{\prime}_{tot}=3$, and for the negative flux, the transition occurs at $\Phi_0$ with $C^{\prime}_{tot}=-1$ to $C^{\prime}_{tot}=-3$ (c). Hence the total Chern number including both valleys $C^{K+K^{\prime}}_{tot}=C_{tot}+C^{\prime}_{tot}$ undergoes changes from $|C^{K+K^{\prime}}_{tot}|=0$ to $|C^{K+K^{\prime}}_{tot}|=4$ at $\Phi_{0}/2$ and $|C^{K+K^{\prime}}_{tot}|=4$ to $|C^{K+K^{\prime}}_{tot}|=6$ at $\Phi_0$, in both directions of the magnetic field, as depicted in (d). We thus note that the asymmetric response in the two field directions is balanced when both valleys are taken into account.

The change in Chern number across $\Phi_c$ can also be deduced by inspecting the energy bands at $\Phi_c$ where the moiré translational invariance is recovered~\cite{herzog-arbeitman2022}.
We find that the gap closings in $\Phi_c$ in (a) and (b) occur through one and two Dirac cones, respectively, as shown in (e, f). Each half of a Dirac cone is associated to Chern numbers of $\pm 0.5$ per valley. Consequently, at each valley $C_{tot}$ changes by $\pm 2$ at $\Phi_c$, as illustrated in the schematic (g).

\begin{figure}
\centering
    \includegraphics[width=0.45\textwidth]{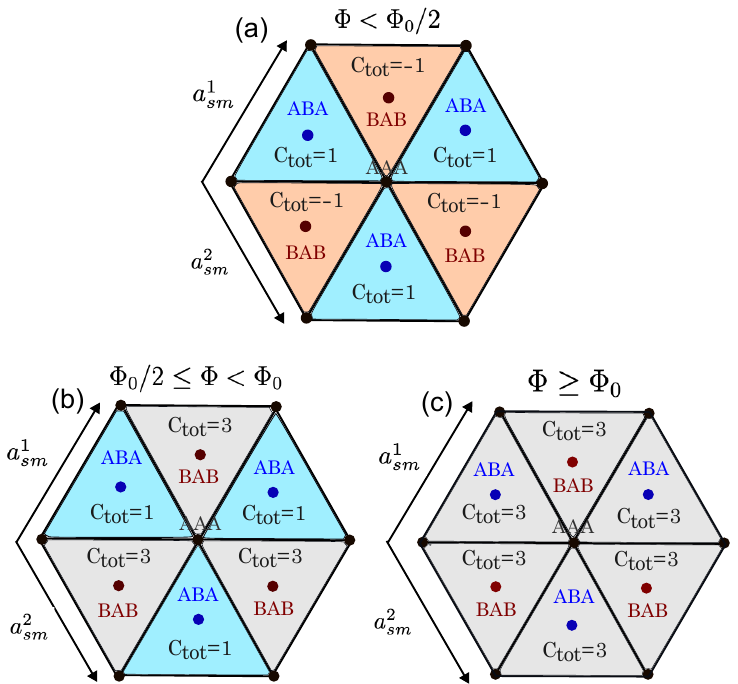}
\caption{Valley $K$ Chern mosaic in the supermoiré unit cell with unit vectors $a^{1}_{sm}$ and $a^{2}_{sm}$ ($|a^1_{sm}|=250$ nm) for positive magnetic flux $\Phi$.}
\label{fig:fig2new}
\end{figure}

So far, we have illustrated our findings at the ABA stacking configuration. The corresponding spectrum and the associated topological transitions at the BAB stacking in the same valley are exactly opposite with respect to the direction of the magnetic flux, as obtained by the $C_{2z}\mathcal{T}$ symmetry. The opposite response can also be physically understood by noting that exchanging the top and bottom layers (equivalent to reversing the magnetic field direction) at the ABA stacking results in a moiré pattern looking identical to the one in the BAB stacking configuration \cite{doi:10.1126/sciadv.adi6063, guerci2023nature}. The sign of the Chern numbers at the BAB stacking is also reversed along with an exchange of the two sublattices: $C_{A/B}\to -C_{B/A}$. Consequently, for a given direction of the magnetic flux, local Chern number pattern in the supermoiré unit cell, i.e., the Chern mosaic evolves with magnetic field. This evolution 
is showcased at different positive magnetic field values for valley $K$ in Fig.~\ref{fig:fig2new}, for $\Phi<\Phi_0/2$ in (a), for $\Phi_{0}/2\leq\Phi<\Phi_0$ in (b), and $\Phi\geq\Phi_0$ in (c). 
The Chern mosaic in (a) is the same as in Refs.~\onlinecite{guerci2023chern, doi:10.1126/sciadv.adi6063} at $\Phi=0$. In (b), while the $C_{tot}$ for ABA stacking remains the same as in (a), $C_{tot}$ of the BAB stacking changes from $C_{tot}=-1$ to $C_{tot}=3$ due to transition at $\Phi_0/2$. In (c) the ABA stacking configuration undergoes a topological transition at $\Phi_0$ whereas the $C_{tot}$ at BAB remains the same as in (b). 
We note that $\Phi_0$ corresponds to a magnetic field of $68$T at the magic angle $\theta=1.687^{\circ}$.

\subsection{Chern/Sublattice polarisation}\label{subpol} 
\begin{figure}
\centering
    \includegraphics[width=0.49\textwidth]{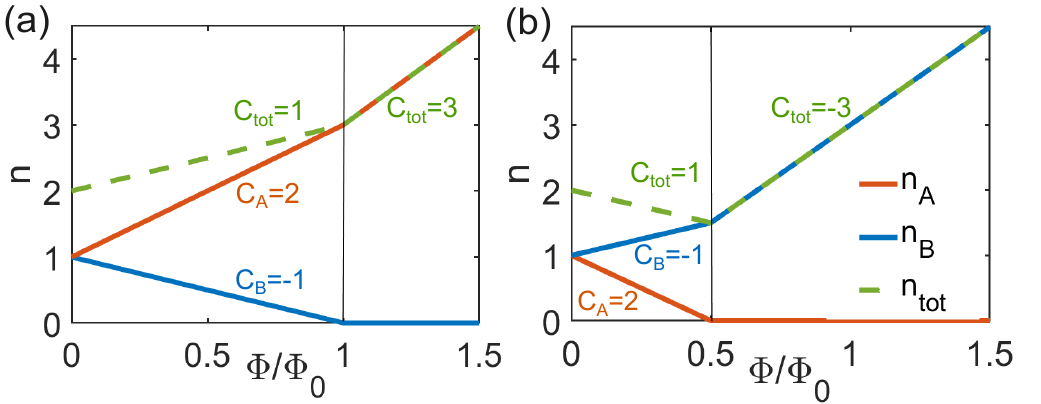}
\caption{Number of zero modes per moiré unit cell in A-sublattice $n_A$ (red lines) and B-sublattice $n_B$ (blue lines), and $n_{tot}=n_{A}+n_{B}$ (green dashed lines) as a function of positive magnetic flux (a) and negative magnetic flux (b), at the ABA stacking [$\phi=(0, 2\pi/3, -2\pi/3)$] and valley $K$. Vertical lines denote topological phase transitions.} 
\label{fig:fig2}
\end{figure}

To gain more insights into the origin of the topological phase transitions, we numerically investigate the counting of zero-energy states at each sublattice.
Fig.~\ref{fig:fig2} depicts the number of zero-energy states per moiré unit cell in the two sublattices $n_{A/B}$  for $\Phi>0$ (a) and $\Phi<0$ (b) at the ABA stacked configuration in valley $K$. We observe that $n_{A/B}$ varies with $\Phi$ as $n_{A/B}=1\pm C_{A/B}(\Phi)\Phi/\Phi_0$ for $\Phi<\Phi_c$ and $n_{A/B}= \pm C_{A/B}(\Phi)\Phi/\Phi_0$ for $\Phi>\Phi_c$, following the Streda formula~\cite{Streda1982}, with $\pm$ indicating the direction of the magnetic flux. 
Since $C_B=-1$ and $C_A=2$ for $\Phi<\Phi_c$, for positive flux (a) the B-sublattice gets emptied at $\Phi_0$, and for negative flux (b) the A-sublattice gets emptied at $\Phi_0/2$, as seen from the Streda formula.
Therefore the topological transitions are attributed to the depletion of one of the two sublattice Chern bands, and two different $\Phi_c$ values reflecting different $C_A$ and $C_B$ at $\Phi=0$.
We further notice that $n_{A}-n_{B}=3\Phi/\Phi_0$ holds for all $\Phi$, a consequence of Atiyah-Singer index theorem for Dirac kernels~\cite{Atiyah1968,sheffer2021,crepel2024topologically} further discussed in Appendix~\ref{sec:ASindextheorem}.
So at $\Phi>\Phi_c$, the sublattice band that remains with a finite number of zero-energy states, acquires a higher Chern number of $|C_{tot}|=3$ to offset the emptied sublattice band and satisfy the index theorem.
The Chern value $C_{tot}=3$ reflects the Landau level zero modes of the three decoupled Dirac cones to which the phase at $\Phi > \Phi_c$ is continuously connected.

\section{Analytical expressions of zero-energy wave functions}\label{analytics}

We now look into the properties of the zero-energy modes analytically by deriving the expressions of the corresponding wave functions. The analytical forms can provide more insights into our numerical results discussed thus far. Chiral TBG exhibits a Landau level correspondence~\cite{PhysRevResearch.2.023237,PhysRevResearch.3.023155,PhysRevLett.127.246403}, {\it i.e.} a mapping between the zero-energy FB and lowest Landau level wavefunctions~\cite{estienne2023}, which survives at non-zero magnetic fields~\cite{sheffer2021}. A similar correspondence is also observed for chiral hTG and extends to the concept of color wavefunctions for Chern $2$ bands~\cite{jiewang2023,dong2023}, corresponding roughly to a sum of Landau levels~\cite{guerci2023chern}. We show below that the Landau level correspondence continues to hold for chiral hTG at finite magnetic fields.

Here we again focus on the ABA stacking for positive and negative magnetic field directions in valley $K$. 
The local Hamiltonian Eq.~\eqref{Ham0} in the chiral limit can be cast to a form which anticommutes with the chiral operator in the sublattice space, by a rearrangement to the ``Chern-sublattice'' basis \cite{PhysRevLett.122.106405,LEDWITH2021168646,PhysRevResearch.2.023237,Liu2019,bultinck2020} $\Psi=\left(\psi_{1}, \psi_{2}, \psi_{3}, \chi_{1}, \chi_{2}, \chi_{3}\right)^{T}$, and is given by:
\begin{equation}
    \frac{\mathcal{H}^{ABA}(\br,{\alpha})}{v_{F}k_{\theta}} = \begin{pmatrix}
        0 & \mathcal D_b (\br) \\ 
        \mathcal D^\dagger_b (\br) & 0 
    \end{pmatrix},
    \label{Ham}
\end{equation}
with
\begin{eqnarray}\label{chiraloperator}
    {\mathcal  D_b} =\begin{pmatrix}
    -2i\partial_b & U_{\omega}(\br) & 0 \\
    U_{0}(-\br) & -2i\partial_b & U_{0}(\br)\\
   0 & U_{\omega}(-\br) & -2i\partial_b
    \end{pmatrix}~,    
\end{eqnarray}
where $\partial_b=\left(\partial - b \Bar{z}/4\right)$, $\Bar{\partial}_b=-\left(\Bar{\partial} + b z/4\right)$, $\partial=(\partial_{x}-i\partial_{y})/2k_{\theta}$, and $z=k_{\theta}(x+iy)$. The moiré potential is $U_0(\br)=\alpha \sum^{3}_{j=1} e^{-i\bq_j\cdot\br}$, $U_\omega(\br)=\alpha \sum^{3}_{j=1} \omega^{j-1}e^{-i\bq_j\cdot\br}$, with $\omega=e^{2 i \pi/3}$. 
$b=\frac{p}{q}\frac{2\pi}{A_M}$ quantifies the magnetic field as inverse square of the magnetic length, and is related to $\bm{A(\br)}$ in the symmetric gauge: 
$\bm{A(\br)}=\frac{b}{2}(y\hat{\mathbf{x}}-x\hat{\mathbf{y}})$, and $p$ and $q$ being coprime integers, and $A_M$ is the moiré unit cell area. The field $b$ generates a magnetic flux of $\Phi=bA_M$ per moiré unit call, and hence $\Phi/\Phi_0 = p/q$. In Eq.~\eqref{Ham}, we introduce a superscript $ABA$ in the Hamiltonian and also an additional parameter $\alpha=w_{AB}/\left(v_{F}k_{\theta}\right)$ which casts the Hamiltonian in a dimensionless form. 

The zero-mode flat band solutions of Eq.~\eqref{Ham} is equivalent to finding solutions to following zero-mode equations: 
\begin{align}
  \label{zeromode-1}
\mathcal D_b (\br) {\bm \chi}_{\bk}(\br) = 0, \\[2mm]  \label{zeromode-2}
\mathcal D^{\dagger}_b (\br) {\bm \psi}_{\bk}(\br) = 0,
\end{align}
for all $\bk$ in the magnetic moiré Brillouin zone (mBZ), where the spinors ${\bm \chi}_{\bk}(\br)$ and ${\bm \psi}_{\bk}(\br)$ correspond to A and B sublattice wave functions, respectively. Additionally, the wave functions must obey the Bloch periodicity with the magnetic translations operators $T_1^q$ and $T_2$ translating by $q\mathbf{a_1}$ and $\mathbf{a_2}$, respectively, where $\mathbf{a_1}=4\pi/3k_{\theta}\left(0,-1\right)$ and $ \mathbf{a_2}=4\pi/3k_{\theta}\left(\sqrt{3}/2,1/2\right)$ are the moiré unit vectors.
The conditions are:
\begin{subequations}\label{boundary-conditions}
\begin{align}
    {\bm \psi}_{\bk}(\br + q {\bf a}_1) & =    U_{\varphi}{\bm \psi}_{\bk}(\br) \, e^{\frac{b q}{4} (\bar{a}_1 z - a_1 \bar{z})} e^{i q {\bf k} \cdot {\bf a}_1}  \\
     {\bm \psi}_{\bk}(\br + {\bf a}_2) & =  U_{\varphi}{\bm \psi}_{\bk}(\br) \, e^{\frac{b}{4} (\bar{a}_2 z - a_2 \bar{z})} e^{i {\bf k} \cdot {\bf a}_2}
\end{align}
\end{subequations} 
with $U_{\varphi}=\rm diag[\omega^{\ast},1,\omega]$.
We note that the zero-mode solutions of Eqs.~\eqref{zeromode-1} and \eqref{zeromode-2} can be cast to the form 
\begin{subequations}\label{finitebwf}
\begin{align}
    {\bm \psi}_{\bk}(\br)=f(z)e^{-b|z|^2/4}\bm \psi_{\bk_{sym}}(\br)  \\
     {\bm\chi}_{\bk}(\br)=\Bar{f}(\bar{z})e^{b|z|^2/4}{\bm\chi}_{\bk_{sym}}(\br)
\end{align}
\end{subequations} 
where $z$ is the complex notation of $\br$, $f(z)$ and $\Bar{f}(\bar{z})$ are functions holomorphic and anti-holomorphic in $z$ respectively, and $\bm\psi_{\bk_{sym}}$, ${\bm\chi}_{\bk_{sym}}$ are zero-mode solutions at $b=0$. It is also possible, and we shall do it for $C_A =2$ bands, to form linear combinations of these solutions. The momenta $\bk_{sym}$ are arbitrary and we find it convenient to use $\Gamma$, $K$, and $K^{\prime}$.

\subsection{$C_{B}=-1$ zero modes}\label{Bsub}
We first focus on the solutions of Eq.~\eqref{zeromode-1} corresponding to the B-sublattice for $b> 0$. We obtain the following explicit analytical form of the zero modes: 
\begin{equation}\label{psiB}
 {\bm \chi}_{n,\bk}(\br)={\Bar{\varphi}}_{n, \bk}(z,\Bar{z}) {\bm \chi}_{\Gamma}(\br),
\end{equation}
where $\mathbf{\chi}_{\Gamma}$ is a zero-mode solution at $b=0$ and ${\Bar{\varphi}}_{n, \bk}(z,\Bar{z})$ is related to the meromorphic function (up to a factor of $e^{b|z|^2/2}$) 
\begin{eqnarray}\label{basis}
{\varphi}_{n, \bk}(z,\Bar{z}) = \frac{\vartheta_1\left[\frac{q-p}{q}\frac{z}{a_1} - \frac{k}{b_2}+\frac{\omega}{q} \left(n+\frac{p}{2} \right),\frac{q-p}{q}\omega\right]}{\vartheta_1\left[\frac{z}{a_1},\omega\right]} \nonumber \\ e^{i k_1 \frac{z}{\Bar{a}_1}} e^{\frac{2 i \pi}{q a_1} \left(n+\frac{p}{2}\right) z} e^{b ({z}^2 + |z|^2)/4} \,
\end{eqnarray}
with ${\Bar{\varphi}}_{n, \bk}(z,\Bar{z})={\varphi}^{\ast}_{n, \bk}(-z,-\Bar{z})$ and solves
\begin{equation}\label{LLmap}
    \left( \partial - \frac{b\bar{z}}{4}\right) \Bar{\varphi}_{n, \bk}(z,\Bar{z}) = 0,
\end{equation}
with $n=1,\ldots , q-p$. Here $k_{1}=\bk\cdot{\bf a}_{1}$, $a_1$ and $b_2$ are the complex number versions of vectors ${\bf a}_1$ and ${\bf b}_2$, respectively, with ${\bf b}_{2}=\sqrt{3}k_{\theta}(1,0)$ being a reciprocal lattice vector, and $\omega=a_{2}/a_{1}=e^{i2\pi/3}$. 
The function $\vartheta_1(z,\omega)$ represents the Jacobi theta functions having the properties 
$\vartheta_{1}[z\pm 1, \omega]=-\vartheta_{1}[z,\omega]$ and $\vartheta_{1}[z\pm\omega, \omega]=-e^{-i\pi\omega}e^{\mp 2i\pi z}\vartheta_1[z,\omega]$ (also see Appendix.~\ref{thetafunc}). Consequently they exhibit zeros at $z=n_1 a_1+n_2 a_2$ with $n_1$ and $n_2$ as integers, ensuring magnetic Bloch periodicity in ${\bar\varphi}_{n, \bk}(z,\Bar{z})$. 
Note that the zeros in $\vartheta_1$ also lead to poles in  ${\bar\varphi}_{n, \bk}(z,\Bar{z})$. However, as seen in previous studies \cite{guerci2023chern,doi:10.1126/sciadv.adi6063}, the $b=0$ spinor $\bm{\chi}_{\Gamma}(n_1 a_1+n_2 a_2)$ vanishes at the magic angle. Hence the poles in ${\bar\varphi}_{n, \bk}(z,\Bar{z})$ get compensated by the spinor ${\bm \chi}_{\Gamma}(\br=n_1 a_1+n_2 a_2)$ and ${\bm \chi}_{\bk, n}(\br)$ become finite at the magic angle for all $\br$.  

Since index $n$ in $\chi_{n,\bk}$ runs from $1$ to $q-p$, the number of FB consisting of zero-energy states per magnetic unit cell is $q-p$. To connect to the count of zero-energy states in moiré unit cell that we discuss in Sec.~\ref{subpol}, we recall that the magnetic unit cells are spanned by $q{\bf a}_1$ and ${\bf a}_2$, and thus include $q$ moiré unit cells. Hence the number of zero-energy states per moiré unit cell is $1-p/q=1-\Phi/\Phi_0$ and they are depleted at topological transition $\Phi = \Phi_0$, consistent with the Streda formula for $n_B$, as described in Fig.~\ref{fig:fig2}a.
 
Interestingly, even though ${\bm \chi}_{\bk, n}(\br)$ is depleted at $\Phi = \Phi_0$, we find one zero-energy solution for a single $\bk$ value (but notably not a flat band) from Eq.~\eqref{zeromode-1} that obeys the magnetic Bloch periodicity with the analytical expression
\begin{equation}\label{psik0}
     {\bm \chi}_{\bk_0}(\br) = \frac{e^{i \bk_0 \cdot {\bf a}_1 \frac{\Bar{z}}{\Bar{a}_1}} e^{\frac{i \pi \Bar{z}}{ \Bar{a}_1}} e^{b (\Bar{z}^2 + |z|^2)/4} }{\vartheta_1\left[-\frac{\Bar{z}}{\Bar{a}_1},\omega^{*}\right]}  \, {\bm\chi}_\Gamma (\br)~,
\end{equation}
at a specific momentum $\bk_0=(\mathbf{b_2}-\mathbf{b_1})/2$, where $\mathbf{b_1}=\sqrt{3}k_{\theta}(1/2, -\sqrt{3}/2)$ is a reciprocal lattice vector. The single zero-mode pairs up with another state from the A-sublattice, as discussed later in Sec.~\ref{hiddenwf}, to form a Dirac cone at $\bk_0$ (also in Fig.~\ref{fig:fig1}e). 

For $b<0$, we find that the analytical form of the wave functions have the same form as Eq.~\eqref{psiB}, but with $b\to -b$, and therefore the number of magnetic FB is $q+p$. Drawing similar counting argument of zero-energy states per moiré unit cell, we have $n_B=1+p/q=1+\Phi/\Phi_0$, agreeing with the Streda formula as in Fig.~\ref{fig:fig2}b. 
We further observe that corresponding magnetic Bloch wave functions $u_{n,\bar{k}}(\br)={\bm \chi}_{n,\bk}(\br)e^{-i\bk.\br}$ of Eq.~\eqref{psiB} are $k$-antiholomorphic, and hence they constitute an ideal flat band \cite{PhysRevResearch.2.023237,PhysRevLett.127.246403,ledwith2022vortexability}.

\subsection{$C_{A}=2$ zero modes}
We now focus on the solutions of Eq.~\eqref{zeromode-2} corresponding to the A-sublattice for $b>0$. In this case, the $b=0$ spinor ${\bm \psi}_{\Gamma}(0)$ is non zero, but ${\bm \psi}_{K}(\br)$ and ${\bm \psi}_{K^{\prime}}(\br)$ becomes asymptotically colinear at $\br=0$ at the magic angle \cite{guerci2023chern}. For $b\neq 0$, the wave functions of zero modes are then found to be
\begin{eqnarray}\label{Chern_2_wfq_1}
{\bm \psi}_{n,\bk}(\br)=a_{n, k+q_{1}}{\varphi}^{1}_{n,\bk-q_{1}} (\br) {\bm \psi}_{K}(\br) -\nonumber \\ a_{n,k-q_{1}} {\varphi}^{1}_{n,\bk+q_{1}} (\br) {\bm \psi}_{K'}(\br),
\end{eqnarray}
Here ${\varphi}^{1}_{n,\bk}(\br)$ is related to the basis function
\begin{eqnarray}\label{function_basis}
    \Tilde{\varphi}_{n,\bk} (\br) &=& \frac{\vartheta_1\left[s_1 \frac{z}{a_1} - \frac{k}{b_2}+\frac{1}{2} + \frac{\omega}{q} \left(n+\frac{p}{2} \right),s_1 \omega\right]} {\vartheta_1\left[s_3 \frac{z}{a_1},s_3 \omega\right]}  \nonumber \\ &\times&\vartheta_1\left[s_2 \frac{z}{a_1},s_2  \omega\right]e^{i k_1 \frac{z}{a_1}} e^{\frac{2 i \pi}{q a_1} \left(n+\frac{p}{2}\right) z} e^{-b (z^2 + |z|^2)/4}\nonumber\\
\end{eqnarray}
with $s_1=s_3=1$ and $s_2=p/q$, that satisfies the magnetic Bloch periodicity and solves
\begin{equation}
 \left( \bar\partial + \frac{b z}{4}\right) \tilde{\varphi}_{n, \bk}(z,\Bar{z}) = 0,
\end{equation}
when $s_1 + s_2 - s_3 = \frac{p}{q}$. 
We set the center of the mBZ to be at $\Gamma$, and thus $K$ and $K^\prime$ are at $\pm \bq_1$.
The coeffcients in Eq.~\eqref{Chern_2_wfq_1} are given by $a_{n,k}=\vartheta_1\left[\frac{k}{b_2}-\frac{1}{2}- \frac{\omega}{q} \left(n+\frac{p}{2} \right),\omega\right]$, and with $\psi_K(0)=\psi_K^{\prime}(0)$ we see that they cancel the singularities of ${\varphi}^{1}_{n,\bk}$ at $z=n_1 {a}_{1}+n_2 {a}_{2}$. 
We further find $2p$ independent solutions obeying the magnetic Bloch periodicity, given by:  
\begin{eqnarray}\label{Chern_2_wfq_2}
    &&{\bm \psi}_{n,\bk}(\br) = {\varphi}^{2}_{n,\bk-q_{1}}(\br) {\bm \psi}_K(\br),\quad \nonumber \\&&{\bm \psi}_{n,\bk}(\br) = {\varphi}^{2}_{n,\bk+q_{1}}(\br) {\bm \psi}_{K'}(\br),
\end{eqnarray}
where the function basis corresponds to Eq.~\eqref{function_basis}, with $s_2=s_3=1$ and $s_1=p/q$. 
We therefore obtain $q+2p$ flat magnetic bands or $1+2p/q=1+2\Phi/\Phi_0$ zero-energy states per moiré unit cell at the A-sublattice, consistent with the Streda formula for $n_A$, featured in Fig.~\ref{fig:fig2}a.

For $b<0$, the A-sublattice wave functions take the form 
\begin{equation}\label{Chen2_wf_nB}
   {\bm \psi}_{\bk}(\br) = \sum_{n=1}^{q-p} \Big[ a_{n,k}  \varphi_{n,k-q_1} (\br) {\bm \psi}_{K}(\br) + b_{n,k} \varphi_{n,k+q_1} (\br) {\bm \psi}_{K'}(\br) \Big],
\end{equation} 
where $\varphi_{n,k}(\br)$ are the same basis functions defined in Eq.~\eqref{basis}. 
The $2 p - 2 q$ coefficients $a_{n,k},b_{n,k}$ can be obtained
with a set of conditions to compensate the poles at $z=n_1 a_1 + n_2 a_2$: $\sum_{n=1}^{q-p} [ a_{n,k} \varphi^{\prime}_{n,k-q_1} (s a_1) e^{i s \bq_1 \cdot {\bf a}_1} + b_{n,k}  \varphi^{\prime}_{n,k+q_1} (s a_1) e^{-i s \bq_1 \cdot {\bf a}_1}] =0$
for $s=1,2,\ldots,q$, where $\varphi^{\prime}(z)=\varphi(z)\vartheta(z/a_1, \omega)$.
The removal of the other poles follows from the periodicity of ${\bm \psi}_{\bk}(\br)$ with $q {\bf a}_1$ and ${\bf a}_2$.  
Although explicit expressions of $a_{n,k},b_{n,k}$ are not found, we can still find the independent solutions corresponding to Eq.~\eqref{Chen2_wf_nB} in the following way.
Forming a vector with all the unknown coefficients, $U = (a_{1,k},a_{2,k},\ldots,a_{q-p,k},b_{1,k},\ldots,b_{q-p,k})$, the pole cancellation condition can be written in a matrix form $ {\mathcal M} U = 0$.
where the matrix elements of ${\mathcal M}$ are given by ${\mathcal M}_{s,m} = {\varphi}^{\prime}_{m,k-q_1} (s a_1)e^{i s \bq_1 \cdot {\bf a}_1}$ for $1\le m \le q-p$ and ${\mathcal M}_{s,m}={\varphi}^{\prime}_{m-q+p,k+q_1} (s a_1)e^{-i s \bq_1 \cdot {\bf a}_1}$ for $q-p+1\le m \le 2 q - 2p$ and it has $q$ lines and $(2 q-2p)$ columns. The size of the kernel of ${\mathcal M}$ determines the number of independent solutions to Eq.~\eqref{Chern_2_wfq_2}, and it is equal to the number of its columns minus the number of lines, i.e.,
\begin{equation}
    2 q - 2 p -q = q- 2p,
\end{equation}
Hence the number of zero-energy states per moiré unit cells $n_A =1- 2p/q=1-2\Phi/\Phi_0$, and the zero-energy FB disappear at the topological transition $\Phi=\Phi_0/2$, in agreement with our numerical results shown in Fig.~\ref{fig:fig2}b.

Despite disappearing FB at $\Phi_0/2$, we find zero-mode solutions at specific $\bk$ points obeying the magnetic boundary condition. These wavefunctions take the following form 
\begin{equation}\label{two_k0}
     {\bm \psi}_{\bk_0}(\br) = \Big[ a_0  \varphi_{k_0-q_1} (\br) {\bm \psi}_{K}(\br) + b_0 \varphi_{k_0+q_1} (\br) {\bm \psi}_{K'}(\br) \Big],
\end{equation}
with $\bk_0=(n_{1}/2){\bb_1}+(n_{2}/2){\bb_2}$ and $\varphi_{k}(z)$ corresponds to Eq.~\eqref{basis} with $q=2$, $p=1$, and $n=0$. Within the magnetic mBZ, $\bk_0$ correspond to two momenta separated by a distance of half the reciprocal lattice vector. These two zero modes combine with two other zero modes in the B-sublattice, as we discuss later in Sec.~\ref{hiddenwf}, to form a pair of Dirac cones that appear at $\Phi_0/2$, in agreement with the numerical band structure shown in Fig.~\ref{fig:fig1}f.
The coefficients $a_0$, $b_0$ can be obtained from the condition of pole cancellations at $z=0$ and $z=a_1$. 
We notice that the Bloch wave functions associated to Eqs.~\eqref{Chern_2_wfq_1} and \eqref{Chen2_wf_nB} are $k$-holomorphic and thus form an ideal quantum geometry \cite{PhysRevResearch.2.023237,PhysRevLett.127.246403}.

We further note that the analytical expressions of wave functions in Eqs.~\eqref{psiB}, \eqref{Chern_2_wfq_1}, \eqref{Chern_2_wfq_2} and \eqref{Chen2_wf_nB} featuring the theta functions resemble to those of quantum Hall wave functions on a torus, highlighting their correspondence with Landau levels~\cite{Haldane1985}.

\begin{table}[t]
\begin{center}
 \begin{tabular}{|c|c|}
 \hline
 \bf{Physical zero modes}
  & \bf{Hidden zero modes} 
   \\
 \hline\hline
 ${\bm\chi}_{\Gamma}(\br)$  &  ${\bm\chi}^{h}_{K}(\br)$,${\bm\chi}^{h}_{K^{\prime}}(\br)$ \\[2mm] ${\bm\chi}_{\Gamma}(\br)\sim A\Bar{z}$  &  ${\bm\chi}^{h}_{K}(\br)\sim \frac{D}{\Bar{z}}$, ${\bm\chi}^{h}_{K^{\prime}}(\br)\sim \frac{D}{\Bar{z}}$  \\[1mm]
 \hline
  ${\bm\psi}_{K}(\br)$, ${\bm\psi}_{K^{\prime}}(\br)$ & $\psi^{h}_{\Gamma}(\br)$ \\[2mm]
  ${\bm\psi}_{K}(\br)\sim D_{2}$, $\bm\psi_{K^{\prime}}(\br)\sim D_{2}$ & ${\bm\psi}^{h}_{\Gamma}(\br)\sim \frac{D^{\prime}}{z}$ \\[1mm]
 \hline
\end{tabular}
\caption{Asymptotic $\br\to 0$ behavior of physical and hidden zero modes at $\Gamma$, $K$ and $K^{\prime}$ points in the absence of magnetic field. $\bm\psi$ and $\bm\chi$ are solutions in the A and B sublattices, respectively. $A$, $D_2$, $D$ and $D^{\prime}$ are constant vectors in the layer space. \label{tab:table1}}
\end{center}
\end{table}

\begin{table}[t]
\begin{center}
 \begin{tabular}{|c|c|}
 \hline
 \bf{Wavefunctions}
  & \bf{Correspondence} \\
 \hline\hline
  ${\bm\psi}^{h}_{\Gamma}(\br), {\bm\chi}_{K}(\br), {\bm\chi}_{K^{\prime}}(\br)$  & ${\bm\psi}^{h}_{\Gamma}(\br)=[{\bm\chi}_{K}(\br)]^{\ast}\times[{\bm\chi}_{K^{\prime}}(\br)]^{\ast}$\\ & $\frac{D^{\prime}}{z}= \frac{D^{\ast}}{z}\times C$ \\ 
 &  \\
 \hline
  ${\bm\psi}_{K}(\br), {\bm\chi}_{\Gamma}(\br), {\bm\chi}^{h}_{K^{\prime}}(\br)$  & ${\bm\psi}_{K}(\br)=[{\bm\chi}_{\Gamma}(\br)]^{\ast}\times[{\bm\chi}^{h}_{K^{\prime}}(\br)]^{\ast}$\\ & $D_{2}= A^{\ast} \times D^{\ast}$ \\ &   \\
 \hline
 ${\bm\chi}^{h}_{K}(\br), {\bm\psi}^{h}_{\Gamma}(\br), {\bm\psi}_{K^{\prime}}(\br)$ &  ${\bm\chi}^{h}_{K}(\br)=[{\bm\psi}^{h}_{\Gamma}(\br)]^{\ast}\times[{\bm\psi}_{K^{\prime}}(\br)]^{\ast}$\\ & $D= D^{\prime \ast} \times D^{\ast}_{2}$\\
 \hline
\end{tabular}
\caption{Relation between hidden and physical solutions and associated constant vectors. $C$, $A$, $D_2$, $D$, and $D^{\prime}$ are vectors in the layer sector.\label{tab:table2}}
\end{center}
\end{table}

\subsection{Hidden wave functions, Dirac cones at criticality and zero modes above flux $\Phi_c$}\label{hiddenwf}
We now demonstrate a new set of zero-energy wave functions that emerge at $\Phi\geq\Phi_c$. 
These wavefunctions are also built upon zero-mode solutions at $b=0$ same as the previous ones. However, unlike in the previous ones, these $b=0$ zero-mode solutions are singular and hence unphysical or hidden when $\Phi < \Phi_c$.

First we demonstrate the nature of the $b=0$ hidden wavefunctions.
As discussed in Ref.~\cite{PhysRevB.103.155150} in the context of TBG, the hidden zero modes solutions live in a subspace perpendicular to the physical zero-mode solutions which exhibit zeros at some $\br_0$.
In the case of hTG, we recall that the physical solution in the B-sublattice at $\Gamma$ point ${\bm\chi}_{\Gamma}$ has the property that it vanishes at $\br=0$. We then find the existence of two singular hidden solutions ${\bm\chi}^{h}_{K}(\br)$ and ${\bm\chi}^{h}_{K^{\prime}}(\br)$ at $K$ and $K^{\prime}$ points at $b=0$: $\mathcal D_{b=0} {\bm\chi}^{h}_{K}(\br)=0$, $\mathcal D_{b=0} {\bm\chi}^{h}_{K^{\prime}}(\br)=0$. To analyze the form of the hidden wavefunctions we look at the Wronskian: 
$W(\br)=\chi_\Gamma (\br) \cdot \left[ \chi^{h}_{K} (\br) \times \chi^{h}_{K^{\prime}} (\br) \right]$, 
which satisfies the relation $\Bar{\partial }W(\br)=0$. Hence as per the Liouville's theorem $W(\br)$ is a constant. 
To ensure $W(\br)$ is constant, the zeros of ${\bm\chi}_{\Gamma}(\br)$ must be cancelled by the poles in ${\bm\chi}^{h}_{K}(\br)$ or ${\bm\chi}^{h}_{K^{\prime}}(\br)$. We note that this can happen when the two singular solutions become asymptotically colinear at $\br\to0$, i.e. 
\begin{equation}
{\bm\chi}^{h}_{K}(\br)\sim D/\Bar{z} + \order{1}; \quad {\bm\chi}^{h}_{K^{\prime}}(\br)\sim D/\Bar{z}+ \order{1}
\end{equation}
where $D$ is a vector in the three dimensional space spanned by the layers. Notably, this behavior is a mirror to the way the physical zero-mode solutions ${\bm\psi}_{K}(\br)$ and ${\bm\psi}_{K^{\prime}}(\br)$ become asymtotically colinear at $\br=0$ (Table.~\ref{tab:table1}). 

Now for the A-sublattice, we
construct the singular hidden solution with ${\bm\chi}^{h}_{K}(\br)$ and ${\bm\chi}^{h}_{K^{\prime}}(\br)$: 
\begin{equation}\label{hidden2}
{\bm\psi}^{h}_{\Gamma}(\br)=[{\bm\chi}^{h}_{K}(\br)]^{\ast}\times [{\bm\chi}^{h}_{K^{\prime}}(\br)]^{\ast},
\end{equation}
noting that a zero-mode of Eq.~\eqref{zeromode-1} can arise from a pair of zero modes from Eq.~\eqref{zeromode-2} when the corresponding momenta sum up to zero~\cite{guerci2023nature}. More generally, the hidden and physical solutions that exhibit zeros are thus related to each other, as shown in Table.~\ref{tab:table2}.
From the properties of ${\bm\chi}^{h}_{K}(\br)$ and ${\bm\chi}^{h}_{K^{\prime}}(\br)$,  we observe that $\bm\psi^{h}_{\Gamma}(\br)$ exhibits a simple pole at $\br=0$, interestingly mirroring again the behavior of the physical solution in the other sublattice, as illustrated in Table.~\ref{tab:table1}. 

Using these singular hidden zero-mode solutions we now construct the wavefunctions at finite $b$. For $b>0$, we build the wavefunctions with ${\bm\psi}^{h}_{\Gamma}(\br)$ and obtain following set of solutions that satisfy the magnetic Bloch periodicity: 
\begin{eqnarray}\label{hidden_Chern2}
{\bm \psi}^{h}_{n,\bk}(\br)=\vartheta_{1}\Big[\frac{p-q}{q}\frac{z}{a_1}-\frac{k}{b_2}+\frac{\omega}{q}\left(n+\frac{p}{2}\right), \frac{p-q}{q}\omega\Big]\nonumber \\\vartheta_{1}\Big[\frac{z}{a_1},\omega\Big]e^{i{k_1}\frac{z}{a_1}}e^{\frac{2i\pi z}{qa_1}(n+\frac{p}{2})}e^{-b(z^{2}+|z|^2)/4}{\bm \psi}^{h}_{\Gamma}(\br),\nonumber\\
\end{eqnarray}
with $n=1,\ldots p-q$ and hence forming $p-q$ magnetic FB. The pole at $\br=0$ in ${\bm \psi}^{h}_\Gamma(\br)$ is compensated by the zero of $\vartheta_{1}\Big[\frac{z}{a_1},\omega\Big]$. Hence the wave functions ${\bm \psi}^{h}_{n,\bk}(\br)$ are finite at all $\br$ and become physical when $p> q$ or $\Phi > \Phi_0$.
Therefore, at $\Phi>\Phi_0$ we have these new $p-q$ FB that add up with the $q+2p$ FB from Eqs.~\eqref{Chern_2_wfq_1} and \eqref{Chern_2_wfq_2} leading to a total of $n_A= 3p/q=3\Phi/\Phi_0$ zero-energy states per moiré unit cell with an enhanced Chern number $C_{A}=3$, consistent with numerical findings as in Fig.~\ref{fig:fig2}a.

In addition to the FB at $\Phi>\Phi_0$, we also find a single physical solution at $\Phi_0$ obeying the magnetic boundary condition:
\begin{equation}
{\bm\psi}^{h}_{\bk_0}(\br)=e^{i{\bk_0}.{\bf a}_{1}\frac{z}{a_1}}e^{\frac{i\pi z}{a_1}}e^{-b(z^{2}+|z|^2)/4}\vartheta_{1}\Big[\frac{z}{a_1},\omega\Big] {\bm\psi}^{h}_{\Gamma}(\br),
\label{dirac1}
\end{equation}
with the same momentum $\bk_0$ as in Eq.~\eqref{psik0}. Hence this resurrected single mode is the mirror of the zero-mode defined in Eq.~\eqref{psik0} to form a Dirac cone crossing the $C_A =2$ flat band at $\bk_0$, as seen in Fig.~\ref{fig:fig1}e. Notably, ${\bm\psi}^{h}_{\bk_0}(\br)$ remains distinct from the $C_A =2$ flat band while forming Dirac cone, as ${\bm\psi}^{h}_{\Gamma}(\br)$ is orthogonal to the physical solutions $\bm\psi_{K}(\br)$ and $\bm\psi_{K}(\br)$ that constitute the flat band (Eq.~\eqref{Chen2_wf_nB}).

For $b<0$, wavefunctions are built by using ${\bm \chi}^{h}_{K}(\br)$, ${\bm \chi}^{h}_{K^{\prime}}(\br)$ and they have the analytical form
\begin{eqnarray}\label{hidden2}
   {\bm \chi}^{h}_{\bk}(\br) = \sum_{n=1}^{p} \Big[ a^{\prime}_{n,k} \Bar{\varphi}^{2}_{n,k-q_1} (z,\Bar{z}) {\bm \chi}^{h}_{K}(\br) \nonumber \\+ b^{\prime}_{n,k} \Bar{\varphi}^{2}_{n,k+q_1} (z,\Bar{z}) {\bm \chi}^{h}_{K'}(\br) \Big],
\end{eqnarray}
with $\Bar{\varphi}^{2}_{n,k}(z,\Bar{z})=\varphi^{2\ast}_{n,k}(-z,-\Bar{z})$, where $\varphi^{2}$ are the same basis functions as in Eq.~\eqref{Chern_2_wfq_2}. 
The magnetic periodicity of $\bm\chi^{h}_{\bk}$ and pole cancellations of $\bm\chi^{h}_K$ and $\bm\chi^{h}_{k^{\prime}}$ at $z=n_1 a_1+n_2 a_2$ lead to $2p-q$ linearly independent zero-mode solutions and hence they become physical for $\Phi/\Phi_0> 1/2$. Therefore, together with the $p+q$ zero-energy wave functions discussed in Sec.~\ref{Bsub}, we have $2p-q$ additional zero modes from Eq.~\eqref{hidden2} for $\Phi>\Phi_0/2$ leading to a total of $n_B = 3p/q$ zero-energy states per moiré unit cell with $C_{B}=-3$, again in agreement with the numerical findings (Fig.~\ref{fig:fig2}b).

Even at $\Phi_c=\Phi_0/2$, we find exactly two physical zero-mode wave functions with the same $\bk_0$ values as in Eq.~\eqref{two_k0}:
\begin{equation}
     {\bm \chi}^{h}_{\bk_0}(\br) = \Big[ a^{\prime}_0  \Bar{\varphi}^{2}_{k_0-q_1} (\br) {\bm \chi}^{h}_{K}(\br) + b^{\prime}_0 \Bar{\varphi}^{2}_{k_0+q_1} (\br) {\bm \chi}^{h}_{K'}(\br) \Big],
\end{equation}
to form a pair of Dirac cones orthogonally crossing the $C_{B}=1$ zero-energy flat band at $\Phi_{0}/2$ (Fig.~\ref{fig:fig1}f). The basis functions are the same as in Eq.~\eqref{hidden2}, with $q=2$, $p=1$, and $n=0$. Coefficients $a^{\prime}_0$, $b^{\prime}_0$ can be obtained from the condition of pole cancellations at $z=0$ and $z=a_1$

We thus identify a set of hidden zero-energy wave functions which mirror the physical zero-energy wave functions to form Dirac cones at the topological transitions. Importantly, these hidden wave functions result in new zero-energy states at fields higher than the topological transitions leading to higher Chern numbers in the FB. 
\begin{figure}
\centering
    \includegraphics[width=0.49\textwidth]{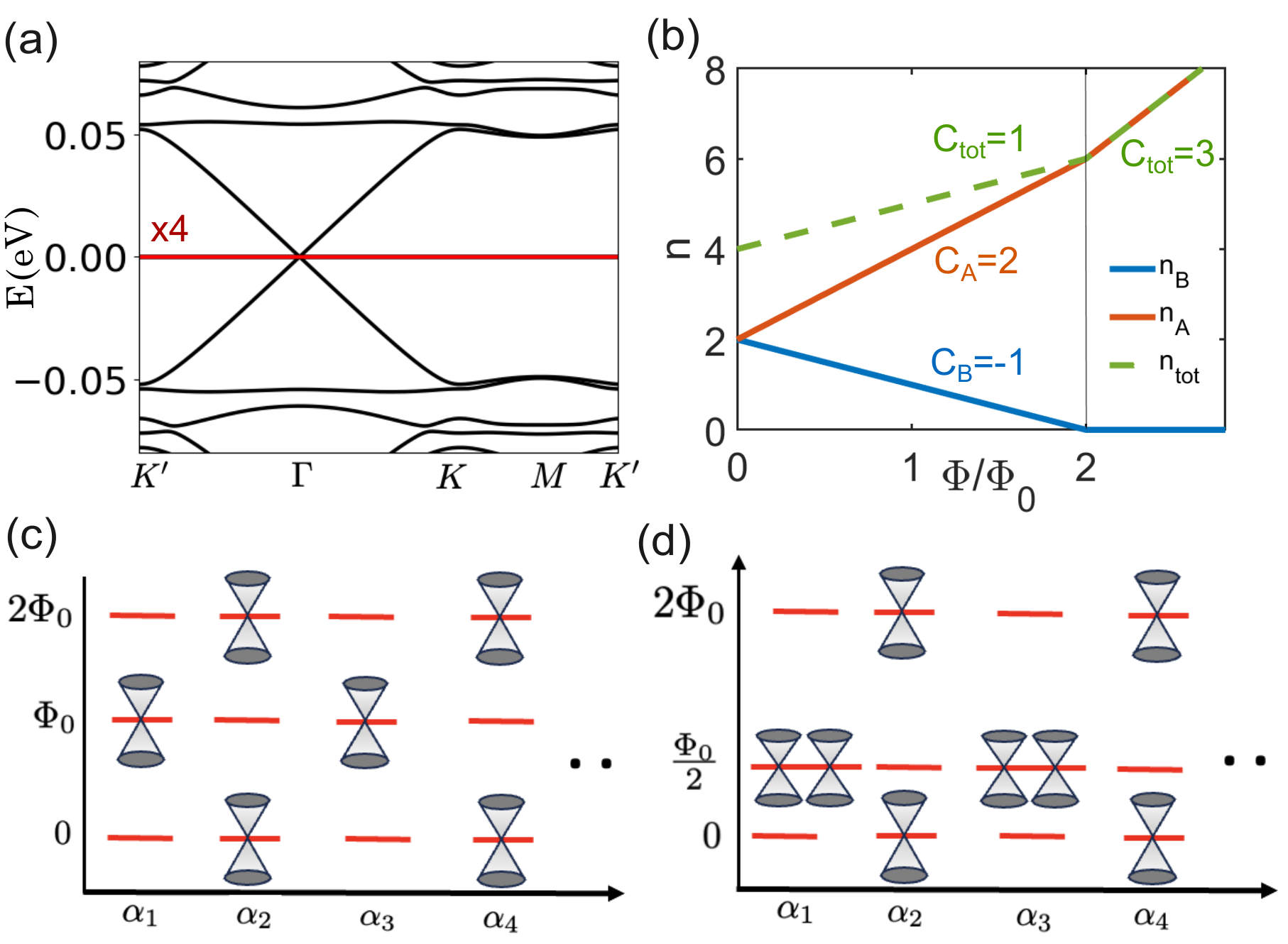}
\caption{(a) Spectrum at zero field at the second magic angle $\theta_{2}=0.532^{\circ}$, at the ABA stacking [$\phi=(0, 2\pi/3, -2\pi/3)$], adapted from Ref.~\cite{guerci2023nature}. (b) Number of zero modes per moiré unit cell in A-sublattice $n_A$ (red lines) and B-sublattice $n_B$ (blue lines), and $n_{tot}=n_{A}+n_{B}$ (green dashed lines) at $\theta_{2}$, at the ABA stacking and valley $K$, as a function of positive magnetic flux. Vertical line denotes phase transition. (c, d) Schematic of Dirac cones at different magic angles $\alpha_n$ at the topological transitions for $\Phi>0$ (c) and $\Phi<0$ (d). Red lines represent the zero-energy flat bands. }
\label{fig:fig3}
\end{figure}
\section{Topological phase transition at the second magic angle}\label{2ndmagic}
The chiral limit of hTG hosts a series of magic angles $\alpha_n$ with $\alpha_{2n+1}-\alpha_{2n-1}=\alpha_{2n+2}-\alpha_{2n}\approx 1.214$ \cite{guerci2023nature}, where $\alpha=w_{AB}/\left(v_{F}k_{\theta}\right)$. 
So far in the paper we have focused on 
the first magic angle $\theta_1 = 1.687^{\circ}$  ($\alpha_{1}\approx 0.38$). We now look into the effect of magnetic fields at the second magic angle $\alpha_{2}\approx 1.197$ ($\theta_2 \approx 0.532^{\circ}$) at the ABA stacking numerically. 
As shown in Fig.~\ref{fig:fig3}a, in the absence of magnetic fields at the second magic angle the spectrum has a zero-energy FB manifold with a dispersive Dirac cone crossing at $\Gamma$. The FB are four-fold degenerate with two FB in each sublattice \cite{popov2023,popov2023magic,guerci2023nature}.
This is in contrast to the first magic angle, where two central zero-energy FB gapped from the higher energy remote bands. 
With finite magnetic fields at the second magic angle, we find that the Dirac cone at $\Gamma$ first gets gapped out from the zero-energy FB at a small magnetic field and then returns again at a magnetic flux of $\Phi=2\Phi_0$. Beyond $2\Phi_0$ the Dirac cone becomes gapped again. The re-emergence of the Dirac cone is suggestive of a topological phase transition at $2\Phi_0$. To gain more insights we look into the sublattice polarization of zero modes in the FB expecting it follow the Streda formula, as seen in the case of first magic angle.
As shown in Fig.~\ref{fig:fig3}b, we observe that for $\Phi<2\Phi_0$, the A-polarised zero modes evolve as $n_A=n(0)+2\Phi/\Phi_0$ and the B-polarised modes change as $n_B=n(0)-\Phi/\Phi_0$, with $n(0)=2$ being the number of zero field zero-energy states per moiré unit cell in each sublattice. This suggests $C_A=2$ and $C_B=-1$. For $\Phi>2\Phi_0$, we have $n_A=3\Phi/\Phi_0$ and $n_B=0$, indicating $C_A=3$. Hence the total Chern number $C_{tot}=C_A + C_B$ changes from $C_{tot}=1$ to $C_{tot}=3$ across $2\Phi_0$, and therefore implying a topological phase transition at $\Phi_c=2\Phi_0$. Furthermore, we find the same $\Phi_c$ for both positive and negative magnetic field directions.

Notably, these findings have crucial differences from the first magic angle. First, in contrast to the first magic angle, $C_{A/B}$ for $\Phi<\Phi_c$ in Fig.~\ref{fig:fig3}b are not the same as their values at $\Phi=0$. At $\Phi=0$ sublattice A and sublattice B are associated to Chern numbers $1$ and $-1$, respectively \cite{guerci2023nature}. The difference arises due to the Dirac cone present at $\Gamma$, i.e. the flat bands are not separated by a gap from the other bands at $\Phi = 0$. At finite $\Phi$, the Dirac cone can form Landau levels, with the zeroth Landau level contributing an additional Chern number $1$ to the A sublattice. Secondly, the same $\Phi_c$ for positive and negative fields in the second magic angle is in contrast to the case of the first magic angle. Lastly, the critical field $\Phi_c$ are different for two magic angles. This can be understood using the Streda formula. For $\Phi=0$, $n(0)=1$ at the first magic angle and $n(0)=2$ at the second magic angle, due to the two and four fold degeneracies of the zero-energy FB, respectively. Hence, $n_{A/B}$ becomes zero for $\Phi_c =\Phi_0$ ($\Phi_c =\Phi_0/2$) for positive (negative) flux in the former and $\Phi_c=2\Phi_0$ in the latter. 

Since the properties of energy spectra at all even magic angles are the same, we expect topological transitions at other even magic angles to be similar to the second one. Similarly the behavior of odd magic angles follows the first one.
The different nature of topological transitions at even and odd $\alpha_n$ results in an intriguing sequence of the Dirac cones at FB, as shown in Fig.~\ref{fig:fig3} for positive $\Phi$ (b) and negative $\Phi$ (c). 

In this paper we do not discuss the analytical forms of the zero modes at the second magic angle, which will be addressed elsewhere.
\begin{figure}
\centering
    \includegraphics[width=0.46\textwidth]{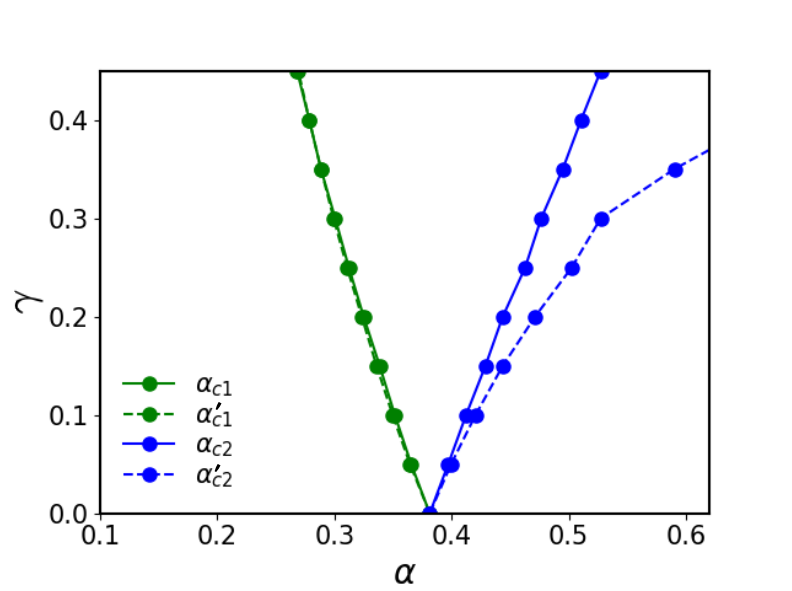}
\caption{Critical twist angles for different values of corrugation $\gamma$ at $\Phi_c$. $\alpha_{c1}$, $\alpha_{c2}$ represent critical twist angles at $\Phi_c =\Phi_0$ for $\Phi>0$ and $\alpha^{\prime}_{c1}$, $\alpha^{\prime}_{c2}$ denote critical twist angles at $\Phi_c =\Phi_0/2$ for $\Phi<0$.} 
\label{fig:fig4}
\end{figure}
\section{Deviation from chiral limit}\label{finitecorr} Until now, we have only discussed the chiral limit. In the presence of finite corrugations, the FB at zero magnetic field acquire a finite dispersion but still remain narrow and separated from the remote bands and with total Chern numbers $C_{tot}=1$ and $C_{tot}=-1$ \cite{guerci2023chern} at ABA and BAB stackings, respectively. We now investigate the topological properties of the low energy FB at finite corrugations $\gamma=w_{AA}/w_{AB}$ at the ABA stacking for finite field. For finite $\gamma$, we observe that the topological transitions at $\Phi_c$ do not occur at the magic angle (however, due to $C_{tot}=1$ in the FB, a gap closing with remote bands occur at $\Phi=2\Phi_0$ for negative field (not shown) following Streda formula, also see Appendix.~\ref{Hofstadtercorr}). Hence we explore a range of twist angles $\alpha$. In Fig.~\ref{fig:fig4} we show the gap closures occurring at $\Phi_c$ for different $\gamma$ and $\alpha$. 
Remarkably, we observe two critical values of the twist-angles $\alpha_{c1}$ ($\alpha^{\prime}_{c1}$) and $\alpha_{c2}$ ($\alpha^{\prime}_{c2}$) where the separation of remote bands with the low energy FB vanishes at $\Phi_c=\Phi_0$ ($\Phi_c=\Phi_0/2$) for positive (negative) flux. Interestingly, at $\alpha_{c1} (\alpha^{\prime}_{c1}) <\alpha<\alpha_{c2} (\alpha^{\prime}_{c2})$ the low energy FB exhibit a total Chern number of $C_{tot}=1$, whereas for $\alpha<\alpha_{c1} (\alpha^{\prime}_{c1})$ and $\alpha>\alpha_{c2} (\alpha^{\prime}_{c2})$ the Chern numbers associated to the low energy FB are found to be $C_{tot}=3$ ($C_{tot}=-3$) at $\Phi= \Phi_c$.
Notably, topological transitions occurring at a smaller twist angle correspond to a smaller flux quantum due to a larger moiré length scale. Therefore, the transitions that take place at smaller twist angles, can be observed even at lower magnetic fields. For example, in the chiral limit the transitions occur at magnetic fields $68$T and $34$T. However with a corrugation $\gamma=0.35$, transitions at lower critical twist angles $\alpha_{c2}$ ($\theta\approx 1.3^{\circ}$) and $\alpha^{\prime}_{c2}$ ($\theta\approx 1.1^{\circ}$) correspond to magnetic fields of $\sim40$T and $\sim14$T, respectively.

\begin{figure}
\centering
    \includegraphics[width=0.46\textwidth]{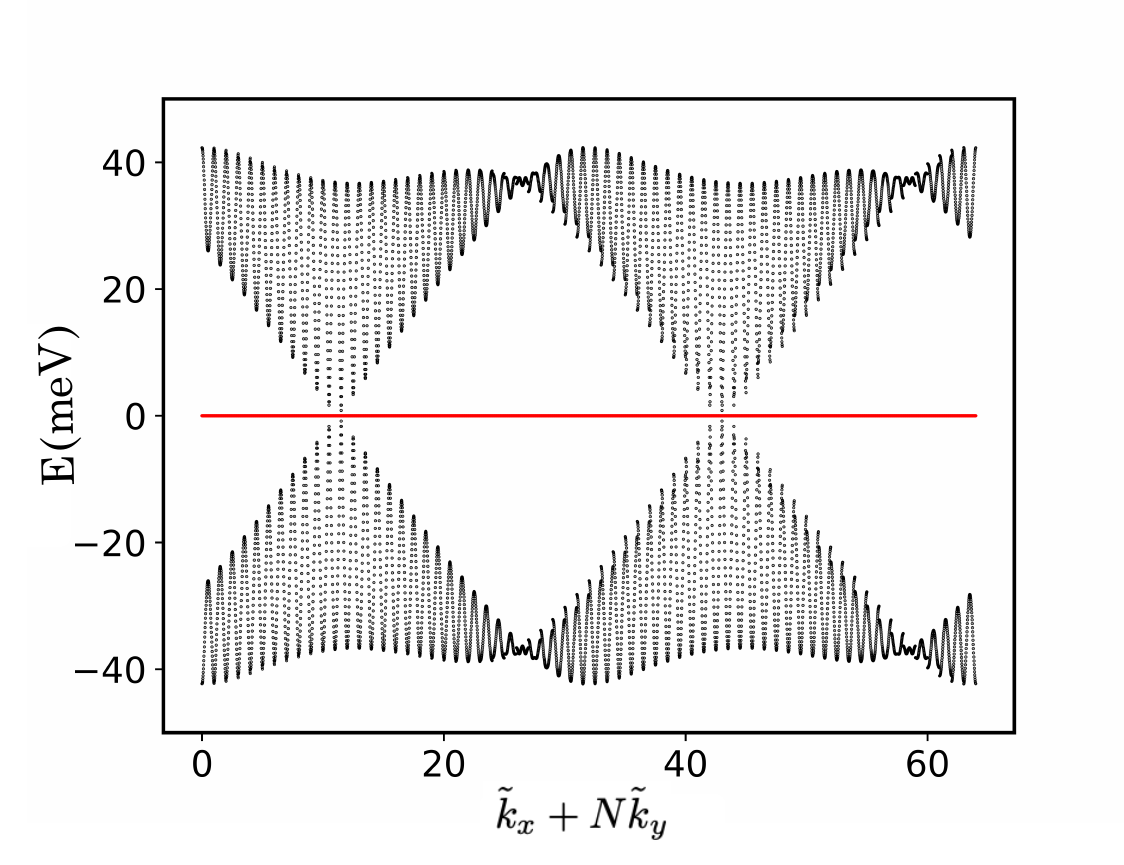}
\caption{Energy bands in the momentum space at the topological transition $\Phi_c =\Phi_0/2$ for twisted monolayer-bilayer graphene at the magic angle $\theta\approx1.08^{\circ}$ featuring two Dirac cones. Red line corresponds to the zero-energy flat band. $\Tilde{k}_{x}=k_{x}/k_{1}$, $\Tilde{k}_{y}=k_{y}/k_{2}$, with $k_{1}=\sqrt{3}k_{\theta}/2$, $k_{2}=3k_{\theta}/2$ defining the magnetic Brillouin zone, and $k_{x}$, $k_{y}$ denoting magnetic Bloch momenta along $x$ and $y$ directions. $N$ is number of grid points in magnetic Brillouin zone along $x$ direction.}
\label{fig:fig6}
\end{figure}
\section{Conclusion}
To summarize, we show a rich phenomenology arising from the interplay of a magnetic field with the flat bands of hTG and their local topology at the ABA/BAB stacking region in the supermoiré lattice. Application of magnetic field leads to two distinct topological phase transitions that modify the Chern mosaic and gives rise to higher Chern number ideal flat bands. The two transitions can be connected to two different zero-field Chern
numbers of flat energy bands. We derive the analytical expressions of zero energy flat-band wave functions and identify a set of hidden wave functions which lead to new physical zero-energy wave functions emerging at topological phase transitions.
We show that these wave functions are responsible for the enhancement of Chern numbers in the flat bands. We also find that the correspondence of wavefunctions with lowest Landau levels in the chiral limit continues to hold at finite magnetic fields. 

While we have focused on the ABA/BAB stacking in this work, we also find that for the AAA stacking, the nature of the topological transitions at the AAA stacking is similar to the transitions at ABA/BAB stackings at the second magic angle, consistent with similar properties of their energy bands in the absence of a magnetic field.
Our findings in the context of hTG also highlight that the finite field topological transitions can be inferred from the Chern numbers of flat bands in the absence of magnetic field following Streda formula. Similar analysis can be extended to other configurations of twisted multilayer graphene, like monolayer-bilayer, double-bilayer, and $n$-layer structures~\cite{makov2024}, which also host exact flat bands in the chiral limit. 
For instance, in monolayer-bilayer twisted graphene at the chiral limit, where the flat bands carry Chern numbers of $2, -1$ or $1, -2$ at zero field, we observe that the topological phase transitions happen at magnetic flux values of $\Phi=\Phi_0$ and $\Phi=\Phi_0/2$ for positive and negative field directions at a given valley. Moreover we find that these transitions involve emergence of one and two Dirac cones at $\Phi=\Phi_0$ and $\Phi=\Phi_0/2$, respectively, similar to hTG. In Fig.~\ref{fig:fig6}, we show the two Dirac cones at $\Phi=\Phi_0/2$.
Interestingly, with more layers, the Chern number in the flat bands can be larger. For instance, in $n$-layer twisted Bernal-stacked graphene, the flat bands have Chern numbers of $\pm n$ at zero magnetic field \cite{PhysRevLett.128.176403}. Subsequently, the topological transitions are expected to occur at much lower magnetic fields as the number of layers increases.

\begin{acknowledgments}
We are grateful to Jie Wang, Ady Stern, Felix von Oppen, Francisco Guinea for fruitful discussions.
We acknowledge financial support by the French National Research Agency (project TWISTGRAPH, ANR-21-CE47-0018).
The Flatiron Institute is a division of the Simons Foundation.    
\end{acknowledgments}

\appendix

\section{Details of numerically obtaining energy spectrum}
\label{sec:numericaldetails}

To find the spectrum of Eq.~\eqref{Ham0} numerically, we project it to a basis with low energy Landau levels (LLs). To this end we employ, the Landau gauge: $\mathbf{A}=-yB(1,0)$. We then use the same methodology that has been introduced for TBG in Refs.~\onlinecite{PhysRevB.84.035440,PhysRevB.100.035115}, but with essential modifications to apply it to the case of hTG.
For convenience of projection to the LL basis, we choose a gauge where the moiré tunneling $T(\br, \bm{\phi})$ is explicitly periodic under moiré translations $T(\br, \bm{\phi})=e^{-i\bq_1.\br}T(\br, \bm{\phi})$, $T(\br,\bm{\phi})=T(\br+\bR,\bm{\phi})$. We thus introduce the following unitary transformation:
\begin{equation}
U(\br)={\rm diag}[e^{i\bq_1.\br},1,e^{-i\bq_1.\br}],
\end{equation}
leading to the following Hamiltonian:
\begin{eqnarray}\label{Ham0app}
    \mathcal{H}(\br,{\bm\phi})&=&U^{\dagger}(\br)H_b(\br)U(\br) =
    \mathbf{1}_{3\times 3}\otimes  v_{F}e {\bm A}(\br)/\hbar \cdot{\bm \sigma}\nonumber \\ &&+ \begin{pmatrix}
     v_{F}\left(\bk+ \bq_{1}\right)\cdot{\bm \sigma}  & T(\br, \bm{\phi}) & 0\\
    h.c. &  v_{F}\bk\cdot{\bm \sigma}  & T(\br, -\bm{\phi}) \\
    0 & h.c. &  v_{F}\left(\bk- \bq_{1}\right)\cdot{\bm \sigma}
    \end{pmatrix},\nonumber \\
\end{eqnarray}
in a basis $\Psi=\left(\psi_{1}, \chi_{1}, \psi_{2}, \chi_{2}, \psi_{3}, \chi_{3}\right)^{T}$ with $\psi_l$ and $\chi_l$ representing A and B sublattices in layer $l$. Here $ T(\br, \bm{\phi})=T_{1}+ T_{2}e^{-i\bb_{1}.\br_j}e^{-i\phi_0}+ T_{3}e^{-i\bb_{2}.\br_j}e^{i\phi_0}$, where $\bb_{1}=\bq_1 - \bq_2$, $\bb_{2}=\bq_1 - \bq_3$. Here, $\phi_0 = 2\pi/3, -2\pi/3,0$ maps out the ABA, BAB and AAA regions, respectively.

Now, the LL basis is given by $\ket{n,\alpha, y,l}$, where $n$ labels the LL index, $\alpha$ denotes the sublattices, $y$ stands for the guiding center coordinate, and $l$ is the layer index. 
Employing the ladder operators $a$, $a^{\dagger}$ the intra-layer part can be written as
\begin{eqnarray}\label{intralayerLL1}
\mathcal{H}^{D}(\pm \bq_1)&=&-\frac{\sqrt{2}v_F}{l_B}\left(\sum^{\infty}_{n=0}\sqrt{n+1} \ket{nAyl}\bra{n+1Byl} + h.c.\right)\nonumber \\
&\mp& i v_{F}k_{\theta}\left(\sum^{\infty}_{n=0}\ket{nAyl}\bra{nByl}- h.c.\right),
\end{eqnarray}
\begin{eqnarray}\label{intralayerLL2}
\mathcal{H}^{D}(0)&=&-\frac{\sqrt{2}v_F}{l_B}\left(\sum^{\infty}_{n=0}\sqrt{n+1} \ket{nAyl}\bra{n+1Byl} + h.c.\right),\nonumber \\
\end{eqnarray}
with $a\ket{n}=\sqrt{n}\ket{n-1}$, $a^{\dagger}\ket{n}=\sqrt{n+1}\ket{n+1}$, and $l_{B}=\sqrt{\hbar/eB}$. While the intra-layer part of the Hamiltonian remains diagonal in $y$, the inter-layer $T_2$ and $T_3$ couple the LLs with guiding centers differed by $\pm \Delta_y$ where $\Delta_y =\sqrt{3}l_{B}^{2}k_{\theta}/2$. 
In order to write the interlayer tunneling in the LL basis, we note the commensurate flux condition which relates $\Delta_y$ to the moiré lattice unit cell generated by ${\bf a}^{1/2}_{M}=L_{\theta}(\sqrt{3}/2,\pm 1/2)$ with $L_{\theta}=4\pi/3 k_{\theta}$ as:
\begin{equation}
\frac{\Delta y}{L_{\theta}}=\frac{3k_{\theta}}{4\pi}\Delta y =\frac{r}{s},
\end{equation}
where $r$ and $s$ are coprime intergers. This also means that
\begin{equation}
\frac{\Phi}{\Phi_0} = \frac{s}{2r},
\end{equation}
where $\Phi$ is the flux piercing through the moiré unit cell $\Phi=B A_M$ and $\Phi_0$ is the unit quantum flux per moiré unit cell.
Moreover we assume the size of the system $L_y= Nr L_{\theta}= Ns\Delta_y$. 
So the position of guiding center of the LLs is given by
\begin{equation}
y_{c}=y_{0}+j\Delta_y + ms\Delta_y,
\end{equation}
where $0\leq y_0=k_x l^{2}_B < \Delta_y $, $j=0,\ldots,s-1$ and $m=0,\ldots N-1$. Here $k_x$ represents magnetic Bloch momenta in  the $x$-direction, where $0<k_x < \sqrt{3}k_{\theta}/2$. With a Fourier transform in $m$, we have
\begin{eqnarray}
&&\ket{n,\alpha,y_{0}+j\Delta_y + ms\Delta_y,l}\nonumber \\&=&\frac{1}{\sqrt{N}}\sum_{k_2}e^{-ik_{2}(ms+j)\Delta_y } \ket{n,\alpha,y_{0},j,k_2, l},
\end{eqnarray}
where $k_y$ is the magnetic Bloch momentum in the $y$-direction given by $k_y = 2\pi(n-1)/N s \Delta_y =(3k_{\theta}/2r)(n-1)/N$, with $n=1,\ldots N$. The matrix elements of interlayer part of the Hamiltonian in this basis are then given by:
\begin{eqnarray}
\bra{n,\alpha,y_0, j, k_2, 1}T_{1}\ket{n^{\prime},\beta,y^{\prime}_0, j^{\prime}, p_2, 2}=\nonumber \\ \delta_{y_0, y^{\prime}_0}\delta_{j,j^{\prime}}\delta_{k_{2},p_{2}}\delta_{n,n^{\prime}}T^{1}_{\alpha\beta},
\end{eqnarray}
\begin{eqnarray}
\bra{n,\alpha,y_0, j, k_2, 1}e^{-i\bb_1.\br}e^{-i \phi_0}T_{2}\ket{n^{\prime},\beta,y^{\prime}_0, j^{\prime}, p_2, 2}=\nonumber \\ \delta_{y_0, y^{\prime}_0}\delta_{j+1,j^{\prime}}\delta_{k_{2},p_{2}}T^{2}_{\alpha\beta}e^{-i \phi_0}e^{ik_{2}\Delta_y }e^{-i\frac{3k_{\theta}y_0}{2}}e^{-i2\pi\frac{r}{s}(j+\frac{1}{2})}\mathcal{F}_{n,n^{\prime}}\nonumber \\\left(-\frac{\tilde{\bb}_1 l_{B}}{\sqrt{2}}\right),\nonumber \\
\end{eqnarray}
\begin{eqnarray}
\bra{n,\alpha,y_0, j, k_2, 1}e^{-i\bb_2.\br}e^{i\phi_0}T_{3}\ket{n^{\prime},\beta,y^{\prime}_0, j^{\prime}, p_2, 2}=\nonumber \\ \delta_{y_0, y^{\prime}_0}\delta_{j-1,j^{\prime}}\delta_{k_{2},p_{2}}T^{3}_{\alpha\beta}e^{i \phi_0}e^{-ik_{2}\Delta_y }e^{-i\frac{3k_{\theta}y_0}{2}}e^{-i2\pi\frac{r}{s}(j-\frac{1}{2})}\mathcal{F}_{n,n^{\prime}}\nonumber \\\left(-\frac{\tilde{\bb}_2 l_{B}}{\sqrt{2}}\right).\nonumber \\
\end{eqnarray}

The interlayer tunneling term in the LL basis then reads as
\begin{eqnarray}\label{interlayerLL}
&&T(\br,\pm\bm{\phi})=\sum_{n,n^{\prime},\alpha,\beta, y_0}\sum^{s-1}_{j=0}\sum^{N-1}_{k_{2}=0}\nonumber\\ &&\Big[\delta_{n,n^{\prime}}T^{1}_{\alpha\beta} \ket{n,\alpha,y_{0},j,k_{2},1}\bra{n,\beta,y_{0},j,k_{2},2}\nonumber \\&& +T^{2}_{\alpha\beta}e^{\mp i\phi_0}e^{ik_{2}\Delta_y }e^{-i2\pi\frac{r}{s}(j+\frac{1}{2})}e^{-i\frac{3}{2}k_{\theta}y_0}\nonumber \\ &&\mathcal{F}_{n,n^{\prime}}(\frac{-{\tilde\bb}_1 l_{B}}{\sqrt{2}})\ket{n,\alpha,y_{0},j,k_{2},1}\bra{n^{\prime},\beta,y_{0},j+1,k_{2},2} \nonumber \\&&+T^{3}_{\alpha\beta}e^{\pm i\phi_0}e^{-ik_{2}\Delta_y }e^{-i2\pi\frac{r}{s}(j-\frac{1}{2})}e^{-i\frac{3}{2}k_{\theta}y_0}\nonumber \\ &&\mathcal{F}_{n,n^{\prime}}(\frac{-\Tilde{\bb}_2 l_{B}}{\sqrt{2}})\ket{n,\alpha,y_{0},j,k_{2},1}\bra{n^{\prime},\beta,y_{0},j-1,k_{2},2}\Big],\nonumber \\
\end{eqnarray}
with ${\tilde \bb_1}=b_{1x} + ib_{1y}$ and ${\tilde \bb_2}=b_{2x} + ib_{2y}$, and 
\begin{equation}\label{LLfunc}
\mathcal{F}_{n,m}(z)=\sqrt{\frac{n!}{m!}}z^{m-n}e^{-zz^{\ast}/2} L^{m-n}_{n}(zz^{\ast})~,
\end{equation}
for $m> n$ and $L^{m-n}_{n}$ is Laguerre polynomial. For $n\geq m$ 
$\mathcal{F}_{n,m}(z)$ can be obtained by $m \leftrightarrow n$ and $z \leftrightarrow -z^{\ast}$. 

The corresponding  LL basis form of Eq.~\eqref{Ham0app} is therefore obtained using Eqs.~\eqref{intralayerLL1}, \eqref{intralayerLL2}, and \eqref{interlayerLL}. 

\section{Evaluation of Berry curvature in magnetic field}

In this appendix, we give the details of obtaining the Chern numbers of the Hamiltonian in Appendix.~\ref{sec:numericaldetails}. Given a set of interconnected energy bands the Berry curvature is obtained first computing the overlap matrix: 
\begin{eqnarray}
\label{overlap}
&&\Lambda_{\eta\eta'}(\bk,\bq)=\sum_{j=0}^{s-1}\sum_{\alpha j}\sum_{n n'}U^*_{\eta,n\alpha l j}(\bk)\nonumber \\&& \left[e^{-iq_2k_1l^2_B}e^{-i\frac{q_1q_2l^2_B}{2}}\mathcal F_{n,n'}\left(\frac{-\tilde \bq l_B}{\sqrt{2}}\right)\right]U_{\eta',n'\alpha lj}(\bk+\bq)\nonumber\\&&=\left[U^\dagger(\bk)\,\lambda(\bk,\bq)\,U(\bk+\bq)\right]_{\eta\eta'},
\end{eqnarray}
where $\tilde\bq=q_x+iq_y$. This result is obtained employing the expression of the eigenstates of the Hamiltonian Eq.~\eqref{Ham0app}: 
\begin{eqnarray}
 &&H_{n\alpha lj,n'\beta l'j'}(\bk)\implies \ket{v_{\eta\bk}} \nonumber \\&=&\sum_{j=0}^{s-1}\sum_\alpha \sum_l\sum_{n=0}^\infty U_{\eta,n\alpha lj}(\bk)\ket{n,\alpha,l,j,\bk},    
\end{eqnarray}
and employing the overlap between two Landau levels with different guiding center. The Berry phase is then obtained as: 
\begin{eqnarray}
\label{Berry_curvature}
\tr[\mathcal F]&=&\frac{1}{i}\log\det\Big[\Lambda (\bk,du\bm G_{1})\Lambda(\bk+du\bm G_1,du\bm G_2) \nonumber \\&& \Lambda(\bk+du\bm G_1+du\bm G_2,-du\bm G_1)\nonumber \\&& \Lambda(\bk+du\bm G_2,-du\bm G_2)\Big],\nonumber \\
\end{eqnarray}
and the Chern number
\begin{equation}
    C=\frac{1}{2\pi}\sum_{\bk}\,\text{Tr}[\mathcal F(\bk)].
\end{equation}

\section{Atiyah-Singer index theorem}\label{sec:ASindextheorem}

In the following we discuss implications of the Atiyah-Singer index theorem on counting of the zero-energy states. Employing the index theorem in Hamiltonian Eq.~\eqref{Ham}, we obtain
\begin{equation}\label{AS}
{\rm dim}[{\rm Ker}~\mathcal{D}_b]-{\rm dim}[{\rm Ker}~\mathcal{D}^{\dagger}_{b}]=\frac{1}{2\pi}\int {\rm Tr}\mathcal{B}_{xy}ds, 
\end{equation}
where $\mathcal{B}_{xy}=\partial_{x}\mathcal{A}_{y}-\partial_{y}\mathcal{A}_{x}+ \frac{1}{2}\Big[\mathcal{A}_{x},\mathcal{A}_{y}]$ 
is a curvature associated to $\mathcal{A}$, the gauge potential in the Hamiltonian. From Eq.~\eqref{Ham}, we have
\begin{equation}
\mathcal{A}={\bm A}{\rm I} + \tilde{\bm A}(\br).
\end{equation}
Here ${\bm A}=\frac{b}{2}(y, -x)$ is the vector potential corresponding to the applied magnetic field $b$, $\rm I$ is identity matrix in layer subspace, and $\tilde{\bm A}(\br)$ is a matrix gauge potential attributed to the interlayer moiré tunneling. It takes the form $\tilde{\bm A}(\br)=\tilde{A}_x +i\tilde{A}_y$, where
\begin{equation}
\tilde{A}_x = \begin{pmatrix}
    0 & a^{1}_{\omega}(\br) & 0 \\
   a^{1}_{0}(-\br) & 0 & a^{1}_{0}(\br)\\
   0 & a^{1}_{\omega}(-\br) & 0
    \end{pmatrix}.   
\end{equation}
\begin{equation}
\tilde{A}_y =\begin{pmatrix}
    0 & a^{2}_{\omega}(\br) & 0 \\
    a^{2}_{0}(-\br) & 0 & a^{2}_{0}(\br)\\
   0 & a^{2}_{\omega}(-\br) & 0
   \end{pmatrix}.
\end{equation}
where $a^{1}_{0}(\br)=\alpha\sum^{3}_{j=1}\cos\left(\bm{q_j}.\br\right)$, $a^{2}_{0}(\br)=\alpha\sum^{3}_{j=1}\sin\left(\bm{q_j}.\br\right)$, $a^{1}_{\omega}(\br)=\alpha\sum^{3}_{j=1}\omega^{ (j-1)}\cos\left(\bm{q_j}.\br\right)$, and $a^{2}_{\omega}(\br)=\alpha\sum^{3}_{j=1}\omega^{(j-1)}\sin\left(\bm{q_j}.\br\right)$.
Both matrix gauge potentials $\tilde{A}_x$ and $\tilde{A}_y$ are traceless and hence they do not contribute to the $\mathcal{B}$. On the other hand the applied field couples to the Dirac components of the three layers in Hamiltonian Eq.~\eqref{Ham} and gives rise to a contribution of $3b$ in $\mathcal{B}$. Since kernels $\mathcal{D}_b$ and $\mathcal{D}^{\dagger}_b$ belong to A and B sublattices, respectively, we finally have
\begin{equation}
n_A -n_B = 3\frac{\Phi}{\Phi_0},
\end{equation}
which holds for all parameter range in the chiral limit.

\section{Symmetry relation between different stacking configurations and valleys}\label{symmetryhTTG}

Here we discuss symmetries of hTG and resulting correspondence between the stacking configurations ABA and BAB at two valleys $K$ and $K^{\prime}$ for finite magnetic fields.

In the absence of magnetic field the Hamiltonian Eq.~\eqref{Ham0} reads as
\begin{eqnarray}\label{HamKb0}
    \mathcal{H}(\br,\bm{\phi}) = \mathbf{1}_{3\times 3}\otimes  v_{F} \hat{\bk}\cdot{\bm \sigma} \nonumber \\+ \begin{pmatrix}
    0   & T(\br, \bm{\phi}) & 0\\
    h.c. & 0 & T(\br, -\bm{\phi}) \\
    0 & h.c. & 0
    \end{pmatrix},\nonumber \\
\end{eqnarray}
in the basis $\Psi=\left(\psi_{1}, \chi_{1}, \psi_{2}, \chi_{2}, \psi_{3}, \chi_{3}\right)^{T}$. Here $\bk=-i\grad_{\br}$ and $\bm\sigma$ denote Pauli matrices in sublattice sector. The corresponding Hamiltonian in valley $K^{\prime}$ is obtained by time-reversal symmetry
\begin{eqnarray}\label{HamKpb0}
    \mathcal{H}^{\prime}(\br,\bm{\phi}) = -\mathbf{1}_{3\times 3}\otimes  v_{F} \hat{\bk}\cdot{\bm \sigma}^* \nonumber\\+ \begin{pmatrix}
    0   & T^{\ast}(\br, \bm{\phi}) & 0\\
    h.c. & 0 & T^{\ast}(\br, -\bm{\phi}) \\
    0 & h.c. & 0
    \end{pmatrix},\nonumber \\
\end{eqnarray}
where ${\bm \sigma}^* = (\sigma_x,-\sigma_y)$.

The $C_{2z}\mathcal{T}$ symmetry acts as $\mathbf{1}_{3\times 3}\otimes \sigma_x {\cal K}$ on the Hamiltonian Eq.~\eqref{HamKb0}, where ${\cal K}$ denotes complex conjugation.
With $C_{2z}\mathcal{T}$ the moiré tunneling matrix in valley $K$ then transforms as
\begin{eqnarray}
&&C_{2z}\mathcal{T} \begin{pmatrix}
    0   & T(\br, \bm{\phi}) & 0\\
    h.c. & 0 & T(\br, -\bm{\phi}) \\
    0 & h.c. & 0
    \end{pmatrix} (C_{2z}\mathcal{T})^{\dagger} \nonumber\\&=& \begin{pmatrix}
    0   & T(-\br, -\bm{\phi}) & 0\\
    h.c. & 0 & T(-\br, \bm{\phi}) \\
    0 & h.c. & 0
    \end{pmatrix}.\nonumber\\
\end{eqnarray}
 Now we couple to magnetic field and focus on the kinetic part of the Hamiltonian. In valley $K$ it transforms as the following
\begin{eqnarray}
(C_{2z}\mathcal{T})\big[\mathbf{1}_{3\times 3}\otimes  v_{F} \left(\hat{\bk}+ e\bm{A (r)}/\hbar\right)\cdot{\bm \sigma}\big] (C_{2z}\mathcal{T})^{\dagger}\nonumber \\=\mathbf{1}_{3\times 3}\otimes  v_{F} \left(-\hat{\bk}+ e\bm{A (r)}/\hbar\right)\cdot{\bm \sigma}\nonumber \\
\end{eqnarray}
Hence the full Hamiltonian transforms as
\begin{equation}
(C_{2z}\mathcal{T}) \mathcal{H}(\br,\bm{\phi},\bm{A})(C_{2z}\mathcal{T})^{\dagger}=\mathcal{H}(-\br,-\bm{\phi}, -\bm{A}),
\end{equation}
since $\bm A(\br)=-\bm A(-\br)$ is odd under inversion. Similarly in valley $K^{\prime}$ we obtain

\begin{equation}
(C_{2z}\mathcal{T}) \mathcal{H}^{\prime}(\br,\bm{\phi},\bm{A})(C_{2z}\mathcal{T})^{\dagger}=\mathcal{H}^{\prime}(-\br,-\bm{\phi},\bm{A}),\\
\end{equation}
Since at the ABA stacking $\phi_{ABA}=(0, 2\pi/3,-2\pi/3)=-\phi_{BAB}$ at the BAB stacking, the $C_{2z}\mathcal{T}$ symmetry maps ABA to BAB in a given valley with opposite magnetic field direction and the resulting energy spectra are identical.
\begin{table}
\begin{center}
\begin{tabular}{ |c|c|c| } 
\hline
 & $C_{2Z}\mathcal{T}$ & $\mathcal{T}$ \\
\hline
${\mathcal H}(\br,\bm{\phi}_{ABA},{\bm A})$ & $\mathcal{H}(-\br,\bm{\phi}_{BAB},-{\bm A})$ & $\mathcal{H}^{\prime}(\br,\bm{\phi}_{ABA},{-\bm A})$ \\ 
${\mathcal H}(\br,\bm{\phi}_{BAB},{\bm A})$ & ${\mathcal H}(-\br,\bm{\phi}_{ABA},{-\bm A})$ & ${\mathcal H}^{\prime}(\br,\bm{\phi}_{BAB},-{\bm A})$ \\ 
\hline
\end{tabular}
\caption{Symmetry relation between different sectors of hTG.\label{tab:table3}}
\end{center}
\end{table}
Under time-reversal operation the kinetic part in valley $K$ takes the form
\begin{eqnarray}
\mathcal{T}\big[\mathbf{1}_{3\times 3}\otimes  v_{F} \left(\hat{\bk}+ e\bm{A (r)}/\hbar\right)\cdot{\bm \sigma}\big] \mathcal{T}^{\dagger}\nonumber \\=\mathbf{1}_{3\times 3}\otimes  v_{F} \left(-\hat{\bk}+ e\bm{A (r)}/\hbar\right)\cdot{\bm \sigma^{\ast}}.\nonumber \\
\end{eqnarray}
Hence the full Hamiltonian with $\mathcal{T}$ operation becomes
\begin{equation}
\label{time_reversal_K}
\mathcal{T} \mathcal{H}(\br,\bm{\phi},\bm{A})\mathcal{T}^{\dagger}=\mathcal{H}^{\prime}(\br,\bm{\phi}, -\bm{A}).
\end{equation}
implying that the time-reversal operation maps the Hamiltonian of one valley to the one in other valley, inverting the magnetic field orientation, but keeping the stacking configuration same.

We summarize the mapping between different sectors at finite fields in Table~\ref{tab:table3}.
\begin{figure*}
\centering
    \includegraphics[width=0.75\textwidth]{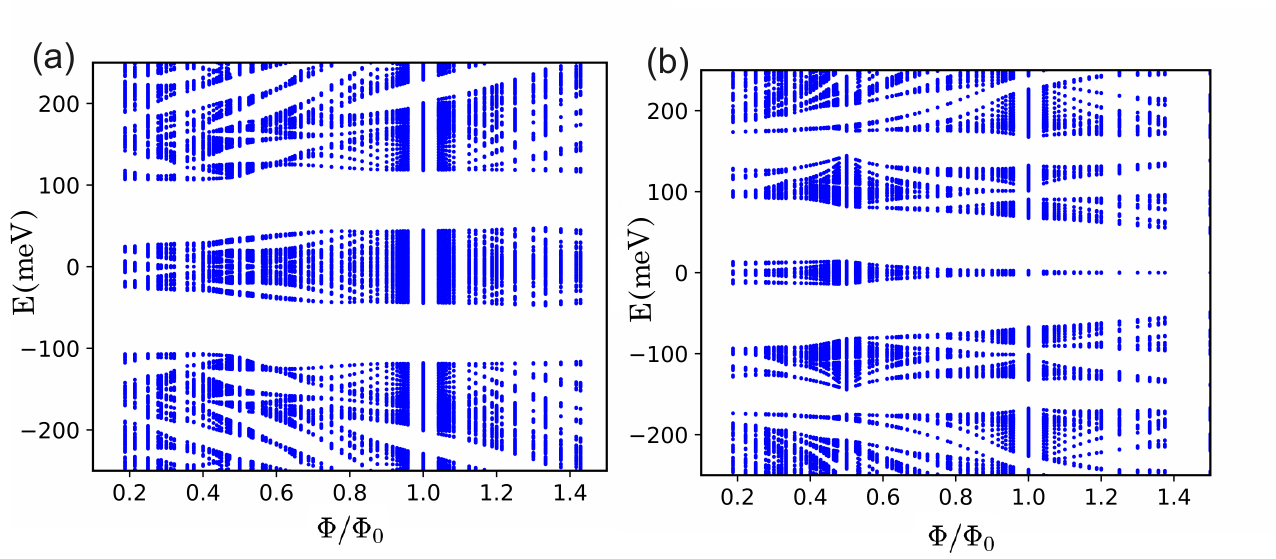}
\caption{Hofstadter spectra at the ABA stacking configuration for positive (a) and negative (b) magnetic flux, with corrugation $\gamma=0.7$ at twist $\theta=1.687^{\circ}$.} 
\label{fig:figapp}
\end{figure*}
\section{Jacobi theta function}\label{thetafunc}

The Jacobi theta functions of the first kind are defined as

\begin{equation}
\vartheta_1(z,\omega)=\sum_{n\in \mathbb{Z}}e^{i\pi\omega(n+1/2)^{2}e^{2i\pi(z-1/2)(n+1/2)}}
\end{equation}

Apart from the properties mentioned in Sec.~\ref{Bsub}, some other useful properties of the theta functions are the following

\begin{equation}
\vartheta_{1}[-z, \omega]=-\vartheta_{1}[z,\omega]
\end{equation}

\begin{equation}
\vartheta_{1}[\Bar{z}\pm 1, -\omega^{\ast}]=-\vartheta_{1}[\Bar{z},-\omega^{\ast}]
\end{equation}

\begin{eqnarray}
&&\vartheta_{1}[\Bar{z}+\omega^{\ast}, -\omega^{\ast}]=-e^{-i\pi\omega^{*}}e^{2i\pi \Bar{z}}\vartheta_1[\Bar{z},-\omega^{\ast}],\nonumber\\
&&\vartheta_{1}[\Bar{z}-\omega^{\ast}, -\omega^{*}]=-e^{-i\pi\omega^{*}}e^{-2i\pi \Bar{z}}\vartheta_1[\Bar{z},-\omega^{*}]
\end{eqnarray}

\section{Hofstadter spectra at finite corrugations}\label{Hofstadtercorr}

In this appendix we demonstrate the characteristics of energy levels obtained from Eq.~\eqref{Ham0} with $w_{AA}/w_{AB}=0.7$ and $\theta=1.687^{\circ}$, numerically. 
In Fig.~\ref{fig:figapp}, we show the Hofstadter spectra at the ABA stacking for positive (a) and negative (b) flux. Clearly, the energy states from the central FB (narrow) bands lying within an energy range of approximately $\pm 50$ meV do not exhibit a gap closing with the remote bands at $\Phi =\Phi_0$ and $\Phi_{0}/2$, the values where the gap closing occur in the chiral limit. 

Now focusing on states in central FB, we observe that their bandwidth increases with $\Phi$ in (a) and interestingly reduces in (b) while still maintaining a large gap with the remote bands, in the range of $\Phi$ shown in the figure. For instance, at $\Phi=\Phi_0$ the FB show a bandwidth of $\approx 5$meV, and the remote band energies remain at $\approx \pm 67$ meV. The change of bandwidth can possibly be due to the finite Chern number associated to FB because of which the number of states increases in (a) and decreases in (b) following Streda formula. 
Interestingly, for negative field, the remote band energies continuously shift towards FB and we find a gap closing between them at $\Phi=2\Phi_0$ (not shown in the figure), which is consistent with the Streda formula $n=n(0)- C_{tot}\Phi/\Phi_0$, with $n(0)=2$ and $C_{tot}=1$.

\bibliography{ref}

\begin{thebibliography}{88}%
\makeatletter
\providecommand \@ifxundefined [1]{%
 \@ifx{#1\undefined}
}%
\providecommand \@ifnum [1]{%
 \ifnum #1\expandafter \@firstoftwo
 \else \expandafter \@secondoftwo
 \fi
}%
\providecommand \@ifx [1]{%
 \ifx #1\expandafter \@firstoftwo
 \else \expandafter \@secondoftwo
 \fi
}%
\providecommand \natexlab [1]{#1}%
\providecommand \enquote  [1]{``#1''}%
\providecommand \bibnamefont  [1]{#1}%
\providecommand \bibfnamefont [1]{#1}%
\providecommand \citenamefont [1]{#1}%
\providecommand \href@noop [0]{\@secondoftwo}%
\providecommand \href [0]{\begingroup \@sanitize@url \@href}%
\providecommand \@href[1]{\@@startlink{#1}\@@href}%
\providecommand \@@href[1]{\endgroup#1\@@endlink}%
\providecommand \@sanitize@url [0]{\catcode `\\12\catcode `\$12\catcode
  `\&12\catcode `\#12\catcode `\^12\catcode `\_12\catcode `\%12\relax}%
\providecommand \@@startlink[1]{}%
\providecommand \@@endlink[0]{}%
\providecommand \url  [0]{\begingroup\@sanitize@url \@url }%
\providecommand \@url [1]{\endgroup\@href {#1}{\urlprefix }}%
\providecommand \urlprefix  [0]{URL }%
\providecommand \Eprint [0]{\href }%
\providecommand \doibase [0]{https://doi.org/}%
\providecommand \selectlanguage [0]{\@gobble}%
\providecommand \bibinfo  [0]{\@secondoftwo}%
\providecommand \bibfield  [0]{\@secondoftwo}%
\providecommand \translation [1]{[#1]}%
\providecommand \BibitemOpen [0]{}%
\providecommand \bibitemStop [0]{}%
\providecommand \bibitemNoStop [0]{.\EOS\space}%
\providecommand \EOS [0]{\spacefactor3000\relax}%
\providecommand \BibitemShut  [1]{\csname bibitem#1\endcsname}%
\let\auto@bib@innerbib\@empty
\bibitem [{\citenamefont {Andrei}\ and\ \citenamefont
  {MacDonald}(2020)}]{Andrei2020}%
  \BibitemOpen
  \bibfield  {author} {\bibinfo {author} {\bibfnamefont {E.~Y.}\ \bibnamefont
  {Andrei}}\ and\ \bibinfo {author} {\bibfnamefont {A.~H.}\ \bibnamefont
  {MacDonald}},\ }\bibfield  {title} {\bibinfo {title} {Graphene bilayers with
  a twist},\ }\href {https://doi.org/10.1038/s41563-020-00840-0} {\bibfield
  {journal} {\bibinfo  {journal} {Nature Materials}\ }\textbf {\bibinfo
  {volume} {19}},\ \bibinfo {pages} {1265} (\bibinfo {year}
  {2020})}\BibitemShut {NoStop}%
\bibitem [{\citenamefont {Cao}\ \emph {et~al.}(2018{\natexlab{a}})\citenamefont
  {Cao}, \citenamefont {Fatemi}, \citenamefont {Fang}, \citenamefont
  {Watanabe}, \citenamefont {Taniguchi}, \citenamefont {Kaxiras},\ and\
  \citenamefont {Jarillo-Herrero}}]{Cao2018_1}%
  \BibitemOpen
  \bibfield  {author} {\bibinfo {author} {\bibfnamefont {Y.}~\bibnamefont
  {Cao}}, \bibinfo {author} {\bibfnamefont {V.}~\bibnamefont {Fatemi}},
  \bibinfo {author} {\bibfnamefont {S.}~\bibnamefont {Fang}}, \bibinfo {author}
  {\bibfnamefont {K.}~\bibnamefont {Watanabe}}, \bibinfo {author}
  {\bibfnamefont {T.}~\bibnamefont {Taniguchi}}, \bibinfo {author}
  {\bibfnamefont {E.}~\bibnamefont {Kaxiras}},\ and\ \bibinfo {author}
  {\bibfnamefont {P.}~\bibnamefont {Jarillo-Herrero}},\ }\bibfield  {title}
  {\bibinfo {title} {Unconventional superconductivity in magic-angle graphene
  superlattices},\ }\href {https://doi.org/10.1038/nature26160} {\bibfield
  {journal} {\bibinfo  {journal} {Nature}\ }\textbf {\bibinfo {volume} {556}},\
  \bibinfo {pages} {43} (\bibinfo {year} {2018}{\natexlab{a}})}\BibitemShut
  {NoStop}%
\bibitem [{\citenamefont {Park}\ \emph {et~al.}(2021)\citenamefont {Park},
  \citenamefont {Cao}, \citenamefont {Watanabe}, \citenamefont {Taniguchi},\
  and\ \citenamefont {Jarillo-Herrero}}]{Park2021}%
  \BibitemOpen
  \bibfield  {author} {\bibinfo {author} {\bibfnamefont {J.~M.}\ \bibnamefont
  {Park}}, \bibinfo {author} {\bibfnamefont {Y.}~\bibnamefont {Cao}}, \bibinfo
  {author} {\bibfnamefont {K.}~\bibnamefont {Watanabe}}, \bibinfo {author}
  {\bibfnamefont {T.}~\bibnamefont {Taniguchi}},\ and\ \bibinfo {author}
  {\bibfnamefont {P.}~\bibnamefont {Jarillo-Herrero}},\ }\bibfield  {title}
  {\bibinfo {title} {Tunable strongly coupled superconductivity in magic-angle
  twisted trilayer graphene},\ }\href
  {https://doi.org/10.1038/s41586-021-03192-0} {\bibfield  {journal} {\bibinfo
  {journal} {Nature}\ }\textbf {\bibinfo {volume} {590}},\ \bibinfo {pages}
  {249} (\bibinfo {year} {2021})}\BibitemShut {NoStop}%
\bibitem [{\citenamefont {Hao}\ \emph {et~al.}(2021)\citenamefont {Hao},
  \citenamefont {Zimmerman}, \citenamefont {Ledwith}, \citenamefont {Khalaf},
  \citenamefont {Najafabadi}, \citenamefont {Watanabe}, \citenamefont
  {Taniguchi}, \citenamefont {Vishwanath},\ and\ \citenamefont
  {Kim}}]{doi:10.1126/science.abg0399}%
  \BibitemOpen
  \bibfield  {author} {\bibinfo {author} {\bibfnamefont {Z.}~\bibnamefont
  {Hao}}, \bibinfo {author} {\bibfnamefont {A.~M.}\ \bibnamefont {Zimmerman}},
  \bibinfo {author} {\bibfnamefont {P.}~\bibnamefont {Ledwith}}, \bibinfo
  {author} {\bibfnamefont {E.}~\bibnamefont {Khalaf}}, \bibinfo {author}
  {\bibfnamefont {D.~H.}\ \bibnamefont {Najafabadi}}, \bibinfo {author}
  {\bibfnamefont {K.}~\bibnamefont {Watanabe}}, \bibinfo {author}
  {\bibfnamefont {T.}~\bibnamefont {Taniguchi}}, \bibinfo {author}
  {\bibfnamefont {A.}~\bibnamefont {Vishwanath}},\ and\ \bibinfo {author}
  {\bibfnamefont {P.}~\bibnamefont {Kim}},\ }\bibfield  {title} {\bibinfo
  {title} {Electric field-tunable superconductivity in alternating-twist
  magic-angle trilayer graphene},\ }\href
  {https://doi.org/10.1126/science.abg0399} {\bibfield  {journal} {\bibinfo
  {journal} {Science}\ }\textbf {\bibinfo {volume} {371}},\ \bibinfo {pages}
  {1133} (\bibinfo {year} {2021})}\BibitemShut {NoStop}%
\bibitem [{\citenamefont {Cao}\ \emph {et~al.}(2021)\citenamefont {Cao},
  \citenamefont {Park}, \citenamefont {Watanabe}, \citenamefont {Taniguchi},\
  and\ \citenamefont {Jarillo-Herrero}}]{Cao2021}%
  \BibitemOpen
  \bibfield  {author} {\bibinfo {author} {\bibfnamefont {Y.}~\bibnamefont
  {Cao}}, \bibinfo {author} {\bibfnamefont {J.~M.}\ \bibnamefont {Park}},
  \bibinfo {author} {\bibfnamefont {K.}~\bibnamefont {Watanabe}}, \bibinfo
  {author} {\bibfnamefont {T.}~\bibnamefont {Taniguchi}},\ and\ \bibinfo
  {author} {\bibfnamefont {P.}~\bibnamefont {Jarillo-Herrero}},\ }\bibfield
  {title} {\bibinfo {title} {Pauli-limit violation and re-entrant
  superconductivity in moiré graphene},\ }\href
  {https://doi.org/10.1038/s41586-021-03685-y} {\bibfield  {journal} {\bibinfo
  {journal} {Nature}\ }\textbf {\bibinfo {volume} {595}},\ \bibinfo {pages}
  {526} (\bibinfo {year} {2021})}\BibitemShut {NoStop}%
\bibitem [{\citenamefont {Park}\ \emph {et~al.}(2022)\citenamefont {Park},
  \citenamefont {Cao}, \citenamefont {Xia}, \citenamefont {Sun}, \citenamefont
  {Watanabe}, \citenamefont {Taniguchi},\ and\ \citenamefont
  {Jarillo-Herrero}}]{Park2022}%
  \BibitemOpen
  \bibfield  {author} {\bibinfo {author} {\bibfnamefont {J.~M.}\ \bibnamefont
  {Park}}, \bibinfo {author} {\bibfnamefont {Y.}~\bibnamefont {Cao}}, \bibinfo
  {author} {\bibfnamefont {L.-Q.}\ \bibnamefont {Xia}}, \bibinfo {author}
  {\bibfnamefont {S.}~\bibnamefont {Sun}}, \bibinfo {author} {\bibfnamefont
  {K.}~\bibnamefont {Watanabe}}, \bibinfo {author} {\bibfnamefont
  {T.}~\bibnamefont {Taniguchi}},\ and\ \bibinfo {author} {\bibfnamefont
  {P.}~\bibnamefont {Jarillo-Herrero}},\ }\bibfield  {title} {\bibinfo {title}
  {Robust superconductivity in magic-angle multilayer graphene family},\ }\href
  {https://doi.org/10.1038/s41563-022-01287-1} {\bibfield  {journal} {\bibinfo
  {journal} {Nature Materials}\ }\textbf {\bibinfo {volume} {21}},\ \bibinfo
  {pages} {877} (\bibinfo {year} {2022})}\BibitemShut {NoStop}%
\bibitem [{\citenamefont {Cao}\ \emph {et~al.}(2018{\natexlab{b}})\citenamefont
  {Cao}, \citenamefont {Fatemi}, \citenamefont {Demir}, \citenamefont {Fang},
  \citenamefont {Tomarken}, \citenamefont {Luo}, \citenamefont
  {Sanchez-Yamagishi}, \citenamefont {Watanabe}, \citenamefont {Taniguchi},
  \citenamefont {Kaxiras}, \citenamefont {Ashoori},\ and\ \citenamefont
  {Jarillo-Herrero}}]{Cao2018_2}%
  \BibitemOpen
  \bibfield  {author} {\bibinfo {author} {\bibfnamefont {Y.}~\bibnamefont
  {Cao}}, \bibinfo {author} {\bibfnamefont {V.}~\bibnamefont {Fatemi}},
  \bibinfo {author} {\bibfnamefont {A.}~\bibnamefont {Demir}}, \bibinfo
  {author} {\bibfnamefont {S.}~\bibnamefont {Fang}}, \bibinfo {author}
  {\bibfnamefont {S.~L.}\ \bibnamefont {Tomarken}}, \bibinfo {author}
  {\bibfnamefont {J.~Y.}\ \bibnamefont {Luo}}, \bibinfo {author} {\bibfnamefont
  {J.~D.}\ \bibnamefont {Sanchez-Yamagishi}}, \bibinfo {author} {\bibfnamefont
  {K.}~\bibnamefont {Watanabe}}, \bibinfo {author} {\bibfnamefont
  {T.}~\bibnamefont {Taniguchi}}, \bibinfo {author} {\bibfnamefont
  {E.}~\bibnamefont {Kaxiras}}, \bibinfo {author} {\bibfnamefont {R.~C.}\
  \bibnamefont {Ashoori}},\ and\ \bibinfo {author} {\bibfnamefont
  {P.}~\bibnamefont {Jarillo-Herrero}},\ }\bibfield  {title} {\bibinfo {title}
  {Correlated insulator behaviour at half-filling in magic-angle graphene
  superlattices},\ }\href {https://doi.org/10.1038/nature26154} {\bibfield
  {journal} {\bibinfo  {journal} {Nature}\ }\textbf {\bibinfo {volume} {556}},\
  \bibinfo {pages} {80} (\bibinfo {year} {2018}{\natexlab{b}})}\BibitemShut
  {NoStop}%
\bibitem [{\citenamefont {Lu}\ \emph {et~al.}(2019)\citenamefont {Lu},
  \citenamefont {Stepanov}, \citenamefont {Yang}, \citenamefont {Xie},
  \citenamefont {Aamir}, \citenamefont {Das}, \citenamefont {Urgell},
  \citenamefont {Watanabe}, \citenamefont {Taniguchi}, \citenamefont {Zhang},
  \citenamefont {Bachtold}, \citenamefont {MacDonald},\ and\ \citenamefont
  {Efetov}}]{Lu2019}%
  \BibitemOpen
  \bibfield  {author} {\bibinfo {author} {\bibfnamefont {X.}~\bibnamefont
  {Lu}}, \bibinfo {author} {\bibfnamefont {P.}~\bibnamefont {Stepanov}},
  \bibinfo {author} {\bibfnamefont {W.}~\bibnamefont {Yang}}, \bibinfo {author}
  {\bibfnamefont {M.}~\bibnamefont {Xie}}, \bibinfo {author} {\bibfnamefont
  {M.~A.}\ \bibnamefont {Aamir}}, \bibinfo {author} {\bibfnamefont
  {I.}~\bibnamefont {Das}}, \bibinfo {author} {\bibfnamefont {C.}~\bibnamefont
  {Urgell}}, \bibinfo {author} {\bibfnamefont {K.}~\bibnamefont {Watanabe}},
  \bibinfo {author} {\bibfnamefont {T.}~\bibnamefont {Taniguchi}}, \bibinfo
  {author} {\bibfnamefont {G.}~\bibnamefont {Zhang}}, \bibinfo {author}
  {\bibfnamefont {A.}~\bibnamefont {Bachtold}}, \bibinfo {author}
  {\bibfnamefont {A.~H.}\ \bibnamefont {MacDonald}},\ and\ \bibinfo {author}
  {\bibfnamefont {D.~K.}\ \bibnamefont {Efetov}},\ }\bibfield  {title}
  {\bibinfo {title} {Superconductors, orbital magnets and correlated states in
  magic-angle bilayer graphene},\ }\href
  {https://doi.org/10.1038/s41586-019-1695-0} {\bibfield  {journal} {\bibinfo
  {journal} {Nature}\ }\textbf {\bibinfo {volume} {574}},\ \bibinfo {pages}
  {653} (\bibinfo {year} {2019})}\BibitemShut {NoStop}%
\bibitem [{\citenamefont {Choi}\ \emph {et~al.}(2019)\citenamefont {Choi},
  \citenamefont {Kemmer}, \citenamefont {Peng}, \citenamefont {Thomson},
  \citenamefont {Arora}, \citenamefont {Polski}, \citenamefont {Zhang},
  \citenamefont {Ren}, \citenamefont {Alicea}, \citenamefont {Refael},
  \citenamefont {von Oppen}, \citenamefont {Watanabe}, \citenamefont
  {Taniguchi},\ and\ \citenamefont {Nadj-Perge}}]{Choi2019}%
  \BibitemOpen
  \bibfield  {author} {\bibinfo {author} {\bibfnamefont {Y.}~\bibnamefont
  {Choi}}, \bibinfo {author} {\bibfnamefont {J.}~\bibnamefont {Kemmer}},
  \bibinfo {author} {\bibfnamefont {Y.}~\bibnamefont {Peng}}, \bibinfo {author}
  {\bibfnamefont {A.}~\bibnamefont {Thomson}}, \bibinfo {author} {\bibfnamefont
  {H.}~\bibnamefont {Arora}}, \bibinfo {author} {\bibfnamefont
  {R.}~\bibnamefont {Polski}}, \bibinfo {author} {\bibfnamefont
  {Y.}~\bibnamefont {Zhang}}, \bibinfo {author} {\bibfnamefont
  {H.}~\bibnamefont {Ren}}, \bibinfo {author} {\bibfnamefont {J.}~\bibnamefont
  {Alicea}}, \bibinfo {author} {\bibfnamefont {G.}~\bibnamefont {Refael}},
  \bibinfo {author} {\bibfnamefont {F.}~\bibnamefont {von Oppen}}, \bibinfo
  {author} {\bibfnamefont {K.}~\bibnamefont {Watanabe}}, \bibinfo {author}
  {\bibfnamefont {T.}~\bibnamefont {Taniguchi}},\ and\ \bibinfo {author}
  {\bibfnamefont {S.}~\bibnamefont {Nadj-Perge}},\ }\bibfield  {title}
  {\bibinfo {title} {Electronic correlations in twisted bilayer graphene near
  the magic angle},\ }\href {https://doi.org/10.1038/s41567-019-0606-5}
  {\bibfield  {journal} {\bibinfo  {journal} {Nature Physics}\ }\textbf
  {\bibinfo {volume} {15}},\ \bibinfo {pages} {1174} (\bibinfo {year}
  {2019})}\BibitemShut {NoStop}%
\bibitem [{\citenamefont {Wong}\ \emph {et~al.}(2020)\citenamefont {Wong},
  \citenamefont {Nuckolls}, \citenamefont {Oh}, \citenamefont {Lian},
  \citenamefont {Xie}, \citenamefont {Jeon}, \citenamefont {Watanabe},
  \citenamefont {Taniguchi}, \citenamefont {Bernevig},\ and\ \citenamefont
  {Yazdani}}]{Wong2020}%
  \BibitemOpen
  \bibfield  {author} {\bibinfo {author} {\bibfnamefont {D.}~\bibnamefont
  {Wong}}, \bibinfo {author} {\bibfnamefont {K.~P.}\ \bibnamefont {Nuckolls}},
  \bibinfo {author} {\bibfnamefont {M.}~\bibnamefont {Oh}}, \bibinfo {author}
  {\bibfnamefont {B.}~\bibnamefont {Lian}}, \bibinfo {author} {\bibfnamefont
  {Y.}~\bibnamefont {Xie}}, \bibinfo {author} {\bibfnamefont {S.}~\bibnamefont
  {Jeon}}, \bibinfo {author} {\bibfnamefont {K.}~\bibnamefont {Watanabe}},
  \bibinfo {author} {\bibfnamefont {T.}~\bibnamefont {Taniguchi}}, \bibinfo
  {author} {\bibfnamefont {B.~A.}\ \bibnamefont {Bernevig}},\ and\ \bibinfo
  {author} {\bibfnamefont {A.}~\bibnamefont {Yazdani}},\ }\bibfield  {title}
  {\bibinfo {title} {Cascade of electronic transitions in magic-angle twisted
  bilayer graphene},\ }\href {https://doi.org/10.1038/s41586-020-2339-0}
  {\bibfield  {journal} {\bibinfo  {journal} {Nature}\ }\textbf {\bibinfo
  {volume} {582}},\ \bibinfo {pages} {198} (\bibinfo {year}
  {2020})}\BibitemShut {NoStop}%
\bibitem [{\citenamefont {Zondiner}\ \emph {et~al.}(2020)\citenamefont
  {Zondiner}, \citenamefont {Rozen}, \citenamefont {Rodan-Legrain},
  \citenamefont {Cao}, \citenamefont {Queiroz}, \citenamefont {Taniguchi},
  \citenamefont {Watanabe}, \citenamefont {Oreg}, \citenamefont {von Oppen},
  \citenamefont {Stern}, \citenamefont {Berg}, \citenamefont
  {Jarillo-Herrero},\ and\ \citenamefont {Ilani}}]{Zondiner2020}%
  \BibitemOpen
  \bibfield  {author} {\bibinfo {author} {\bibfnamefont {U.}~\bibnamefont
  {Zondiner}}, \bibinfo {author} {\bibfnamefont {A.}~\bibnamefont {Rozen}},
  \bibinfo {author} {\bibfnamefont {D.}~\bibnamefont {Rodan-Legrain}}, \bibinfo
  {author} {\bibfnamefont {Y.}~\bibnamefont {Cao}}, \bibinfo {author}
  {\bibfnamefont {R.}~\bibnamefont {Queiroz}}, \bibinfo {author} {\bibfnamefont
  {T.}~\bibnamefont {Taniguchi}}, \bibinfo {author} {\bibfnamefont
  {K.}~\bibnamefont {Watanabe}}, \bibinfo {author} {\bibfnamefont
  {Y.}~\bibnamefont {Oreg}}, \bibinfo {author} {\bibfnamefont {F.}~\bibnamefont
  {von Oppen}}, \bibinfo {author} {\bibfnamefont {A.}~\bibnamefont {Stern}},
  \bibinfo {author} {\bibfnamefont {E.}~\bibnamefont {Berg}}, \bibinfo {author}
  {\bibfnamefont {P.}~\bibnamefont {Jarillo-Herrero}},\ and\ \bibinfo {author}
  {\bibfnamefont {S.}~\bibnamefont {Ilani}},\ }\bibfield  {title} {\bibinfo
  {title} {Cascade of phase transitions and dirac revivals in magic-angle
  graphene},\ }\href {https://doi.org/10.1038/s41586-020-2373-y} {\bibfield
  {journal} {\bibinfo  {journal} {Nature}\ }\textbf {\bibinfo {volume} {582}},\
  \bibinfo {pages} {203} (\bibinfo {year} {2020})}\BibitemShut {NoStop}%
\bibitem [{\citenamefont {Nuckolls}\ \emph {et~al.}(2020)\citenamefont
  {Nuckolls}, \citenamefont {Oh}, \citenamefont {Wong}, \citenamefont {Lian},
  \citenamefont {Watanabe}, \citenamefont {Taniguchi}, \citenamefont
  {Bernevig},\ and\ \citenamefont {Yazdani}}]{Nuckolls2020}%
  \BibitemOpen
  \bibfield  {author} {\bibinfo {author} {\bibfnamefont {K.~P.}\ \bibnamefont
  {Nuckolls}}, \bibinfo {author} {\bibfnamefont {M.}~\bibnamefont {Oh}},
  \bibinfo {author} {\bibfnamefont {D.}~\bibnamefont {Wong}}, \bibinfo {author}
  {\bibfnamefont {B.}~\bibnamefont {Lian}}, \bibinfo {author} {\bibfnamefont
  {K.}~\bibnamefont {Watanabe}}, \bibinfo {author} {\bibfnamefont
  {T.}~\bibnamefont {Taniguchi}}, \bibinfo {author} {\bibfnamefont {B.~A.}\
  \bibnamefont {Bernevig}},\ and\ \bibinfo {author} {\bibfnamefont
  {A.}~\bibnamefont {Yazdani}},\ }\bibfield  {title} {\bibinfo {title}
  {Strongly correlated chern insulators in magic-angle twisted bilayer
  graphene},\ }\href {https://doi.org/10.1038/s41586-020-3028-8} {\bibfield
  {journal} {\bibinfo  {journal} {Nature}\ }\textbf {\bibinfo {volume} {588}},\
  \bibinfo {pages} {610} (\bibinfo {year} {2020})}\BibitemShut {NoStop}%
\bibitem [{\citenamefont {Liu}\ \emph {et~al.}(2022)\citenamefont {Liu},
  \citenamefont {Zhang}, \citenamefont {Watanabe}, \citenamefont {Taniguchi},\
  and\ \citenamefont {Li}}]{Liu2022}%
  \BibitemOpen
  \bibfield  {author} {\bibinfo {author} {\bibfnamefont {X.}~\bibnamefont
  {Liu}}, \bibinfo {author} {\bibfnamefont {N.~J.}\ \bibnamefont {Zhang}},
  \bibinfo {author} {\bibfnamefont {K.}~\bibnamefont {Watanabe}}, \bibinfo
  {author} {\bibfnamefont {T.}~\bibnamefont {Taniguchi}},\ and\ \bibinfo
  {author} {\bibfnamefont {J.~I.~A.}\ \bibnamefont {Li}},\ }\bibfield  {title}
  {\bibinfo {title} {Isospin order in superconducting magic-angle twisted
  trilayer graphene},\ }\href {https://doi.org/10.1038/s41567-022-01515-0}
  {\bibfield  {journal} {\bibinfo  {journal} {Nature Physics}\ }\textbf
  {\bibinfo {volume} {18}},\ \bibinfo {pages} {522} (\bibinfo {year}
  {2022})}\BibitemShut {NoStop}%
\bibitem [{\citenamefont {Song}\ and\ \citenamefont
  {Bernevig}(2022)}]{PhysRevLett.129.047601}%
  \BibitemOpen
  \bibfield  {author} {\bibinfo {author} {\bibfnamefont {Z.-D.}\ \bibnamefont
  {Song}}\ and\ \bibinfo {author} {\bibfnamefont {B.~A.}\ \bibnamefont
  {Bernevig}},\ }\bibfield  {title} {\bibinfo {title} {Magic-angle twisted
  bilayer graphene as a topological heavy fermion problem},\ }\href
  {https://doi.org/10.1103/PhysRevLett.129.047601} {\bibfield  {journal}
  {\bibinfo  {journal} {Phys. Rev. Lett.}\ }\textbf {\bibinfo {volume} {129}},\
  \bibinfo {pages} {047601} (\bibinfo {year} {2022})}\BibitemShut {NoStop}%
\bibitem [{\citenamefont {Datta}\ \emph {et~al.}(2023)\citenamefont {Datta},
  \citenamefont {Calderón}, \citenamefont {Camjayi},\ and\ \citenamefont
  {Bascones}}]{Datta2023}%
  \BibitemOpen
  \bibfield  {author} {\bibinfo {author} {\bibfnamefont {A.}~\bibnamefont
  {Datta}}, \bibinfo {author} {\bibfnamefont {M.~J.}\ \bibnamefont
  {Calderón}}, \bibinfo {author} {\bibfnamefont {A.}~\bibnamefont {Camjayi}},\
  and\ \bibinfo {author} {\bibfnamefont {E.}~\bibnamefont {Bascones}},\
  }\bibfield  {title} {\bibinfo {title} {Heavy quasiparticles and cascades
  without symmetry breaking in twisted bilayer graphene},\ }\href
  {https://doi.org/10.1038/s41467-023-40754-4} {\bibfield  {journal} {\bibinfo
  {journal} {Nature Communications}\ }\textbf {\bibinfo {volume} {14}},\
  \bibinfo {pages} {5036} (\bibinfo {year} {2023})}\BibitemShut {NoStop}%
\bibitem [{\citenamefont {Hu}\ \emph {et~al.}(2023)\citenamefont {Hu},
  \citenamefont {Rai}, \citenamefont {Crippa}, \citenamefont
  {Herzog-Arbeitman}, \citenamefont {C\ifmmode \u{a}\else
  \u{a}\fi{}lug\ifmmode~\u{a}\else \u{a}\fi{}ru}, \citenamefont {Wehling},
  \citenamefont {Sangiovanni}, \citenamefont {Valent\'{\i}}, \citenamefont
  {Tsvelik},\ and\ \citenamefont {Bernevig}}]{PhysRevLett.131.166501}%
  \BibitemOpen
  \bibfield  {author} {\bibinfo {author} {\bibfnamefont {H.}~\bibnamefont
  {Hu}}, \bibinfo {author} {\bibfnamefont {G.}~\bibnamefont {Rai}}, \bibinfo
  {author} {\bibfnamefont {L.}~\bibnamefont {Crippa}}, \bibinfo {author}
  {\bibfnamefont {J.}~\bibnamefont {Herzog-Arbeitman}}, \bibinfo {author}
  {\bibfnamefont {D.}~\bibnamefont {C\ifmmode \u{a}\else
  \u{a}\fi{}lug\ifmmode~\u{a}\else \u{a}\fi{}ru}}, \bibinfo {author}
  {\bibfnamefont {T.}~\bibnamefont {Wehling}}, \bibinfo {author} {\bibfnamefont
  {G.}~\bibnamefont {Sangiovanni}}, \bibinfo {author} {\bibfnamefont
  {R.}~\bibnamefont {Valent\'{\i}}}, \bibinfo {author} {\bibfnamefont {A.~M.}\
  \bibnamefont {Tsvelik}},\ and\ \bibinfo {author} {\bibfnamefont {B.~A.}\
  \bibnamefont {Bernevig}},\ }\bibfield  {title} {\bibinfo {title} {Symmetric
  kondo lattice states in doped strained twisted bilayer graphene},\ }\href
  {https://doi.org/10.1103/PhysRevLett.131.166501} {\bibfield  {journal}
  {\bibinfo  {journal} {Phys. Rev. Lett.}\ }\textbf {\bibinfo {volume} {131}},\
  \bibinfo {pages} {166501} (\bibinfo {year} {2023})}\BibitemShut {NoStop}%
\bibitem [{\citenamefont {Rai}\ \emph {et~al.}(2023)\citenamefont {Rai},
  \citenamefont {Crippa}, \citenamefont {Călugăru}, \citenamefont {Hu},
  \citenamefont {de' Medici}, \citenamefont {Georges}, \citenamefont
  {Bernevig}, \citenamefont {Valentí}, \citenamefont {Sangiovanni},\ and\
  \citenamefont {Wehling}}]{rai2023dynamical}%
  \BibitemOpen
  \bibfield  {author} {\bibinfo {author} {\bibfnamefont {G.}~\bibnamefont
  {Rai}}, \bibinfo {author} {\bibfnamefont {L.}~\bibnamefont {Crippa}},
  \bibinfo {author} {\bibfnamefont {D.}~\bibnamefont {Călugăru}}, \bibinfo
  {author} {\bibfnamefont {H.}~\bibnamefont {Hu}}, \bibinfo {author}
  {\bibfnamefont {L.}~\bibnamefont {de' Medici}}, \bibinfo {author}
  {\bibfnamefont {A.}~\bibnamefont {Georges}}, \bibinfo {author} {\bibfnamefont
  {B.~A.}\ \bibnamefont {Bernevig}}, \bibinfo {author} {\bibfnamefont
  {R.}~\bibnamefont {Valentí}}, \bibinfo {author} {\bibfnamefont
  {G.}~\bibnamefont {Sangiovanni}},\ and\ \bibinfo {author} {\bibfnamefont
  {T.}~\bibnamefont {Wehling}},\ }\href@noop {} {\bibinfo {title} {Dynamical
  correlations and order in magic-angle twisted bilayer graphene}} (\bibinfo
  {year} {2023}),\ \Eprint {https://arxiv.org/abs/2309.08529} {arXiv:2309.08529
  [cond-mat.str-el]} \BibitemShut {NoStop}%
\bibitem [{\citenamefont {Guerci}\ \emph
  {et~al.}(2023{\natexlab{a}})\citenamefont {Guerci}, \citenamefont {Wang},
  \citenamefont {Zang}, \citenamefont {Cano}, \citenamefont {Pixley},\ and\
  \citenamefont {Millis}}]{Guerci_2023}%
  \BibitemOpen
  \bibfield  {author} {\bibinfo {author} {\bibfnamefont {D.}~\bibnamefont
  {Guerci}}, \bibinfo {author} {\bibfnamefont {J.}~\bibnamefont {Wang}},
  \bibinfo {author} {\bibfnamefont {J.}~\bibnamefont {Zang}}, \bibinfo {author}
  {\bibfnamefont {J.}~\bibnamefont {Cano}}, \bibinfo {author} {\bibfnamefont
  {J.~H.}\ \bibnamefont {Pixley}},\ and\ \bibinfo {author} {\bibfnamefont
  {A.}~\bibnamefont {Millis}},\ }\bibfield  {title} {\bibinfo {title} {Chiral
  kondo lattice in doped mote2/wse2 bilayers},\ }\bibfield  {journal} {\bibinfo
   {journal} {Science Advances}\ }\textbf {\bibinfo {volume} {9}},\ \href
  {https://doi.org/10.1126/sciadv.ade7701} {10.1126/sciadv.ade7701} (\bibinfo
  {year} {2023}{\natexlab{a}})\BibitemShut {NoStop}%
\bibitem [{\citenamefont {Ramires}\ and\ \citenamefont
  {Lado}(2021)}]{Ramires2021}%
  \BibitemOpen
  \bibfield  {author} {\bibinfo {author} {\bibfnamefont {A.}~\bibnamefont
  {Ramires}}\ and\ \bibinfo {author} {\bibfnamefont {J.~L.}\ \bibnamefont
  {Lado}},\ }\bibfield  {title} {\bibinfo {title} {Emulating heavy fermions in
  twisted trilayer graphene},\ }\href
  {https://doi.org/10.1103/PhysRevLett.127.026401} {\bibfield  {journal}
  {\bibinfo  {journal} {Phys. Rev. Lett.}\ }\textbf {\bibinfo {volume} {127}},\
  \bibinfo {pages} {026401} (\bibinfo {year} {2021})}\BibitemShut {NoStop}%
\bibitem [{\citenamefont {Yu}\ \emph {et~al.}(2023)\citenamefont {Yu},
  \citenamefont {Xie}, \citenamefont {Bernevig},\ and\ \citenamefont
  {Das~Sarma}}]{Jiabin2023}%
  \BibitemOpen
  \bibfield  {author} {\bibinfo {author} {\bibfnamefont {J.}~\bibnamefont
  {Yu}}, \bibinfo {author} {\bibfnamefont {M.}~\bibnamefont {Xie}}, \bibinfo
  {author} {\bibfnamefont {B.~A.}\ \bibnamefont {Bernevig}},\ and\ \bibinfo
  {author} {\bibfnamefont {S.}~\bibnamefont {Das~Sarma}},\ }\bibfield  {title}
  {\bibinfo {title} {Magic-angle twisted symmetric trilayer graphene as a
  topological heavy-fermion problem},\ }\href
  {https://doi.org/10.1103/PhysRevB.108.035129} {\bibfield  {journal} {\bibinfo
   {journal} {Phys. Rev. B}\ }\textbf {\bibinfo {volume} {108}},\ \bibinfo
  {pages} {035129} (\bibinfo {year} {2023})}\BibitemShut {NoStop}%
\bibitem [{\citenamefont {Merino}\ \emph {et~al.}(2024)\citenamefont {Merino},
  \citenamefont {Calugaru}, \citenamefont {Hu}, \citenamefont {Diez-Merida},
  \citenamefont {Diez-Carlon}, \citenamefont {Taniguchi}, \citenamefont
  {Watanabe}, \citenamefont {Seifert}, \citenamefont {Bernevig},\ and\
  \citenamefont {Efetov}}]{merino2024evidence}%
  \BibitemOpen
  \bibfield  {author} {\bibinfo {author} {\bibfnamefont {R.~L.}\ \bibnamefont
  {Merino}}, \bibinfo {author} {\bibfnamefont {D.}~\bibnamefont {Calugaru}},
  \bibinfo {author} {\bibfnamefont {H.}~\bibnamefont {Hu}}, \bibinfo {author}
  {\bibfnamefont {J.}~\bibnamefont {Diez-Merida}}, \bibinfo {author}
  {\bibfnamefont {A.}~\bibnamefont {Diez-Carlon}}, \bibinfo {author}
  {\bibfnamefont {T.}~\bibnamefont {Taniguchi}}, \bibinfo {author}
  {\bibfnamefont {K.}~\bibnamefont {Watanabe}}, \bibinfo {author}
  {\bibfnamefont {P.}~\bibnamefont {Seifert}}, \bibinfo {author} {\bibfnamefont
  {B.~A.}\ \bibnamefont {Bernevig}},\ and\ \bibinfo {author} {\bibfnamefont
  {D.~K.}\ \bibnamefont {Efetov}},\ }\href@noop {} {\bibinfo {title} {Evidence
  of heavy fermion physics in the thermoelectric transport of magic angle
  twisted bilayer graphene}} (\bibinfo {year} {2024}),\ \Eprint
  {https://arxiv.org/abs/2402.11749} {arXiv:2402.11749 [cond-mat.mes-hall]}
  \BibitemShut {NoStop}%
\bibitem [{\citenamefont {Polshyn}\ \emph {et~al.}(2022)\citenamefont
  {Polshyn}, \citenamefont {Zhang}, \citenamefont {Kumar}, \citenamefont
  {Soejima}, \citenamefont {Ledwith}, \citenamefont {Watanabe}, \citenamefont
  {Taniguchi}, \citenamefont {Vishwanath}, \citenamefont {Zaletel},\ and\
  \citenamefont {Young}}]{Polshyn2022}%
  \BibitemOpen
  \bibfield  {author} {\bibinfo {author} {\bibfnamefont {H.}~\bibnamefont
  {Polshyn}}, \bibinfo {author} {\bibfnamefont {Y.}~\bibnamefont {Zhang}},
  \bibinfo {author} {\bibfnamefont {M.~A.}\ \bibnamefont {Kumar}}, \bibinfo
  {author} {\bibfnamefont {T.}~\bibnamefont {Soejima}}, \bibinfo {author}
  {\bibfnamefont {P.}~\bibnamefont {Ledwith}}, \bibinfo {author} {\bibfnamefont
  {K.}~\bibnamefont {Watanabe}}, \bibinfo {author} {\bibfnamefont
  {T.}~\bibnamefont {Taniguchi}}, \bibinfo {author} {\bibfnamefont
  {A.}~\bibnamefont {Vishwanath}}, \bibinfo {author} {\bibfnamefont {M.~P.}\
  \bibnamefont {Zaletel}},\ and\ \bibinfo {author} {\bibfnamefont {A.~F.}\
  \bibnamefont {Young}},\ }\bibfield  {title} {\bibinfo {title} {Topological
  charge density waves at half-integer filling of a moiré superlattice},\
  }\href {https://doi.org/10.1038/s41567-021-01418-6} {\bibfield  {journal}
  {\bibinfo  {journal} {Nature Physics}\ }\textbf {\bibinfo {volume} {18}},\
  \bibinfo {pages} {42} (\bibinfo {year} {2022})}\BibitemShut {NoStop}%
\bibitem [{\citenamefont {Jiang}\ \emph {et~al.}(2019)\citenamefont {Jiang},
  \citenamefont {Lai}, \citenamefont {Watanabe}, \citenamefont {Taniguchi},
  \citenamefont {Haule}, \citenamefont {Mao},\ and\ \citenamefont
  {Andrei}}]{Jiang2019}%
  \BibitemOpen
  \bibfield  {author} {\bibinfo {author} {\bibfnamefont {Y.}~\bibnamefont
  {Jiang}}, \bibinfo {author} {\bibfnamefont {X.}~\bibnamefont {Lai}}, \bibinfo
  {author} {\bibfnamefont {K.}~\bibnamefont {Watanabe}}, \bibinfo {author}
  {\bibfnamefont {T.}~\bibnamefont {Taniguchi}}, \bibinfo {author}
  {\bibfnamefont {K.}~\bibnamefont {Haule}}, \bibinfo {author} {\bibfnamefont
  {J.}~\bibnamefont {Mao}},\ and\ \bibinfo {author} {\bibfnamefont {E.~Y.}\
  \bibnamefont {Andrei}},\ }\bibfield  {title} {\bibinfo {title} {Charge order
  and broken rotational symmetry in magic-angle twisted bilayer graphene},\
  }\href {https://doi.org/10.1038/s41586-019-1460-4} {\bibfield  {journal}
  {\bibinfo  {journal} {Nature}\ }\textbf {\bibinfo {volume} {573}},\ \bibinfo
  {pages} {91} (\bibinfo {year} {2019})}\BibitemShut {NoStop}%
\bibitem [{\citenamefont {Zeng}\ \emph {et~al.}(2023)\citenamefont {Zeng},
  \citenamefont {Xia}, \citenamefont {Kang}, \citenamefont {Zhu}, \citenamefont
  {Kn{\textsurd}{\textordmasculine}ppel}, \citenamefont {Vaswani},
  \citenamefont {Watanabe}, \citenamefont {Taniguchi}, \citenamefont {Mak},\
  and\ \citenamefont {Shan}}]{Zeng2023}%
  \BibitemOpen
  \bibfield  {author} {\bibinfo {author} {\bibfnamefont {Y.}~\bibnamefont
  {Zeng}}, \bibinfo {author} {\bibfnamefont {Z.}~\bibnamefont {Xia}}, \bibinfo
  {author} {\bibfnamefont {K.}~\bibnamefont {Kang}}, \bibinfo {author}
  {\bibfnamefont {J.}~\bibnamefont {Zhu}}, \bibinfo {author} {\bibfnamefont
  {P.}~\bibnamefont {Kn{\textsurd}{\textordmasculine}ppel}}, \bibinfo {author}
  {\bibfnamefont {C.}~\bibnamefont {Vaswani}}, \bibinfo {author} {\bibfnamefont
  {K.}~\bibnamefont {Watanabe}}, \bibinfo {author} {\bibfnamefont
  {T.}~\bibnamefont {Taniguchi}}, \bibinfo {author} {\bibfnamefont {K.~F.}\
  \bibnamefont {Mak}},\ and\ \bibinfo {author} {\bibfnamefont {J.}~\bibnamefont
  {Shan}},\ }\bibfield  {title} {\bibinfo {title} {Thermodynamic evidence of
  fractional chern insulator in moiré mote2},\ }\href
  {https://doi.org/10.1038/s41586-023-06452-3} {\bibfield  {journal} {\bibinfo
  {journal} {Nature}\ }\textbf {\bibinfo {volume} {622}},\ \bibinfo {pages}
  {69} (\bibinfo {year} {2023})}\BibitemShut {NoStop}%
\bibitem [{\citenamefont {Cai}\ \emph {et~al.}(2023)\citenamefont {Cai},
  \citenamefont {Anderson}, \citenamefont {Wang}, \citenamefont {Zhang},
  \citenamefont {Liu}, \citenamefont {Holtzmann}, \citenamefont {Zhang},
  \citenamefont {Fan}, \citenamefont {Taniguchi}, \citenamefont {Watanabe},
  \citenamefont {Ran}, \citenamefont {Cao}, \citenamefont {Fu}, \citenamefont
  {Xiao}, \citenamefont {Yao},\ and\ \citenamefont {Xu}}]{Cai2023}%
  \BibitemOpen
  \bibfield  {author} {\bibinfo {author} {\bibfnamefont {J.}~\bibnamefont
  {Cai}}, \bibinfo {author} {\bibfnamefont {E.}~\bibnamefont {Anderson}},
  \bibinfo {author} {\bibfnamefont {C.}~\bibnamefont {Wang}}, \bibinfo {author}
  {\bibfnamefont {X.}~\bibnamefont {Zhang}}, \bibinfo {author} {\bibfnamefont
  {X.}~\bibnamefont {Liu}}, \bibinfo {author} {\bibfnamefont {W.}~\bibnamefont
  {Holtzmann}}, \bibinfo {author} {\bibfnamefont {Y.}~\bibnamefont {Zhang}},
  \bibinfo {author} {\bibfnamefont {F.}~\bibnamefont {Fan}}, \bibinfo {author}
  {\bibfnamefont {T.}~\bibnamefont {Taniguchi}}, \bibinfo {author}
  {\bibfnamefont {K.}~\bibnamefont {Watanabe}}, \bibinfo {author}
  {\bibfnamefont {Y.}~\bibnamefont {Ran}}, \bibinfo {author} {\bibfnamefont
  {T.}~\bibnamefont {Cao}}, \bibinfo {author} {\bibfnamefont {L.}~\bibnamefont
  {Fu}}, \bibinfo {author} {\bibfnamefont {D.}~\bibnamefont {Xiao}}, \bibinfo
  {author} {\bibfnamefont {W.}~\bibnamefont {Yao}},\ and\ \bibinfo {author}
  {\bibfnamefont {X.}~\bibnamefont {Xu}},\ }\bibfield  {title} {\bibinfo
  {title} {Signatures of fractional quantum anomalous hall states in twisted
  mote2},\ }\href {https://doi.org/10.1038/s41586-023-06289-w} {\bibfield
  {journal} {\bibinfo  {journal} {Nature}\ }\textbf {\bibinfo {volume} {622}},\
  \bibinfo {pages} {63} (\bibinfo {year} {2023})}\BibitemShut {NoStop}%
\bibitem [{\citenamefont {Xie}\ \emph {et~al.}(2021)\citenamefont {Xie},
  \citenamefont {Pierce}, \citenamefont {Park}, \citenamefont {Parker},
  \citenamefont {Khalaf}, \citenamefont {Ledwith}, \citenamefont {Cao},
  \citenamefont {Lee}, \citenamefont {Chen}, \citenamefont {Forrester},
  \citenamefont {Watanabe}, \citenamefont {Taniguchi}, \citenamefont
  {Vishwanath}, \citenamefont {Jarillo-Herrero},\ and\ \citenamefont
  {Yacoby}}]{Xie2021}%
  \BibitemOpen
  \bibfield  {author} {\bibinfo {author} {\bibfnamefont {Y.}~\bibnamefont
  {Xie}}, \bibinfo {author} {\bibfnamefont {A.~T.}\ \bibnamefont {Pierce}},
  \bibinfo {author} {\bibfnamefont {J.~M.}\ \bibnamefont {Park}}, \bibinfo
  {author} {\bibfnamefont {D.~E.}\ \bibnamefont {Parker}}, \bibinfo {author}
  {\bibfnamefont {E.}~\bibnamefont {Khalaf}}, \bibinfo {author} {\bibfnamefont
  {P.}~\bibnamefont {Ledwith}}, \bibinfo {author} {\bibfnamefont
  {Y.}~\bibnamefont {Cao}}, \bibinfo {author} {\bibfnamefont {S.~H.}\
  \bibnamefont {Lee}}, \bibinfo {author} {\bibfnamefont {S.}~\bibnamefont
  {Chen}}, \bibinfo {author} {\bibfnamefont {P.~R.}\ \bibnamefont {Forrester}},
  \bibinfo {author} {\bibfnamefont {K.}~\bibnamefont {Watanabe}}, \bibinfo
  {author} {\bibfnamefont {T.}~\bibnamefont {Taniguchi}}, \bibinfo {author}
  {\bibfnamefont {A.}~\bibnamefont {Vishwanath}}, \bibinfo {author}
  {\bibfnamefont {P.}~\bibnamefont {Jarillo-Herrero}},\ and\ \bibinfo {author}
  {\bibfnamefont {A.}~\bibnamefont {Yacoby}},\ }\bibfield  {title} {\bibinfo
  {title} {Fractional chern insulators in magic-angle twisted bilayer
  graphene},\ }\href {https://doi.org/10.1038/s41586-021-04002-3} {\bibfield
  {journal} {\bibinfo  {journal} {Nature}\ }\textbf {\bibinfo {volume} {600}},\
  \bibinfo {pages} {439} (\bibinfo {year} {2021})}\BibitemShut {NoStop}%
\bibitem [{\citenamefont {Yu}\ \emph {et~al.}(2024)\citenamefont {Yu},
  \citenamefont {Herzog-Arbeitman}, \citenamefont {Wang}, \citenamefont
  {Vafek}, \citenamefont {Bernevig},\ and\ \citenamefont
  {Regnault}}]{Jiabin2024}%
  \BibitemOpen
  \bibfield  {author} {\bibinfo {author} {\bibfnamefont {J.}~\bibnamefont
  {Yu}}, \bibinfo {author} {\bibfnamefont {J.}~\bibnamefont
  {Herzog-Arbeitman}}, \bibinfo {author} {\bibfnamefont {M.}~\bibnamefont
  {Wang}}, \bibinfo {author} {\bibfnamefont {O.}~\bibnamefont {Vafek}},
  \bibinfo {author} {\bibfnamefont {B.~A.}\ \bibnamefont {Bernevig}},\ and\
  \bibinfo {author} {\bibfnamefont {N.}~\bibnamefont {Regnault}},\ }\bibfield
  {title} {\bibinfo {title} {Fractional chern insulators versus nonmagnetic
  states in twisted bilayer ${\mathrm{mote}}_{2}$},\ }\href
  {https://doi.org/10.1103/PhysRevB.109.045147} {\bibfield  {journal} {\bibinfo
   {journal} {Phys. Rev. B}\ }\textbf {\bibinfo {volume} {109}},\ \bibinfo
  {pages} {045147} (\bibinfo {year} {2024})}\BibitemShut {NoStop}%
\bibitem [{\citenamefont {Morales-Dur\'an}\ \emph {et~al.}(2024)\citenamefont
  {Morales-Dur\'an}, \citenamefont {Wei}, \citenamefont {Shi},\ and\
  \citenamefont {MacDonald}}]{moralesduran2024}%
  \BibitemOpen
  \bibfield  {author} {\bibinfo {author} {\bibfnamefont {N.}~\bibnamefont
  {Morales-Dur\'an}}, \bibinfo {author} {\bibfnamefont {N.}~\bibnamefont
  {Wei}}, \bibinfo {author} {\bibfnamefont {J.}~\bibnamefont {Shi}},\ and\
  \bibinfo {author} {\bibfnamefont {A.~H.}\ \bibnamefont {MacDonald}},\
  }\bibfield  {title} {\bibinfo {title} {Magic angles and fractional chern
  insulators in twisted homobilayer transition metal dichalcogenides},\ }\href
  {https://doi.org/10.1103/PhysRevLett.132.096602} {\bibfield  {journal}
  {\bibinfo  {journal} {Phys. Rev. Lett.}\ }\textbf {\bibinfo {volume} {132}},\
  \bibinfo {pages} {096602} (\bibinfo {year} {2024})}\BibitemShut {NoStop}%
\bibitem [{\citenamefont {Sharpe}\ \emph {et~al.}(2019)\citenamefont {Sharpe},
  \citenamefont {Fox}, \citenamefont {Barnard}, \citenamefont {Finney},
  \citenamefont {Watanabe}, \citenamefont {Taniguchi}, \citenamefont
  {Kastner},\ and\ \citenamefont
  {Goldhaber-Gordon}}]{doi:10.1126/science.aaw3780}%
  \BibitemOpen
  \bibfield  {author} {\bibinfo {author} {\bibfnamefont {A.~L.}\ \bibnamefont
  {Sharpe}}, \bibinfo {author} {\bibfnamefont {E.~J.}\ \bibnamefont {Fox}},
  \bibinfo {author} {\bibfnamefont {A.~W.}\ \bibnamefont {Barnard}}, \bibinfo
  {author} {\bibfnamefont {J.}~\bibnamefont {Finney}}, \bibinfo {author}
  {\bibfnamefont {K.}~\bibnamefont {Watanabe}}, \bibinfo {author}
  {\bibfnamefont {T.}~\bibnamefont {Taniguchi}}, \bibinfo {author}
  {\bibfnamefont {M.~A.}\ \bibnamefont {Kastner}},\ and\ \bibinfo {author}
  {\bibfnamefont {D.}~\bibnamefont {Goldhaber-Gordon}},\ }\bibfield  {title}
  {\bibinfo {title} {Emergent ferromagnetism near three-quarters filling in
  twisted bilayer graphene},\ }\href {https://doi.org/10.1126/science.aaw3780}
  {\bibfield  {journal} {\bibinfo  {journal} {Science}\ }\textbf {\bibinfo
  {volume} {365}},\ \bibinfo {pages} {605} (\bibinfo {year}
  {2019})}\BibitemShut {NoStop}%
\bibitem [{\citenamefont {Chen}\ \emph {et~al.}(2020)\citenamefont {Chen},
  \citenamefont {Sharpe}, \citenamefont {Fox}, \citenamefont {Zhang},
  \citenamefont {Wang}, \citenamefont {Jiang}, \citenamefont {Lyu},
  \citenamefont {Li}, \citenamefont {Watanabe}, \citenamefont {Taniguchi},
  \citenamefont {Shi}, \citenamefont {Senthil}, \citenamefont
  {Goldhaber-Gordon}, \citenamefont {Zhang},\ and\ \citenamefont
  {Wang}}]{Chen2020}%
  \BibitemOpen
  \bibfield  {author} {\bibinfo {author} {\bibfnamefont {G.}~\bibnamefont
  {Chen}}, \bibinfo {author} {\bibfnamefont {A.~L.}\ \bibnamefont {Sharpe}},
  \bibinfo {author} {\bibfnamefont {E.~J.}\ \bibnamefont {Fox}}, \bibinfo
  {author} {\bibfnamefont {Y.-H.}\ \bibnamefont {Zhang}}, \bibinfo {author}
  {\bibfnamefont {S.}~\bibnamefont {Wang}}, \bibinfo {author} {\bibfnamefont
  {L.}~\bibnamefont {Jiang}}, \bibinfo {author} {\bibfnamefont
  {B.}~\bibnamefont {Lyu}}, \bibinfo {author} {\bibfnamefont {H.}~\bibnamefont
  {Li}}, \bibinfo {author} {\bibfnamefont {K.}~\bibnamefont {Watanabe}},
  \bibinfo {author} {\bibfnamefont {T.}~\bibnamefont {Taniguchi}}, \bibinfo
  {author} {\bibfnamefont {Z.}~\bibnamefont {Shi}}, \bibinfo {author}
  {\bibfnamefont {T.}~\bibnamefont {Senthil}}, \bibinfo {author} {\bibfnamefont
  {D.}~\bibnamefont {Goldhaber-Gordon}}, \bibinfo {author} {\bibfnamefont
  {Y.}~\bibnamefont {Zhang}},\ and\ \bibinfo {author} {\bibfnamefont
  {F.}~\bibnamefont {Wang}},\ }\bibfield  {title} {\bibinfo {title} {Tunable
  correlated chern insulator and ferromagnetism in a moiré superlattice},\
  }\href {https://doi.org/10.1038/s41586-020-2049-7} {\bibfield  {journal}
  {\bibinfo  {journal} {Nature}\ }\textbf {\bibinfo {volume} {579}},\ \bibinfo
  {pages} {56} (\bibinfo {year} {2020})}\BibitemShut {NoStop}%
\bibitem [{\citenamefont {Repellin}\ \emph {et~al.}(2020)\citenamefont
  {Repellin}, \citenamefont {Dong}, \citenamefont {Zhang},\ and\ \citenamefont
  {Senthil}}]{Repellin2020}%
  \BibitemOpen
  \bibfield  {author} {\bibinfo {author} {\bibfnamefont {C.}~\bibnamefont
  {Repellin}}, \bibinfo {author} {\bibfnamefont {Z.}~\bibnamefont {Dong}},
  \bibinfo {author} {\bibfnamefont {Y.-H.}\ \bibnamefont {Zhang}},\ and\
  \bibinfo {author} {\bibfnamefont {T.}~\bibnamefont {Senthil}},\ }\bibfield
  {title} {\bibinfo {title} {Ferromagnetism in narrow bands of moir\'e
  superlattices},\ }\href {https://doi.org/10.1103/PhysRevLett.124.187601}
  {\bibfield  {journal} {\bibinfo  {journal} {Phys. Rev. Lett.}\ }\textbf
  {\bibinfo {volume} {124}},\ \bibinfo {pages} {187601} (\bibinfo {year}
  {2020})}\BibitemShut {NoStop}%
\bibitem [{\citenamefont {Uri}\ \emph {et~al.}(2023)\citenamefont {Uri},
  \citenamefont {de~la Barrera}, \citenamefont {Randeria}, \citenamefont
  {Rodan-Legrain}, \citenamefont {Devakul}, \citenamefont {Crowley},
  \citenamefont {Paul}, \citenamefont {Watanabe}, \citenamefont {Taniguchi},
  \citenamefont {Lifshitz}, \citenamefont {Fu}, \citenamefont {Ashoori},\ and\
  \citenamefont {Jarillo-Herrero}}]{Uri2023}%
  \BibitemOpen
  \bibfield  {author} {\bibinfo {author} {\bibfnamefont {A.}~\bibnamefont
  {Uri}}, \bibinfo {author} {\bibfnamefont {S.~C.}\ \bibnamefont {de~la
  Barrera}}, \bibinfo {author} {\bibfnamefont {M.~T.}\ \bibnamefont
  {Randeria}}, \bibinfo {author} {\bibfnamefont {D.}~\bibnamefont
  {Rodan-Legrain}}, \bibinfo {author} {\bibfnamefont {T.}~\bibnamefont
  {Devakul}}, \bibinfo {author} {\bibfnamefont {P.~J.~D.}\ \bibnamefont
  {Crowley}}, \bibinfo {author} {\bibfnamefont {N.}~\bibnamefont {Paul}},
  \bibinfo {author} {\bibfnamefont {K.}~\bibnamefont {Watanabe}}, \bibinfo
  {author} {\bibfnamefont {T.}~\bibnamefont {Taniguchi}}, \bibinfo {author}
  {\bibfnamefont {R.}~\bibnamefont {Lifshitz}}, \bibinfo {author}
  {\bibfnamefont {L.}~\bibnamefont {Fu}}, \bibinfo {author} {\bibfnamefont
  {R.~C.}\ \bibnamefont {Ashoori}},\ and\ \bibinfo {author} {\bibfnamefont
  {P.}~\bibnamefont {Jarillo-Herrero}},\ }\bibfield  {title} {\bibinfo {title}
  {Superconductivity and strong interactions in a tunable moiré
  quasicrystal},\ }\href {https://doi.org/10.1038/s41586-023-06294-z}
  {\bibfield  {journal} {\bibinfo  {journal} {Nature}\ }\textbf {\bibinfo
  {volume} {620}},\ \bibinfo {pages} {762} (\bibinfo {year}
  {2023})}\BibitemShut {NoStop}%
\bibitem [{\citenamefont {Chen}\ \emph {et~al.}(2019)\citenamefont {Chen},
  \citenamefont {Jiang}, \citenamefont {Wu}, \citenamefont {Lyu}, \citenamefont
  {Li}, \citenamefont {Chittari}, \citenamefont {Watanabe}, \citenamefont
  {Taniguchi}, \citenamefont {Shi}, \citenamefont {Jung}, \citenamefont
  {Zhang},\ and\ \citenamefont {Wang}}]{Chen2019}%
  \BibitemOpen
  \bibfield  {author} {\bibinfo {author} {\bibfnamefont {G.}~\bibnamefont
  {Chen}}, \bibinfo {author} {\bibfnamefont {L.}~\bibnamefont {Jiang}},
  \bibinfo {author} {\bibfnamefont {S.}~\bibnamefont {Wu}}, \bibinfo {author}
  {\bibfnamefont {B.}~\bibnamefont {Lyu}}, \bibinfo {author} {\bibfnamefont
  {H.}~\bibnamefont {Li}}, \bibinfo {author} {\bibfnamefont {B.~L.}\
  \bibnamefont {Chittari}}, \bibinfo {author} {\bibfnamefont {K.}~\bibnamefont
  {Watanabe}}, \bibinfo {author} {\bibfnamefont {T.}~\bibnamefont {Taniguchi}},
  \bibinfo {author} {\bibfnamefont {Z.}~\bibnamefont {Shi}}, \bibinfo {author}
  {\bibfnamefont {J.}~\bibnamefont {Jung}}, \bibinfo {author} {\bibfnamefont
  {Y.}~\bibnamefont {Zhang}},\ and\ \bibinfo {author} {\bibfnamefont
  {F.}~\bibnamefont {Wang}},\ }\bibfield  {title} {\bibinfo {title} {Evidence
  of a gate-tunable mott insulator in a trilayer graphene moiré
  superlattice},\ }\href {https://doi.org/10.1038/s41567-018-0387-2} {\bibfield
   {journal} {\bibinfo  {journal} {Nature Physics}\ }\textbf {\bibinfo {volume}
  {15}},\ \bibinfo {pages} {237} (\bibinfo {year} {2019})}\BibitemShut
  {NoStop}%
\bibitem [{\citenamefont {Zhu}\ \emph {et~al.}(2020)\citenamefont {Zhu},
  \citenamefont {Carr}, \citenamefont {Massatt}, \citenamefont {Luskin},\ and\
  \citenamefont {Kaxiras}}]{PhysRevLett.125.116404}%
  \BibitemOpen
  \bibfield  {author} {\bibinfo {author} {\bibfnamefont {Z.}~\bibnamefont
  {Zhu}}, \bibinfo {author} {\bibfnamefont {S.}~\bibnamefont {Carr}}, \bibinfo
  {author} {\bibfnamefont {D.}~\bibnamefont {Massatt}}, \bibinfo {author}
  {\bibfnamefont {M.}~\bibnamefont {Luskin}},\ and\ \bibinfo {author}
  {\bibfnamefont {E.}~\bibnamefont {Kaxiras}},\ }\bibfield  {title} {\bibinfo
  {title} {Twisted trilayer graphene: A precisely tunable platform for
  correlated electrons},\ }\href
  {https://doi.org/10.1103/PhysRevLett.125.116404} {\bibfield  {journal}
  {\bibinfo  {journal} {Phys. Rev. Lett.}\ }\textbf {\bibinfo {volume} {125}},\
  \bibinfo {pages} {116404} (\bibinfo {year} {2020})}\BibitemShut {NoStop}%
\bibitem [{\citenamefont {Lai}\ \emph {et~al.}(2023)\citenamefont {Lai},
  \citenamefont {Guerci}, \citenamefont {Li}, \citenamefont {Watanabe},
  \citenamefont {Taniguchi}, \citenamefont {Wilson}, \citenamefont {Pixley},\
  and\ \citenamefont {Andrei}}]{lai2023imaging}%
  \BibitemOpen
  \bibfield  {author} {\bibinfo {author} {\bibfnamefont {X.}~\bibnamefont
  {Lai}}, \bibinfo {author} {\bibfnamefont {D.}~\bibnamefont {Guerci}},
  \bibinfo {author} {\bibfnamefont {G.}~\bibnamefont {Li}}, \bibinfo {author}
  {\bibfnamefont {K.}~\bibnamefont {Watanabe}}, \bibinfo {author}
  {\bibfnamefont {T.}~\bibnamefont {Taniguchi}}, \bibinfo {author}
  {\bibfnamefont {J.}~\bibnamefont {Wilson}}, \bibinfo {author} {\bibfnamefont
  {J.~H.}\ \bibnamefont {Pixley}},\ and\ \bibinfo {author} {\bibfnamefont
  {E.~Y.}\ \bibnamefont {Andrei}},\ }\href@noop {} {\bibinfo {title} {Imaging
  self-aligned moir\'e crystals and quasicrystals in magic-angle bilayer
  graphene on hbn heterostructures}} (\bibinfo {year} {2023}),\ \Eprint
  {https://arxiv.org/abs/2311.07819} {arXiv:2311.07819 [cond-mat.mes-hall]}
  \BibitemShut {NoStop}%
\bibitem [{\citenamefont {Chen}\ \emph {et~al.}(2021)\citenamefont {Chen},
  \citenamefont {He}, \citenamefont {Zhang}, \citenamefont {Hsieh},
  \citenamefont {Fei}, \citenamefont {Watanabe}, \citenamefont {Taniguchi},
  \citenamefont {Cobden}, \citenamefont {Xu}, \citenamefont {Dean},\ and\
  \citenamefont {Yankowitz}}]{Chen2021}%
  \BibitemOpen
  \bibfield  {author} {\bibinfo {author} {\bibfnamefont {S.}~\bibnamefont
  {Chen}}, \bibinfo {author} {\bibfnamefont {M.}~\bibnamefont {He}}, \bibinfo
  {author} {\bibfnamefont {Y.-H.}\ \bibnamefont {Zhang}}, \bibinfo {author}
  {\bibfnamefont {V.}~\bibnamefont {Hsieh}}, \bibinfo {author} {\bibfnamefont
  {Z.}~\bibnamefont {Fei}}, \bibinfo {author} {\bibfnamefont {K.}~\bibnamefont
  {Watanabe}}, \bibinfo {author} {\bibfnamefont {T.}~\bibnamefont {Taniguchi}},
  \bibinfo {author} {\bibfnamefont {D.~H.}\ \bibnamefont {Cobden}}, \bibinfo
  {author} {\bibfnamefont {X.}~\bibnamefont {Xu}}, \bibinfo {author}
  {\bibfnamefont {C.~R.}\ \bibnamefont {Dean}},\ and\ \bibinfo {author}
  {\bibfnamefont {M.}~\bibnamefont {Yankowitz}},\ }\bibfield  {title} {\bibinfo
  {title} {Electrically tunable correlated and topological states in twisted
  monolayer-bilayer graphene},\ }\href
  {https://doi.org/10.1038/s41567-020-01062-6} {\bibfield  {journal} {\bibinfo
  {journal} {Nature Physics}\ }\textbf {\bibinfo {volume} {17}},\ \bibinfo
  {pages} {374} (\bibinfo {year} {2021})}\BibitemShut {NoStop}%
\bibitem [{\citenamefont {Mora}\ \emph {et~al.}(2019)\citenamefont {Mora},
  \citenamefont {Regnault},\ and\ \citenamefont
  {Bernevig}}]{PhysRevLett.123.026402}%
  \BibitemOpen
  \bibfield  {author} {\bibinfo {author} {\bibfnamefont {C.}~\bibnamefont
  {Mora}}, \bibinfo {author} {\bibfnamefont {N.}~\bibnamefont {Regnault}},\
  and\ \bibinfo {author} {\bibfnamefont {B.~A.}\ \bibnamefont {Bernevig}},\
  }\bibfield  {title} {\bibinfo {title} {Flatbands and perfect metal in
  trilayer moir\'e graphene},\ }\href
  {https://doi.org/10.1103/PhysRevLett.123.026402} {\bibfield  {journal}
  {\bibinfo  {journal} {Phys. Rev. Lett.}\ }\textbf {\bibinfo {volume} {123}},\
  \bibinfo {pages} {026402} (\bibinfo {year} {2019})}\BibitemShut {NoStop}%
\bibitem [{\citenamefont {Mao}\ \emph {et~al.}(2023)\citenamefont {Mao},
  \citenamefont {Guerci},\ and\ \citenamefont {Mora}}]{PhysRevB.107.125423}%
  \BibitemOpen
  \bibfield  {author} {\bibinfo {author} {\bibfnamefont {Y.}~\bibnamefont
  {Mao}}, \bibinfo {author} {\bibfnamefont {D.}~\bibnamefont {Guerci}},\ and\
  \bibinfo {author} {\bibfnamefont {C.}~\bibnamefont {Mora}},\ }\bibfield
  {title} {\bibinfo {title} {Supermoir\'e low-energy effective theory of
  twisted trilayer graphene},\ }\href
  {https://doi.org/10.1103/PhysRevB.107.125423} {\bibfield  {journal} {\bibinfo
   {journal} {Phys. Rev. B}\ }\textbf {\bibinfo {volume} {107}},\ \bibinfo
  {pages} {125423} (\bibinfo {year} {2023})}\BibitemShut {NoStop}%
\bibitem [{\citenamefont {Wang}\ and\ \citenamefont
  {Liu}(2022)}]{PhysRevLett.128.176403}%
  \BibitemOpen
  \bibfield  {author} {\bibinfo {author} {\bibfnamefont {J.}~\bibnamefont
  {Wang}}\ and\ \bibinfo {author} {\bibfnamefont {Z.}~\bibnamefont {Liu}},\
  }\bibfield  {title} {\bibinfo {title} {Hierarchy of ideal flatbands in chiral
  twisted multilayer graphene models},\ }\href
  {https://doi.org/10.1103/PhysRevLett.128.176403} {\bibfield  {journal}
  {\bibinfo  {journal} {Phys. Rev. Lett.}\ }\textbf {\bibinfo {volume} {128}},\
  \bibinfo {pages} {176403} (\bibinfo {year} {2022})}\BibitemShut {NoStop}%
\bibitem [{\citenamefont {Khalaf}\ \emph {et~al.}(2019)\citenamefont {Khalaf},
  \citenamefont {Kruchkov}, \citenamefont {Tarnopolsky},\ and\ \citenamefont
  {Vishwanath}}]{PhysRevB.100.085109}%
  \BibitemOpen
  \bibfield  {author} {\bibinfo {author} {\bibfnamefont {E.}~\bibnamefont
  {Khalaf}}, \bibinfo {author} {\bibfnamefont {A.~J.}\ \bibnamefont
  {Kruchkov}}, \bibinfo {author} {\bibfnamefont {G.}~\bibnamefont
  {Tarnopolsky}},\ and\ \bibinfo {author} {\bibfnamefont {A.}~\bibnamefont
  {Vishwanath}},\ }\bibfield  {title} {\bibinfo {title} {Magic angle hierarchy
  in twisted graphene multilayers},\ }\href
  {https://doi.org/10.1103/PhysRevB.100.085109} {\bibfield  {journal} {\bibinfo
   {journal} {Phys. Rev. B}\ }\textbf {\bibinfo {volume} {100}},\ \bibinfo
  {pages} {085109} (\bibinfo {year} {2019})}\BibitemShut {NoStop}%
\bibitem [{\citenamefont {Foo}\ \emph {et~al.}(2024)\citenamefont {Foo},
  \citenamefont {Zhan}, \citenamefont {Al~Ezzi}, \citenamefont {Peng},
  \citenamefont {Adam},\ and\ \citenamefont
  {Guinea}}]{PhysRevResearch.6.013165}%
  \BibitemOpen
  \bibfield  {author} {\bibinfo {author} {\bibfnamefont {D.~C.~W.}\
  \bibnamefont {Foo}}, \bibinfo {author} {\bibfnamefont {Z.}~\bibnamefont
  {Zhan}}, \bibinfo {author} {\bibfnamefont {M.~M.}\ \bibnamefont {Al~Ezzi}},
  \bibinfo {author} {\bibfnamefont {L.}~\bibnamefont {Peng}}, \bibinfo {author}
  {\bibfnamefont {S.}~\bibnamefont {Adam}},\ and\ \bibinfo {author}
  {\bibfnamefont {F.}~\bibnamefont {Guinea}},\ }\bibfield  {title} {\bibinfo
  {title} {Extended magic phase in twisted graphene multilayers},\ }\href
  {https://doi.org/10.1103/PhysRevResearch.6.013165} {\bibfield  {journal}
  {\bibinfo  {journal} {Phys. Rev. Res.}\ }\textbf {\bibinfo {volume} {6}},\
  \bibinfo {pages} {013165} (\bibinfo {year} {2024})}\BibitemShut {NoStop}%
\bibitem [{\citenamefont {Guerci}\ \emph {et~al.}(2022)\citenamefont {Guerci},
  \citenamefont {Simon},\ and\ \citenamefont
  {Mora}}]{PhysRevResearch.4.L012013}%
  \BibitemOpen
  \bibfield  {author} {\bibinfo {author} {\bibfnamefont {D.}~\bibnamefont
  {Guerci}}, \bibinfo {author} {\bibfnamefont {P.}~\bibnamefont {Simon}},\ and\
  \bibinfo {author} {\bibfnamefont {C.}~\bibnamefont {Mora}},\ }\bibfield
  {title} {\bibinfo {title} {Higher-order van hove singularity in magic-angle
  twisted trilayer graphene},\ }\href
  {https://doi.org/10.1103/PhysRevResearch.4.L012013} {\bibfield  {journal}
  {\bibinfo  {journal} {Phys. Rev. Res.}\ }\textbf {\bibinfo {volume} {4}},\
  \bibinfo {pages} {L012013} (\bibinfo {year} {2022})}\BibitemShut {NoStop}%
\bibitem [{\citenamefont {Zhang}\ \emph {et~al.}(2021)\citenamefont {Zhang},
  \citenamefont {Tsai}, \citenamefont {Zhu}, \citenamefont {Ren}, \citenamefont
  {Luo}, \citenamefont {Carr}, \citenamefont {Luskin}, \citenamefont
  {Kaxiras},\ and\ \citenamefont {Wang}}]{PhysRevLett.127.166802}%
  \BibitemOpen
  \bibfield  {author} {\bibinfo {author} {\bibfnamefont {X.}~\bibnamefont
  {Zhang}}, \bibinfo {author} {\bibfnamefont {K.-T.}\ \bibnamefont {Tsai}},
  \bibinfo {author} {\bibfnamefont {Z.}~\bibnamefont {Zhu}}, \bibinfo {author}
  {\bibfnamefont {W.}~\bibnamefont {Ren}}, \bibinfo {author} {\bibfnamefont
  {Y.}~\bibnamefont {Luo}}, \bibinfo {author} {\bibfnamefont {S.}~\bibnamefont
  {Carr}}, \bibinfo {author} {\bibfnamefont {M.}~\bibnamefont {Luskin}},
  \bibinfo {author} {\bibfnamefont {E.}~\bibnamefont {Kaxiras}},\ and\ \bibinfo
  {author} {\bibfnamefont {K.}~\bibnamefont {Wang}},\ }\bibfield  {title}
  {\bibinfo {title} {Correlated insulating states and transport signature of
  superconductivity in twisted trilayer graphene superlattices},\ }\href
  {https://doi.org/10.1103/PhysRevLett.127.166802} {\bibfield  {journal}
  {\bibinfo  {journal} {Phys. Rev. Lett.}\ }\textbf {\bibinfo {volume} {127}},\
  \bibinfo {pages} {166802} (\bibinfo {year} {2021})}\BibitemShut {NoStop}%
\bibitem [{\citenamefont {Ren}\ \emph {et~al.}(2023)\citenamefont {Ren},
  \citenamefont {Davydov}, \citenamefont {Zhu}, \citenamefont {Ma},
  \citenamefont {Watanabe}, \citenamefont {Taniguchi}, \citenamefont {Kaxiras},
  \citenamefont {Luskin},\ and\ \citenamefont {Wang}}]{ren2023tunable}%
  \BibitemOpen
  \bibfield  {author} {\bibinfo {author} {\bibfnamefont {W.}~\bibnamefont
  {Ren}}, \bibinfo {author} {\bibfnamefont {K.}~\bibnamefont {Davydov}},
  \bibinfo {author} {\bibfnamefont {Z.}~\bibnamefont {Zhu}}, \bibinfo {author}
  {\bibfnamefont {J.}~\bibnamefont {Ma}}, \bibinfo {author} {\bibfnamefont
  {K.}~\bibnamefont {Watanabe}}, \bibinfo {author} {\bibfnamefont
  {T.}~\bibnamefont {Taniguchi}}, \bibinfo {author} {\bibfnamefont
  {E.}~\bibnamefont {Kaxiras}}, \bibinfo {author} {\bibfnamefont
  {M.}~\bibnamefont {Luskin}},\ and\ \bibinfo {author} {\bibfnamefont
  {K.}~\bibnamefont {Wang}},\ }\href@noop {} {\bibinfo {title} {Tunable
  inter-moir\'e physics in consecutively-twisted trilayer graphene}} (\bibinfo
  {year} {2023}),\ \Eprint {https://arxiv.org/abs/2311.10313} {arXiv:2311.10313
  [cond-mat.mes-hall]} \BibitemShut {NoStop}%
\bibitem [{\citenamefont {Xie}\ \emph {et~al.}(2024)\citenamefont {Xie},
  \citenamefont {Pierce}, \citenamefont {Park}, \citenamefont {Parker},
  \citenamefont {Wang}, \citenamefont {Ledwith}, \citenamefont {Cai},
  \citenamefont {Watanabe}, \citenamefont {Taniguchi}, \citenamefont {Khalaf},
  \citenamefont {Vishwanath}, \citenamefont {Jarillo-Herrero},\ and\
  \citenamefont {Yacoby}}]{xie2024strong}%
  \BibitemOpen
  \bibfield  {author} {\bibinfo {author} {\bibfnamefont {Y.}~\bibnamefont
  {Xie}}, \bibinfo {author} {\bibfnamefont {A.~T.}\ \bibnamefont {Pierce}},
  \bibinfo {author} {\bibfnamefont {J.~M.}\ \bibnamefont {Park}}, \bibinfo
  {author} {\bibfnamefont {D.~E.}\ \bibnamefont {Parker}}, \bibinfo {author}
  {\bibfnamefont {J.}~\bibnamefont {Wang}}, \bibinfo {author} {\bibfnamefont
  {P.}~\bibnamefont {Ledwith}}, \bibinfo {author} {\bibfnamefont
  {Z.}~\bibnamefont {Cai}}, \bibinfo {author} {\bibfnamefont {K.}~\bibnamefont
  {Watanabe}}, \bibinfo {author} {\bibfnamefont {T.}~\bibnamefont {Taniguchi}},
  \bibinfo {author} {\bibfnamefont {E.}~\bibnamefont {Khalaf}}, \bibinfo
  {author} {\bibfnamefont {A.}~\bibnamefont {Vishwanath}}, \bibinfo {author}
  {\bibfnamefont {P.}~\bibnamefont {Jarillo-Herrero}},\ and\ \bibinfo {author}
  {\bibfnamefont {A.}~\bibnamefont {Yacoby}},\ }\href@noop {} {\bibinfo {title}
  {Strong interactions and isospin symmetry breaking in a supermoir\'e
  lattice}} (\bibinfo {year} {2024}),\ \Eprint
  {https://arxiv.org/abs/2404.01372} {arXiv:2404.01372 [cond-mat.mes-hall]}
  \BibitemShut {NoStop}%
\bibitem [{\citenamefont {Zhang}\ \emph {et~al.}(2022)\citenamefont {Zhang},
  \citenamefont {Polski}, \citenamefont {Lewandowski}, \citenamefont {Thomson},
  \citenamefont {Peng}, \citenamefont {Choi}, \citenamefont {Kim},
  \citenamefont {Watanabe}, \citenamefont {Taniguchi}, \citenamefont {Alicea},
  \citenamefont {von Oppen}, \citenamefont {Refael},\ and\ \citenamefont
  {Nadj-Perge}}]{Zhang2022}%
  \BibitemOpen
  \bibfield  {author} {\bibinfo {author} {\bibfnamefont {Y.}~\bibnamefont
  {Zhang}}, \bibinfo {author} {\bibfnamefont {R.}~\bibnamefont {Polski}},
  \bibinfo {author} {\bibfnamefont {C.}~\bibnamefont {Lewandowski}}, \bibinfo
  {author} {\bibfnamefont {A.}~\bibnamefont {Thomson}}, \bibinfo {author}
  {\bibfnamefont {Y.}~\bibnamefont {Peng}}, \bibinfo {author} {\bibfnamefont
  {Y.}~\bibnamefont {Choi}}, \bibinfo {author} {\bibfnamefont {H.}~\bibnamefont
  {Kim}}, \bibinfo {author} {\bibfnamefont {K.}~\bibnamefont {Watanabe}},
  \bibinfo {author} {\bibfnamefont {T.}~\bibnamefont {Taniguchi}}, \bibinfo
  {author} {\bibfnamefont {J.}~\bibnamefont {Alicea}}, \bibinfo {author}
  {\bibfnamefont {F.}~\bibnamefont {von Oppen}}, \bibinfo {author}
  {\bibfnamefont {G.}~\bibnamefont {Refael}},\ and\ \bibinfo {author}
  {\bibfnamefont {S.}~\bibnamefont {Nadj-Perge}},\ }\bibfield  {title}
  {\bibinfo {title} {Promotion of superconductivity in magic-angle graphene
  multilayers},\ }\href {https://doi.org/10.1126/science.abn8585} {\bibfield
  {journal} {\bibinfo  {journal} {Science}\ }\textbf {\bibinfo {volume}
  {377}},\ \bibinfo {pages} {1538} (\bibinfo {year} {2022})}\BibitemShut
  {NoStop}%
\bibitem [{\citenamefont {Devakul}\ \emph {et~al.}(2023)\citenamefont
  {Devakul}, \citenamefont {Ledwith}, \citenamefont {Xia}, \citenamefont {Uri},
  \citenamefont {de~la Barrera}, \citenamefont {Jarillo-Herrero},\ and\
  \citenamefont {Fu}}]{doi:10.1126/sciadv.adi6063}%
  \BibitemOpen
  \bibfield  {author} {\bibinfo {author} {\bibfnamefont {T.}~\bibnamefont
  {Devakul}}, \bibinfo {author} {\bibfnamefont {P.~J.}\ \bibnamefont
  {Ledwith}}, \bibinfo {author} {\bibfnamefont {L.-Q.}\ \bibnamefont {Xia}},
  \bibinfo {author} {\bibfnamefont {A.}~\bibnamefont {Uri}}, \bibinfo {author}
  {\bibfnamefont {S.~C.}\ \bibnamefont {de~la Barrera}}, \bibinfo {author}
  {\bibfnamefont {P.}~\bibnamefont {Jarillo-Herrero}},\ and\ \bibinfo {author}
  {\bibfnamefont {L.}~\bibnamefont {Fu}},\ }\bibfield  {title} {\bibinfo
  {title} {Magic-angle helical trilayer graphene},\ }\href
  {https://doi.org/10.1126/sciadv.adi6063} {\bibfield  {journal} {\bibinfo
  {journal} {Science Advances}\ }\textbf {\bibinfo {volume} {9}},\ \bibinfo
  {pages} {eadi6063} (\bibinfo {year} {2023})}\BibitemShut {NoStop}%
\bibitem [{\citenamefont {Guerci}\ \emph {et~al.}(2024)\citenamefont {Guerci},
  \citenamefont {Mao},\ and\ \citenamefont {Mora}}]{guerci2023chern}%
  \BibitemOpen
  \bibfield  {author} {\bibinfo {author} {\bibfnamefont {D.}~\bibnamefont
  {Guerci}}, \bibinfo {author} {\bibfnamefont {Y.}~\bibnamefont {Mao}},\ and\
  \bibinfo {author} {\bibfnamefont {C.}~\bibnamefont {Mora}},\ }\bibfield
  {title} {\bibinfo {title} {Chern mosaic and ideal flat bands in equal-twist
  trilayer graphene},\ }\href
  {https://doi.org/10.1103/PhysRevResearch.6.L022025} {\bibfield  {journal}
  {\bibinfo  {journal} {Phys. Rev. Res.}\ }\textbf {\bibinfo {volume} {6}},\
  \bibinfo {pages} {L022025} (\bibinfo {year} {2024})}\BibitemShut {NoStop}%
\bibitem [{\citenamefont {Popov}\ and\ \citenamefont
  {Tarnopolsky}(2023{\natexlab{a}})}]{popov2023}%
  \BibitemOpen
  \bibfield  {author} {\bibinfo {author} {\bibfnamefont {F.~K.}\ \bibnamefont
  {Popov}}\ and\ \bibinfo {author} {\bibfnamefont {G.}~\bibnamefont
  {Tarnopolsky}},\ }\bibfield  {title} {\bibinfo {title} {Magic angles in
  equal-twist trilayer graphene},\ }\href
  {https://doi.org/10.1103/PhysRevB.108.L081124} {\bibfield  {journal}
  {\bibinfo  {journal} {Phys. Rev. B}\ }\textbf {\bibinfo {volume} {108}},\
  \bibinfo {pages} {L081124} (\bibinfo {year}
  {2023}{\natexlab{a}})}\BibitemShut {NoStop}%
\bibitem [{\citenamefont {Kwan}\ \emph {et~al.}(2024)\citenamefont {Kwan},
  \citenamefont {Ledwith}, \citenamefont {Lo},\ and\ \citenamefont
  {Devakul}}]{PhysRevB.109.125141}%
  \BibitemOpen
  \bibfield  {author} {\bibinfo {author} {\bibfnamefont {Y.~H.}\ \bibnamefont
  {Kwan}}, \bibinfo {author} {\bibfnamefont {P.~J.}\ \bibnamefont {Ledwith}},
  \bibinfo {author} {\bibfnamefont {C.~F.~B.}\ \bibnamefont {Lo}},\ and\
  \bibinfo {author} {\bibfnamefont {T.}~\bibnamefont {Devakul}},\ }\bibfield
  {title} {\bibinfo {title} {Strong-coupling topological states and phase
  transitions in helical trilayer graphene},\ }\href
  {https://doi.org/10.1103/PhysRevB.109.125141} {\bibfield  {journal} {\bibinfo
   {journal} {Phys. Rev. B}\ }\textbf {\bibinfo {volume} {109}},\ \bibinfo
  {pages} {125141} (\bibinfo {year} {2024})}\BibitemShut {NoStop}%
\bibitem [{\citenamefont {Xia}\ \emph {et~al.}(2023)\citenamefont {Xia},
  \citenamefont {de~la Barrera}, \citenamefont {Uri}, \citenamefont {Sharpe},
  \citenamefont {Kwan}, \citenamefont {Zhu}, \citenamefont {Watanabe},
  \citenamefont {Taniguchi}, \citenamefont {Goldhaber-Gordon}, \citenamefont
  {Fu}, \citenamefont {Devakul},\ and\ \citenamefont
  {Jarillo-Herrero}}]{xia2023helical}%
  \BibitemOpen
  \bibfield  {author} {\bibinfo {author} {\bibfnamefont {L.-Q.}\ \bibnamefont
  {Xia}}, \bibinfo {author} {\bibfnamefont {S.~C.}\ \bibnamefont {de~la
  Barrera}}, \bibinfo {author} {\bibfnamefont {A.}~\bibnamefont {Uri}},
  \bibinfo {author} {\bibfnamefont {A.}~\bibnamefont {Sharpe}}, \bibinfo
  {author} {\bibfnamefont {Y.~H.}\ \bibnamefont {Kwan}}, \bibinfo {author}
  {\bibfnamefont {Z.}~\bibnamefont {Zhu}}, \bibinfo {author} {\bibfnamefont
  {K.}~\bibnamefont {Watanabe}}, \bibinfo {author} {\bibfnamefont
  {T.}~\bibnamefont {Taniguchi}}, \bibinfo {author} {\bibfnamefont
  {D.}~\bibnamefont {Goldhaber-Gordon}}, \bibinfo {author} {\bibfnamefont
  {L.}~\bibnamefont {Fu}}, \bibinfo {author} {\bibfnamefont {T.}~\bibnamefont
  {Devakul}},\ and\ \bibinfo {author} {\bibfnamefont {P.}~\bibnamefont
  {Jarillo-Herrero}},\ }\href@noop {} {\bibinfo {title} {Helical trilayer
  graphene: a moir\'e platform for strongly-interacting topological bands}}
  (\bibinfo {year} {2023}),\ \Eprint {https://arxiv.org/abs/2310.12204}
  {arXiv:2310.12204 [cond-mat.mes-hall]} \BibitemShut {NoStop}%
\bibitem [{\citenamefont {Nakatsuji}\ \emph {et~al.}(2023)\citenamefont
  {Nakatsuji}, \citenamefont {Kawakami},\ and\ \citenamefont
  {Koshino}}]{nakatsuji2023multiscale}%
  \BibitemOpen
  \bibfield  {author} {\bibinfo {author} {\bibfnamefont {N.}~\bibnamefont
  {Nakatsuji}}, \bibinfo {author} {\bibfnamefont {T.}~\bibnamefont
  {Kawakami}},\ and\ \bibinfo {author} {\bibfnamefont {M.}~\bibnamefont
  {Koshino}},\ }\bibfield  {title} {\bibinfo {title} {Multiscale lattice
  relaxation in general twisted trilayer graphenes},\ }\href
  {https://doi.org/10.1103/PhysRevX.13.041007} {\bibfield  {journal} {\bibinfo
  {journal} {Phys. Rev. X}\ }\textbf {\bibinfo {volume} {13}},\ \bibinfo
  {pages} {041007} (\bibinfo {year} {2023})}\BibitemShut {NoStop}%
\bibitem [{\citenamefont {Yang}\ \emph {et~al.}(2023)\citenamefont {Yang},
  \citenamefont {May-Mann}, \citenamefont {Zhu},\ and\ \citenamefont
  {Devakul}}]{yang2023multimoire}%
  \BibitemOpen
  \bibfield  {author} {\bibinfo {author} {\bibfnamefont {C.}~\bibnamefont
  {Yang}}, \bibinfo {author} {\bibfnamefont {J.}~\bibnamefont {May-Mann}},
  \bibinfo {author} {\bibfnamefont {Z.}~\bibnamefont {Zhu}},\ and\ \bibinfo
  {author} {\bibfnamefont {T.}~\bibnamefont {Devakul}},\ }\href@noop {}
  {\bibinfo {title} {Multi-moir\'{e} trilayer graphene: lattice relaxation,
  electronic structure, and magic angles}} (\bibinfo {year} {2023}),\ \Eprint
  {https://arxiv.org/abs/2310.12961} {arXiv:2310.12961 [cond-mat.str-el]}
  \BibitemShut {NoStop}%
\bibitem [{\citenamefont {Grover}\ \emph {et~al.}(2022)\citenamefont {Grover},
  \citenamefont {Bocarsly}, \citenamefont {Uri}, \citenamefont {Stepanov},
  \citenamefont {Di~Battista}, \citenamefont {Roy}, \citenamefont {Xiao},
  \citenamefont {Meltzer}, \citenamefont {Myasoedov}, \citenamefont {Pareek},
  \citenamefont {Watanabe}, \citenamefont {Taniguchi}, \citenamefont {Yan},
  \citenamefont {Stern}, \citenamefont {Berg}, \citenamefont {Efetov},\ and\
  \citenamefont {Zeldov}}]{Grover2022}%
  \BibitemOpen
  \bibfield  {author} {\bibinfo {author} {\bibfnamefont {S.}~\bibnamefont
  {Grover}}, \bibinfo {author} {\bibfnamefont {M.}~\bibnamefont {Bocarsly}},
  \bibinfo {author} {\bibfnamefont {A.}~\bibnamefont {Uri}}, \bibinfo {author}
  {\bibfnamefont {P.}~\bibnamefont {Stepanov}}, \bibinfo {author}
  {\bibfnamefont {G.}~\bibnamefont {Di~Battista}}, \bibinfo {author}
  {\bibfnamefont {I.}~\bibnamefont {Roy}}, \bibinfo {author} {\bibfnamefont
  {J.}~\bibnamefont {Xiao}}, \bibinfo {author} {\bibfnamefont {A.~Y.}\
  \bibnamefont {Meltzer}}, \bibinfo {author} {\bibfnamefont {Y.}~\bibnamefont
  {Myasoedov}}, \bibinfo {author} {\bibfnamefont {K.}~\bibnamefont {Pareek}},
  \bibinfo {author} {\bibfnamefont {K.}~\bibnamefont {Watanabe}}, \bibinfo
  {author} {\bibfnamefont {T.}~\bibnamefont {Taniguchi}}, \bibinfo {author}
  {\bibfnamefont {B.}~\bibnamefont {Yan}}, \bibinfo {author} {\bibfnamefont
  {A.}~\bibnamefont {Stern}}, \bibinfo {author} {\bibfnamefont
  {E.}~\bibnamefont {Berg}}, \bibinfo {author} {\bibfnamefont {D.~K.}\
  \bibnamefont {Efetov}},\ and\ \bibinfo {author} {\bibfnamefont
  {E.}~\bibnamefont {Zeldov}},\ }\bibfield  {title} {\bibinfo {title} {Chern
  mosaic and berry-curvature magnetism in magic-angle graphene},\ }\href
  {https://doi.org/10.1038/s41567-022-01635-7} {\bibfield  {journal} {\bibinfo
  {journal} {Nature Physics}\ }\textbf {\bibinfo {volume} {18}},\ \bibinfo
  {pages} {885} (\bibinfo {year} {2022})}\BibitemShut {NoStop}%
\bibitem [{\citenamefont {Li}\ \emph {et~al.}(2020)\citenamefont {Li},
  \citenamefont {Zhang}, \citenamefont {Ren}, \citenamefont {Liu},
  \citenamefont {Dai},\ and\ \citenamefont {He}}]{PhysRevB.102.121406}%
  \BibitemOpen
  \bibfield  {author} {\bibinfo {author} {\bibfnamefont {S.-Y.}\ \bibnamefont
  {Li}}, \bibinfo {author} {\bibfnamefont {Y.}~\bibnamefont {Zhang}}, \bibinfo
  {author} {\bibfnamefont {Y.-N.}\ \bibnamefont {Ren}}, \bibinfo {author}
  {\bibfnamefont {J.}~\bibnamefont {Liu}}, \bibinfo {author} {\bibfnamefont
  {X.}~\bibnamefont {Dai}},\ and\ \bibinfo {author} {\bibfnamefont
  {L.}~\bibnamefont {He}},\ }\bibfield  {title} {\bibinfo {title} {Experimental
  evidence for orbital magnetic moments generated by moir\'e-scale current
  loops in twisted bilayer graphene},\ }\href
  {https://doi.org/10.1103/PhysRevB.102.121406} {\bibfield  {journal} {\bibinfo
   {journal} {Phys. Rev. B}\ }\textbf {\bibinfo {volume} {102}},\ \bibinfo
  {pages} {121406} (\bibinfo {year} {2020})}\BibitemShut {NoStop}%
\bibitem [{\citenamefont {Cea}\ \emph {et~al.}(2020)\citenamefont {Cea},
  \citenamefont {Pantale\'on},\ and\ \citenamefont {Guinea}}]{Cea2020}%
  \BibitemOpen
  \bibfield  {author} {\bibinfo {author} {\bibfnamefont {T.}~\bibnamefont
  {Cea}}, \bibinfo {author} {\bibfnamefont {P.~A.}\ \bibnamefont
  {Pantale\'on}},\ and\ \bibinfo {author} {\bibfnamefont {F.}~\bibnamefont
  {Guinea}},\ }\bibfield  {title} {\bibinfo {title} {Band structure of twisted
  bilayer graphene on hexagonal boron nitride},\ }\href
  {https://doi.org/10.1103/PhysRevB.102.155136} {\bibfield  {journal} {\bibinfo
   {journal} {Phys. Rev. B}\ }\textbf {\bibinfo {volume} {102}},\ \bibinfo
  {pages} {155136} (\bibinfo {year} {2020})}\BibitemShut {NoStop}%
\bibitem [{\citenamefont {Ledwith}\ \emph {et~al.}(2020)\citenamefont
  {Ledwith}, \citenamefont {Tarnopolsky}, \citenamefont {Khalaf},\ and\
  \citenamefont {Vishwanath}}]{PhysRevResearch.2.023237}%
  \BibitemOpen
  \bibfield  {author} {\bibinfo {author} {\bibfnamefont {P.~J.}\ \bibnamefont
  {Ledwith}}, \bibinfo {author} {\bibfnamefont {G.}~\bibnamefont
  {Tarnopolsky}}, \bibinfo {author} {\bibfnamefont {E.}~\bibnamefont
  {Khalaf}},\ and\ \bibinfo {author} {\bibfnamefont {A.}~\bibnamefont
  {Vishwanath}},\ }\bibfield  {title} {\bibinfo {title} {Fractional chern
  insulator states in twisted bilayer graphene: An analytical approach},\
  }\href {https://doi.org/10.1103/PhysRevResearch.2.023237} {\bibfield
  {journal} {\bibinfo  {journal} {Phys. Rev. Res.}\ }\textbf {\bibinfo {volume}
  {2}},\ \bibinfo {pages} {023237} (\bibinfo {year} {2020})}\BibitemShut
  {NoStop}%
\bibitem [{\citenamefont {Wang}\ \emph
  {et~al.}(2021{\natexlab{a}})\citenamefont {Wang}, \citenamefont {Cano},
  \citenamefont {Millis}, \citenamefont {Liu},\ and\ \citenamefont
  {Yang}}]{PhysRevLett.127.246403}%
  \BibitemOpen
  \bibfield  {author} {\bibinfo {author} {\bibfnamefont {J.}~\bibnamefont
  {Wang}}, \bibinfo {author} {\bibfnamefont {J.}~\bibnamefont {Cano}}, \bibinfo
  {author} {\bibfnamefont {A.~J.}\ \bibnamefont {Millis}}, \bibinfo {author}
  {\bibfnamefont {Z.}~\bibnamefont {Liu}},\ and\ \bibinfo {author}
  {\bibfnamefont {B.}~\bibnamefont {Yang}},\ }\bibfield  {title} {\bibinfo
  {title} {Exact landau level description of geometry and interaction in a
  flatband},\ }\href {https://doi.org/10.1103/PhysRevLett.127.246403}
  {\bibfield  {journal} {\bibinfo  {journal} {Phys. Rev. Lett.}\ }\textbf
  {\bibinfo {volume} {127}},\ \bibinfo {pages} {246403} (\bibinfo {year}
  {2021}{\natexlab{a}})}\BibitemShut {NoStop}%
\bibitem [{\citenamefont {Wang}\ \emph
  {et~al.}(2021{\natexlab{b}})\citenamefont {Wang}, \citenamefont {Zheng},
  \citenamefont {Millis},\ and\ \citenamefont
  {Cano}}]{PhysRevResearch.3.023155}%
  \BibitemOpen
  \bibfield  {author} {\bibinfo {author} {\bibfnamefont {J.}~\bibnamefont
  {Wang}}, \bibinfo {author} {\bibfnamefont {Y.}~\bibnamefont {Zheng}},
  \bibinfo {author} {\bibfnamefont {A.~J.}\ \bibnamefont {Millis}},\ and\
  \bibinfo {author} {\bibfnamefont {J.}~\bibnamefont {Cano}},\ }\bibfield
  {title} {\bibinfo {title} {Chiral approximation to twisted bilayer graphene:
  Exact intravalley inversion symmetry, nodal structure, and implications for
  higher magic angles},\ }\href
  {https://doi.org/10.1103/PhysRevResearch.3.023155} {\bibfield  {journal}
  {\bibinfo  {journal} {Phys. Rev. Res.}\ }\textbf {\bibinfo {volume} {3}},\
  \bibinfo {pages} {023155} (\bibinfo {year} {2021}{\natexlab{b}})}\BibitemShut
  {NoStop}%
\bibitem [{\citenamefont {Ledwith}\ \emph {et~al.}(2023)\citenamefont
  {Ledwith}, \citenamefont {Vishwanath},\ and\ \citenamefont
  {Parker}}]{ledwith2022vortexability}%
  \BibitemOpen
  \bibfield  {author} {\bibinfo {author} {\bibfnamefont {P.~J.}\ \bibnamefont
  {Ledwith}}, \bibinfo {author} {\bibfnamefont {A.}~\bibnamefont
  {Vishwanath}},\ and\ \bibinfo {author} {\bibfnamefont {D.~E.}\ \bibnamefont
  {Parker}},\ }\bibfield  {title} {\bibinfo {title} {Vortexability: A unifying
  criterion for ideal fractional chern insulators},\ }\href
  {https://doi.org/10.1103/PhysRevB.108.205144} {\bibfield  {journal} {\bibinfo
   {journal} {Phys. Rev. B}\ }\textbf {\bibinfo {volume} {108}},\ \bibinfo
  {pages} {205144} (\bibinfo {year} {2023})}\BibitemShut {NoStop}%
\bibitem [{\citenamefont {Wang}\ \emph {et~al.}(2023)\citenamefont {Wang},
  \citenamefont {Klevtsov},\ and\ \citenamefont {Liu}}]{jiewang2023}%
  \BibitemOpen
  \bibfield  {author} {\bibinfo {author} {\bibfnamefont {J.}~\bibnamefont
  {Wang}}, \bibinfo {author} {\bibfnamefont {S.}~\bibnamefont {Klevtsov}},\
  and\ \bibinfo {author} {\bibfnamefont {Z.}~\bibnamefont {Liu}},\ }\bibfield
  {title} {\bibinfo {title} {Origin of model fractional chern insulators in all
  topological ideal flatbands: Explicit color-entangled wave function and exact
  density algebra},\ }\href {https://doi.org/10.1103/PhysRevResearch.5.023167}
  {\bibfield  {journal} {\bibinfo  {journal} {Phys. Rev. Res.}\ }\textbf
  {\bibinfo {volume} {5}},\ \bibinfo {pages} {023167} (\bibinfo {year}
  {2023})}\BibitemShut {NoStop}%
\bibitem [{\citenamefont {Dong}\ \emph {et~al.}(2023)\citenamefont {Dong},
  \citenamefont {Ledwith}, \citenamefont {Khalaf}, \citenamefont {Lee},\ and\
  \citenamefont {Vishwanath}}]{dong2023}%
  \BibitemOpen
  \bibfield  {author} {\bibinfo {author} {\bibfnamefont {J.}~\bibnamefont
  {Dong}}, \bibinfo {author} {\bibfnamefont {P.~J.}\ \bibnamefont {Ledwith}},
  \bibinfo {author} {\bibfnamefont {E.}~\bibnamefont {Khalaf}}, \bibinfo
  {author} {\bibfnamefont {J.~Y.}\ \bibnamefont {Lee}},\ and\ \bibinfo {author}
  {\bibfnamefont {A.}~\bibnamefont {Vishwanath}},\ }\bibfield  {title}
  {\bibinfo {title} {Many-body ground states from decomposition of ideal higher
  chern bands: Applications to chirally twisted graphene multilayers},\ }\href
  {https://doi.org/10.1103/PhysRevResearch.5.023166} {\bibfield  {journal}
  {\bibinfo  {journal} {Phys. Rev. Res.}\ }\textbf {\bibinfo {volume} {5}},\
  \bibinfo {pages} {023166} (\bibinfo {year} {2023})}\BibitemShut {NoStop}%
\bibitem [{\citenamefont {Yu}\ \emph {et~al.}(2022)\citenamefont {Yu},
  \citenamefont {Foutty}, \citenamefont {Han}, \citenamefont {Barber},
  \citenamefont {Schattner}, \citenamefont {Watanabe}, \citenamefont
  {Taniguchi}, \citenamefont {Phillips}, \citenamefont {Shen}, \citenamefont
  {Kivelson},\ and\ \citenamefont {Feldman}}]{Yu2022}%
  \BibitemOpen
  \bibfield  {author} {\bibinfo {author} {\bibfnamefont {J.}~\bibnamefont
  {Yu}}, \bibinfo {author} {\bibfnamefont {B.~A.}\ \bibnamefont {Foutty}},
  \bibinfo {author} {\bibfnamefont {Z.}~\bibnamefont {Han}}, \bibinfo {author}
  {\bibfnamefont {M.~E.}\ \bibnamefont {Barber}}, \bibinfo {author}
  {\bibfnamefont {Y.}~\bibnamefont {Schattner}}, \bibinfo {author}
  {\bibfnamefont {K.}~\bibnamefont {Watanabe}}, \bibinfo {author}
  {\bibfnamefont {T.}~\bibnamefont {Taniguchi}}, \bibinfo {author}
  {\bibfnamefont {P.}~\bibnamefont {Phillips}}, \bibinfo {author}
  {\bibfnamefont {Z.-X.}\ \bibnamefont {Shen}}, \bibinfo {author}
  {\bibfnamefont {S.~A.}\ \bibnamefont {Kivelson}},\ and\ \bibinfo {author}
  {\bibfnamefont {B.~E.}\ \bibnamefont {Feldman}},\ }\bibfield  {title}
  {\bibinfo {title} {Correlated hofstadter spectrum and flavour phase diagram
  in magic-angle twisted bilayer graphene},\ }\href
  {https://doi.org/10.1038/s41567-022-01589-w} {\bibfield  {journal} {\bibinfo
  {journal} {Nature Physics}\ }\textbf {\bibinfo {volume} {18}},\ \bibinfo
  {pages} {825} (\bibinfo {year} {2022})}\BibitemShut {NoStop}%
\bibitem [{\citenamefont {Saito}\ \emph {et~al.}(2021)\citenamefont {Saito},
  \citenamefont {Ge}, \citenamefont {Rademaker}, \citenamefont {Watanabe},
  \citenamefont {Taniguchi}, \citenamefont {Abanin},\ and\ \citenamefont
  {Young}}]{Saito2021}%
  \BibitemOpen
  \bibfield  {author} {\bibinfo {author} {\bibfnamefont {Y.}~\bibnamefont
  {Saito}}, \bibinfo {author} {\bibfnamefont {J.}~\bibnamefont {Ge}}, \bibinfo
  {author} {\bibfnamefont {L.}~\bibnamefont {Rademaker}}, \bibinfo {author}
  {\bibfnamefont {K.}~\bibnamefont {Watanabe}}, \bibinfo {author}
  {\bibfnamefont {T.}~\bibnamefont {Taniguchi}}, \bibinfo {author}
  {\bibfnamefont {D.~A.}\ \bibnamefont {Abanin}},\ and\ \bibinfo {author}
  {\bibfnamefont {A.~F.}\ \bibnamefont {Young}},\ }\bibfield  {title} {\bibinfo
  {title} {Hofstadter subband ferromagnetism and symmetry-broken chern
  insulators in twisted bilayer graphene},\ }\href
  {https://doi.org/10.1038/s41567-020-01129-4} {\bibfield  {journal} {\bibinfo
  {journal} {Nature Physics}\ }\textbf {\bibinfo {volume} {17}},\ \bibinfo
  {pages} {478} (\bibinfo {year} {2021})}\BibitemShut {NoStop}%
\bibitem [{\citenamefont {Pierce}\ \emph {et~al.}(2021)\citenamefont {Pierce},
  \citenamefont {Xie}, \citenamefont {Park}, \citenamefont {Khalaf},
  \citenamefont {Lee}, \citenamefont {Cao}, \citenamefont {Parker},
  \citenamefont {Forrester}, \citenamefont {Chen}, \citenamefont {Watanabe},
  \citenamefont {Taniguchi}, \citenamefont {Vishwanath}, \citenamefont
  {Jarillo-Herrero},\ and\ \citenamefont {Yacoby}}]{Pierce2021}%
  \BibitemOpen
  \bibfield  {author} {\bibinfo {author} {\bibfnamefont {A.~T.}\ \bibnamefont
  {Pierce}}, \bibinfo {author} {\bibfnamefont {Y.}~\bibnamefont {Xie}},
  \bibinfo {author} {\bibfnamefont {J.~M.}\ \bibnamefont {Park}}, \bibinfo
  {author} {\bibfnamefont {E.}~\bibnamefont {Khalaf}}, \bibinfo {author}
  {\bibfnamefont {S.~H.}\ \bibnamefont {Lee}}, \bibinfo {author} {\bibfnamefont
  {Y.}~\bibnamefont {Cao}}, \bibinfo {author} {\bibfnamefont {D.~E.}\
  \bibnamefont {Parker}}, \bibinfo {author} {\bibfnamefont {P.~R.}\
  \bibnamefont {Forrester}}, \bibinfo {author} {\bibfnamefont {S.}~\bibnamefont
  {Chen}}, \bibinfo {author} {\bibfnamefont {K.}~\bibnamefont {Watanabe}},
  \bibinfo {author} {\bibfnamefont {T.}~\bibnamefont {Taniguchi}}, \bibinfo
  {author} {\bibfnamefont {A.}~\bibnamefont {Vishwanath}}, \bibinfo {author}
  {\bibfnamefont {P.}~\bibnamefont {Jarillo-Herrero}},\ and\ \bibinfo {author}
  {\bibfnamefont {A.}~\bibnamefont {Yacoby}},\ }\bibfield  {title} {\bibinfo
  {title} {Unconventional sequence of correlated chern insulators in
  magic-angle twisted bilayer graphene},\ }\href
  {https://doi.org/10.1038/s41567-021-01347-4} {\bibfield  {journal} {\bibinfo
  {journal} {Nature Physics}\ }\textbf {\bibinfo {volume} {17}},\ \bibinfo
  {pages} {1210} (\bibinfo {year} {2021})}\BibitemShut {NoStop}%
\bibitem [{\citenamefont {Kometter}\ \emph {et~al.}(2023)\citenamefont
  {Kometter}, \citenamefont {Yu}, \citenamefont {Devakul}, \citenamefont
  {Reddy}, \citenamefont {Zhang}, \citenamefont {Foutty}, \citenamefont
  {Watanabe}, \citenamefont {Taniguchi}, \citenamefont {Fu},\ and\
  \citenamefont {Feldman}}]{Kometter2023}%
  \BibitemOpen
  \bibfield  {author} {\bibinfo {author} {\bibfnamefont {C.~R.}\ \bibnamefont
  {Kometter}}, \bibinfo {author} {\bibfnamefont {J.}~\bibnamefont {Yu}},
  \bibinfo {author} {\bibfnamefont {T.}~\bibnamefont {Devakul}}, \bibinfo
  {author} {\bibfnamefont {A.~P.}\ \bibnamefont {Reddy}}, \bibinfo {author}
  {\bibfnamefont {Y.}~\bibnamefont {Zhang}}, \bibinfo {author} {\bibfnamefont
  {B.~A.}\ \bibnamefont {Foutty}}, \bibinfo {author} {\bibfnamefont
  {K.}~\bibnamefont {Watanabe}}, \bibinfo {author} {\bibfnamefont
  {T.}~\bibnamefont {Taniguchi}}, \bibinfo {author} {\bibfnamefont
  {L.}~\bibnamefont {Fu}},\ and\ \bibinfo {author} {\bibfnamefont {B.~E.}\
  \bibnamefont {Feldman}},\ }\bibfield  {title} {\bibinfo {title} {Hofstadter
  states and re-entrant charge order in a semiconductor moiré lattice},\
  }\href {https://doi.org/10.1038/s41567-023-02195-0} {\bibfield  {journal}
  {\bibinfo  {journal} {Nature Physics}\ }\textbf {\bibinfo {volume} {19}},\
  \bibinfo {pages} {1861} (\bibinfo {year} {2023})}\BibitemShut {NoStop}%
\bibitem [{\citenamefont {Dean}\ \emph {et~al.}(2013)\citenamefont {Dean},
  \citenamefont {Wang}, \citenamefont {Maher}, \citenamefont {Forsythe},
  \citenamefont {Ghahari}, \citenamefont {Gao}, \citenamefont {Katoch},
  \citenamefont {Ishigami}, \citenamefont {Moon}, \citenamefont {Koshino},
  \citenamefont {Taniguchi}, \citenamefont {Watanabe}, \citenamefont {Shepard},
  \citenamefont {Hone},\ and\ \citenamefont {Kim}}]{Dean2013}%
  \BibitemOpen
  \bibfield  {author} {\bibinfo {author} {\bibfnamefont {C.~R.}\ \bibnamefont
  {Dean}}, \bibinfo {author} {\bibfnamefont {L.}~\bibnamefont {Wang}}, \bibinfo
  {author} {\bibfnamefont {P.}~\bibnamefont {Maher}}, \bibinfo {author}
  {\bibfnamefont {C.}~\bibnamefont {Forsythe}}, \bibinfo {author}
  {\bibfnamefont {F.}~\bibnamefont {Ghahari}}, \bibinfo {author} {\bibfnamefont
  {Y.}~\bibnamefont {Gao}}, \bibinfo {author} {\bibfnamefont {J.}~\bibnamefont
  {Katoch}}, \bibinfo {author} {\bibfnamefont {M.}~\bibnamefont {Ishigami}},
  \bibinfo {author} {\bibfnamefont {P.}~\bibnamefont {Moon}}, \bibinfo {author}
  {\bibfnamefont {M.}~\bibnamefont {Koshino}}, \bibinfo {author} {\bibfnamefont
  {T.}~\bibnamefont {Taniguchi}}, \bibinfo {author} {\bibfnamefont
  {K.}~\bibnamefont {Watanabe}}, \bibinfo {author} {\bibfnamefont {K.~L.}\
  \bibnamefont {Shepard}}, \bibinfo {author} {\bibfnamefont {J.}~\bibnamefont
  {Hone}},\ and\ \bibinfo {author} {\bibfnamefont {P.}~\bibnamefont {Kim}},\
  }\bibfield  {title} {\bibinfo {title} {Hofstadter butterfly and
  the fractal quantum hall effect in moire superlattices},\ }\href
  {https://doi.org/10.1038/nature12186} {\bibfield  {journal} {\bibinfo
  {journal} {Nature}\ }\textbf {\bibinfo {volume} {497}},\ \bibinfo {pages}
  {598} (\bibinfo {year} {2013})}\BibitemShut {NoStop}%
\bibitem [{\citenamefont {Sheffer}\ and\ \citenamefont
  {Stern}(2021)}]{sheffer2021}%
  \BibitemOpen
  \bibfield  {author} {\bibinfo {author} {\bibfnamefont {Y.}~\bibnamefont
  {Sheffer}}\ and\ \bibinfo {author} {\bibfnamefont {A.}~\bibnamefont
  {Stern}},\ }\bibfield  {title} {\bibinfo {title} {Chiral magic-angle twisted
  bilayer graphene in a magnetic field: Landau level correspondence, exact wave
  functions, and fractional chern insulators},\ }\href
  {https://doi.org/10.1103/PhysRevB.104.L121405} {\bibfield  {journal}
  {\bibinfo  {journal} {Phys. Rev. B}\ }\textbf {\bibinfo {volume} {104}},\
  \bibinfo {pages} {L121405} (\bibinfo {year} {2021})}\BibitemShut {NoStop}%
\bibitem [{\citenamefont {Wang}\ and\ \citenamefont
  {Vafek}(2022)}]{PhysRevB.106.L121111}%
  \BibitemOpen
  \bibfield  {author} {\bibinfo {author} {\bibfnamefont {X.}~\bibnamefont
  {Wang}}\ and\ \bibinfo {author} {\bibfnamefont {O.}~\bibnamefont {Vafek}},\
  }\bibfield  {title} {\bibinfo {title} {Narrow bands in magnetic field and
  strong-coupling hofstadter spectra},\ }\href
  {https://doi.org/10.1103/PhysRevB.106.L121111} {\bibfield  {journal}
  {\bibinfo  {journal} {Phys. Rev. B}\ }\textbf {\bibinfo {volume} {106}},\
  \bibinfo {pages} {L121111} (\bibinfo {year} {2022})}\BibitemShut {NoStop}%
\bibitem [{\citenamefont {Singh}\ \emph {et~al.}()\citenamefont {Singh},
  \citenamefont {Chew}, \citenamefont {Herzog-Arbeitman}, \citenamefont
  {Bernevig},\ and\ \citenamefont {Vafek}}]{singh2024topological}%
  \BibitemOpen
  \bibfield  {author} {\bibinfo {author} {\bibfnamefont {K.}~\bibnamefont
  {Singh}}, \bibinfo {author} {\bibfnamefont {A.}~\bibnamefont {Chew}},
  \bibinfo {author} {\bibfnamefont {J.}~\bibnamefont {Herzog-Arbeitman}},
  \bibinfo {author} {\bibfnamefont {B.~A.}\ \bibnamefont {Bernevig}},\ and\
  \bibinfo {author} {\bibfnamefont {O.}~\bibnamefont {Vafek}},\ }\href@noop {}
  {\bibinfo {title} {Topological heavy fermions in magnetic field}},\ \Eprint
  {https://arxiv.org/abs/2305.08171} {arXiv:2305.08171 [cond-mat.str-el]}
  \BibitemShut {NoStop}%
\bibitem [{\citenamefont {Herzog-Arbeitman}\ \emph
  {et~al.}(2022{\natexlab{a}})\citenamefont {Herzog-Arbeitman}, \citenamefont
  {Chew}, \citenamefont {Efetov},\ and\ \citenamefont
  {Bernevig}}]{PhysRevLett.129.076401}%
  \BibitemOpen
  \bibfield  {author} {\bibinfo {author} {\bibfnamefont {J.}~\bibnamefont
  {Herzog-Arbeitman}}, \bibinfo {author} {\bibfnamefont {A.}~\bibnamefont
  {Chew}}, \bibinfo {author} {\bibfnamefont {D.~K.}\ \bibnamefont {Efetov}},\
  and\ \bibinfo {author} {\bibfnamefont {B.~A.}\ \bibnamefont {Bernevig}},\
  }\bibfield  {title} {\bibinfo {title} {Reentrant correlated insulators in
  twisted bilayer graphene at 25 t ($2\ensuremath{\pi}$ flux)},\ }\href
  {https://doi.org/10.1103/PhysRevLett.129.076401} {\bibfield  {journal}
  {\bibinfo  {journal} {Phys. Rev. Lett.}\ }\textbf {\bibinfo {volume} {129}},\
  \bibinfo {pages} {076401} (\bibinfo {year} {2022}{\natexlab{a}})}\BibitemShut
  {NoStop}%
\bibitem [{\citenamefont {Herzog-Arbeitman}\ \emph
  {et~al.}(2022{\natexlab{b}})\citenamefont {Herzog-Arbeitman}, \citenamefont
  {Chew},\ and\ \citenamefont {Bernevig}}]{herzog-arbeitman2022}%
  \BibitemOpen
  \bibfield  {author} {\bibinfo {author} {\bibfnamefont {J.}~\bibnamefont
  {Herzog-Arbeitman}}, \bibinfo {author} {\bibfnamefont {A.}~\bibnamefont
  {Chew}},\ and\ \bibinfo {author} {\bibfnamefont {B.~A.}\ \bibnamefont
  {Bernevig}},\ }\bibfield  {title} {\bibinfo {title} {Magnetic bloch theorem
  and reentrant flat bands in twisted bilayer graphene at $2\ensuremath{\pi}$
  flux},\ }\href {https://doi.org/10.1103/PhysRevB.106.085140} {\bibfield
  {journal} {\bibinfo  {journal} {Phys. Rev. B}\ }\textbf {\bibinfo {volume}
  {106}},\ \bibinfo {pages} {085140} (\bibinfo {year}
  {2022}{\natexlab{b}})}\BibitemShut {NoStop}%
\bibitem [{\citenamefont {Popov}\ and\ \citenamefont
  {Milekhin}(2021)}]{PhysRevB.103.155150}%
  \BibitemOpen
  \bibfield  {author} {\bibinfo {author} {\bibfnamefont {F.~K.}\ \bibnamefont
  {Popov}}\ and\ \bibinfo {author} {\bibfnamefont {A.}~\bibnamefont
  {Milekhin}},\ }\bibfield  {title} {\bibinfo {title} {Hidden wave function of
  twisted bilayer graphene: The flat band as a landau level},\ }\href
  {https://doi.org/10.1103/PhysRevB.103.155150} {\bibfield  {journal} {\bibinfo
   {journal} {Phys. Rev. B}\ }\textbf {\bibinfo {volume} {103}},\ \bibinfo
  {pages} {155150} (\bibinfo {year} {2021})}\BibitemShut {NoStop}%
\bibitem [{\citenamefont {Tarnopolsky}\ \emph {et~al.}(2019)\citenamefont
  {Tarnopolsky}, \citenamefont {Kruchkov},\ and\ \citenamefont
  {Vishwanath}}]{PhysRevLett.122.106405}%
  \BibitemOpen
  \bibfield  {author} {\bibinfo {author} {\bibfnamefont {G.}~\bibnamefont
  {Tarnopolsky}}, \bibinfo {author} {\bibfnamefont {A.~J.}\ \bibnamefont
  {Kruchkov}},\ and\ \bibinfo {author} {\bibfnamefont {A.}~\bibnamefont
  {Vishwanath}},\ }\bibfield  {title} {\bibinfo {title} {Origin of magic angles
  in twisted bilayer graphene},\ }\href
  {https://doi.org/10.1103/PhysRevLett.122.106405} {\bibfield  {journal}
  {\bibinfo  {journal} {Phys. Rev. Lett.}\ }\textbf {\bibinfo {volume} {122}},\
  \bibinfo {pages} {106405} (\bibinfo {year} {2019})}\BibitemShut {NoStop}%
\bibitem [{\citenamefont {Vafek}\ and\ \citenamefont
  {Kang}(2020)}]{Oskar_RG_2020}%
  \BibitemOpen
  \bibfield  {author} {\bibinfo {author} {\bibfnamefont {O.}~\bibnamefont
  {Vafek}}\ and\ \bibinfo {author} {\bibfnamefont {J.}~\bibnamefont {Kang}},\
  }\bibfield  {title} {\bibinfo {title} {Renormalization group study of hidden
  symmetry in twisted bilayer graphene with coulomb interactions},\ }\href
  {https://doi.org/10.1103/PhysRevLett.125.257602} {\bibfield  {journal}
  {\bibinfo  {journal} {Phys. Rev. Lett.}\ }\textbf {\bibinfo {volume} {125}},\
  \bibinfo {pages} {257602} (\bibinfo {year} {2020})}\BibitemShut {NoStop}%
\bibitem [{\citenamefont {Guerci}\ \emph
  {et~al.}\citenamefont {Guerci}, \citenamefont {Mao},\
  and\ \citenamefont {Mora}}]{guerci2023nature}%
  \BibitemOpen
  \bibfield  {author} {\bibinfo {author} {\bibfnamefont {D.}~\bibnamefont
  {Guerci}}, \bibinfo {author} {\bibfnamefont {Y.}~\bibnamefont {Mao}},\ and\
  \bibinfo {author} {\bibfnamefont {C.}~\bibnamefont {Mora}},\ }\bibfield  {title} {\bibinfo {title} {Nature of even and odd magic angles in helical twisted trilayer graphene},\ }\href
  {https://doi.org/10.1103/PhysRevB.109.205411} {\bibfield  {journal} {\bibinfo
   {journal} {Phys. Rev. B}\ }\textbf {\bibinfo {volume} {109}},\ \bibinfo
  {pages} {205411} (\bibinfo {year} {2024})}\BibitemShut {NoStop}%
\bibitem [{\citenamefont {Popov}\ and\ \citenamefont
  {Tarnopolsky}(2023{\natexlab{b}})}]{popov2023magic}%
  \BibitemOpen
  \bibfield  {author} {\bibinfo {author} {\bibfnamefont {F.~K.}\ \bibnamefont
  {Popov}}\ and\ \bibinfo {author} {\bibfnamefont {G.}~\bibnamefont
  {Tarnopolsky}},\ }\bibfield  {title} {\bibinfo {title} {Magic angle butterfly
  in twisted trilayer graphene},\ }\href
  {https://doi.org/10.1103/PhysRevResearch.5.043079} {\bibfield  {journal}
  {\bibinfo  {journal} {Phys. Rev. Res.}\ }\textbf {\bibinfo {volume} {5}},\
  \bibinfo {pages} {043079} (\bibinfo {year} {2023}{\natexlab{b}})}\BibitemShut
  {NoStop}%
\bibitem [{\citenamefont {Bistritzer}\ and\ \citenamefont
  {MacDonald}(2011)}]{PhysRevB.84.035440}%
  \BibitemOpen
  \bibfield  {author} {\bibinfo {author} {\bibfnamefont {R.}~\bibnamefont
  {Bistritzer}}\ and\ \bibinfo {author} {\bibfnamefont {A.~H.}\ \bibnamefont
  {MacDonald}},\ }\bibfield  {title} {\bibinfo {title} {Moir\'e butterflies in
  twisted bilayer graphene},\ }\href
  {https://doi.org/10.1103/PhysRevB.84.035440} {\bibfield  {journal} {\bibinfo
  {journal} {Phys. Rev. B}\ }\textbf {\bibinfo {volume} {84}},\ \bibinfo
  {pages} {035440} (\bibinfo {year} {2011})}\BibitemShut {NoStop}%
\bibitem [{\citenamefont {Hejazi}\ \emph {et~al.}(2019)\citenamefont {Hejazi},
  \citenamefont {Liu},\ and\ \citenamefont {Balents}}]{PhysRevB.100.035115}%
  \BibitemOpen
  \bibfield  {author} {\bibinfo {author} {\bibfnamefont {K.}~\bibnamefont
  {Hejazi}}, \bibinfo {author} {\bibfnamefont {C.}~\bibnamefont {Liu}},\ and\
  \bibinfo {author} {\bibfnamefont {L.}~\bibnamefont {Balents}},\ }\bibfield
  {title} {\bibinfo {title} {Landau levels in twisted bilayer graphene and
  semiclassical orbits},\ }\href {https://doi.org/10.1103/PhysRevB.100.035115}
  {\bibfield  {journal} {\bibinfo  {journal} {Phys. Rev. B}\ }\textbf {\bibinfo
  {volume} {100}},\ \bibinfo {pages} {035115} (\bibinfo {year}
  {2019})}\BibitemShut {NoStop}%
\bibitem [{\citenamefont {Streda}(1982)}]{Streda1982}%
  \BibitemOpen
  \bibfield  {author} {\bibinfo {author} {\bibfnamefont {P.}~\bibnamefont
  {Streda}},\ }\bibfield  {title} {\bibinfo {title} {Theory of quantised hall
  conductivity in two dimensions},\ }\href
  {https://doi.org/10.1088/0022-3719/15/22/005} {\bibfield  {journal} {\bibinfo
   {journal} {Journal of Physics C: Solid State Physics}\ }\textbf {\bibinfo
  {volume} {15}},\ \bibinfo {pages} {L717} (\bibinfo {year}
  {1982})}\BibitemShut {NoStop}%
\bibitem [{\citenamefont {Atiyah}\ and\ \citenamefont
  {Singer}(1968)}]{Atiyah1968}%
  \BibitemOpen
  \bibfield  {author} {\bibinfo {author} {\bibfnamefont {M.~F.}\ \bibnamefont
  {Atiyah}}\ and\ \bibinfo {author} {\bibfnamefont {I.~M.}\ \bibnamefont
  {Singer}},\ }\bibfield  {title} {\bibinfo {title} {The index of elliptic
  operators: Iii},\ }\href {https://doi.org/10.2307/1970717} {\bibfield
  {journal} {\bibinfo  {journal} {Annals of Mathematics}\ }\textbf {\bibinfo
  {volume} {87}},\ \bibinfo {pages} {546} (\bibinfo {year} {1968})}\BibitemShut
  {NoStop}%
\bibitem [{\citenamefont {Crépel}\ \emph {et~al.}(2024)\citenamefont
  {Crépel}, \citenamefont {Ding}, \citenamefont {Verma}, \citenamefont
  {Regnault},\ and\ \citenamefont {Queiroz}}]{crepel2024topologically}%
  \BibitemOpen
  \bibfield  {author} {\bibinfo {author} {\bibfnamefont {V.}~\bibnamefont
  {Crépel}}, \bibinfo {author} {\bibfnamefont {P.}~\bibnamefont {Ding}},
  \bibinfo {author} {\bibfnamefont {N.}~\bibnamefont {Verma}}, \bibinfo
  {author} {\bibfnamefont {N.}~\bibnamefont {Regnault}},\ and\ \bibinfo
  {author} {\bibfnamefont {R.}~\bibnamefont {Queiroz}},\ }\href@noop {}
  {\bibinfo {title} {Topologically protected flatness in chiral moir\'e
  heterostructures}} (\bibinfo {year} {2024}),\ \Eprint
  {https://arxiv.org/abs/2403.19656} {arXiv:2403.19656 [cond-mat.mes-hall]}
  \BibitemShut {NoStop}%
\bibitem [{\citenamefont {Estienne}\ \emph {et~al.}(2023)\citenamefont
  {Estienne}, \citenamefont {Regnault},\ and\ \citenamefont
  {Cr\'epel}}]{estienne2023}%
  \BibitemOpen
  \bibfield  {author} {\bibinfo {author} {\bibfnamefont {B.}~\bibnamefont
  {Estienne}}, \bibinfo {author} {\bibfnamefont {N.}~\bibnamefont {Regnault}},\
  and\ \bibinfo {author} {\bibfnamefont {V.}~\bibnamefont {Cr\'epel}},\
  }\bibfield  {title} {\bibinfo {title} {Ideal chern bands as landau levels in
  curved space},\ }\href {https://doi.org/10.1103/PhysRevResearch.5.L032048}
  {\bibfield  {journal} {\bibinfo  {journal} {Phys. Rev. Res.}\ }\textbf
  {\bibinfo {volume} {5}},\ \bibinfo {pages} {L032048} (\bibinfo {year}
  {2023})}\BibitemShut {NoStop}%
\bibitem [{\citenamefont {Ledwith}\ \emph {et~al.}(2021)\citenamefont
  {Ledwith}, \citenamefont {Khalaf},\ and\ \citenamefont
  {Vishwanath}}]{LEDWITH2021168646}%
  \BibitemOpen
  \bibfield  {author} {\bibinfo {author} {\bibfnamefont {P.~J.}\ \bibnamefont
  {Ledwith}}, \bibinfo {author} {\bibfnamefont {E.}~\bibnamefont {Khalaf}},\
  and\ \bibinfo {author} {\bibfnamefont {A.}~\bibnamefont {Vishwanath}},\
  }\bibfield  {title} {\bibinfo {title} {Strong coupling theory of magic-angle
  graphene: A pedagogical introduction},\ }\href
  {https://doi.org/https://doi.org/10.1016/j.aop.2021.168646} {\bibfield
  {journal} {\bibinfo  {journal} {Annals of Physics}\ }\textbf {\bibinfo
  {volume} {435}},\ \bibinfo {pages} {168646} (\bibinfo {year} {2021})},\
  \bibinfo {note} {special issue on Philip W. Anderson}\BibitemShut {NoStop}%
\bibitem [{\citenamefont {Liu}\ \emph {et~al.}(2019)\citenamefont {Liu},
  \citenamefont {Liu},\ and\ \citenamefont {Dai}}]{Liu2019}%
  \BibitemOpen
  \bibfield  {author} {\bibinfo {author} {\bibfnamefont {J.}~\bibnamefont
  {Liu}}, \bibinfo {author} {\bibfnamefont {J.}~\bibnamefont {Liu}},\ and\
  \bibinfo {author} {\bibfnamefont {X.}~\bibnamefont {Dai}},\ }\bibfield
  {title} {\bibinfo {title} {Pseudo landau level representation of twisted
  bilayer graphene: Band topology and implications on the correlated insulating
  phase},\ }\href {https://doi.org/10.1103/PhysRevB.99.155415} {\bibfield
  {journal} {\bibinfo  {journal} {Phys. Rev. B}\ }\textbf {\bibinfo {volume}
  {99}},\ \bibinfo {pages} {155415} (\bibinfo {year} {2019})}\BibitemShut
  {NoStop}%
\bibitem [{\citenamefont {Bultinck}\ \emph {et~al.}(2020)\citenamefont
  {Bultinck}, \citenamefont {Khalaf}, \citenamefont {Liu}, \citenamefont
  {Chatterjee}, \citenamefont {Vishwanath},\ and\ \citenamefont
  {Zaletel}}]{bultinck2020}%
  \BibitemOpen
  \bibfield  {author} {\bibinfo {author} {\bibfnamefont {N.}~\bibnamefont
  {Bultinck}}, \bibinfo {author} {\bibfnamefont {E.}~\bibnamefont {Khalaf}},
  \bibinfo {author} {\bibfnamefont {S.}~\bibnamefont {Liu}}, \bibinfo {author}
  {\bibfnamefont {S.}~\bibnamefont {Chatterjee}}, \bibinfo {author}
  {\bibfnamefont {A.}~\bibnamefont {Vishwanath}},\ and\ \bibinfo {author}
  {\bibfnamefont {M.~P.}\ \bibnamefont {Zaletel}},\ }\bibfield  {title}
  {\bibinfo {title} {Ground state and hidden symmetry of magic-angle graphene
  at even integer filling},\ }\href
  {https://doi.org/10.1103/PhysRevX.10.031034} {\bibfield  {journal} {\bibinfo
  {journal} {Phys. Rev. X}\ }\textbf {\bibinfo {volume} {10}},\ \bibinfo
  {pages} {031034} (\bibinfo {year} {2020})}\BibitemShut {NoStop}%
\bibitem [{\citenamefont {Haldane}\ and\ \citenamefont
  {Rezayi}(1985)}]{Haldane1985}%
  \BibitemOpen
  \bibfield  {author} {\bibinfo {author} {\bibfnamefont {F.~D.~M.}\
  \bibnamefont {Haldane}}\ and\ \bibinfo {author} {\bibfnamefont {E.~H.}\
  \bibnamefont {Rezayi}},\ }\bibfield  {title} {\bibinfo {title} {Periodic
  laughlin-jastrow wave functions for the fractional quantized hall effect},\
  }\href {https://doi.org/10.1103/PhysRevB.31.2529} {\bibfield  {journal}
  {\bibinfo  {journal} {Phys. Rev. B}\ }\textbf {\bibinfo {volume} {31}},\
  \bibinfo {pages} {2529} (\bibinfo {year} {1985})}\BibitemShut {NoStop}%
\bibitem [{\citenamefont {Makov}\ \emph {et~al.}(2024)\citenamefont {Makov},
  \citenamefont {Guinea},\ and\ \citenamefont {Stern}}]{makov2024}%
  \BibitemOpen
  \bibfield  {author} {\bibinfo {author} {\bibfnamefont {R.}~\bibnamefont
  {Makov}}, \bibinfo {author} {\bibfnamefont {F.}~\bibnamefont {Guinea}},\ and\
  \bibinfo {author} {\bibfnamefont {A.}~\bibnamefont {Stern}},\ }\href@noop {}
  {\bibinfo {title} {Flat bands in chiral multilayer graphene}} (\bibinfo
  {year} {2024}),\ \Eprint {https://arxiv.org/abs/2404.15759} {arXiv:2404.15759
  [cond-mat.mes-hall]} \BibitemShut {NoStop}%
\end{thebibliography}%

\end{document}